\documentclass[11pt]{article}
\pdfoutput=1 
\usepackage{amssymb}
\usepackage{amsmath}
\usepackage{amstext}
\usepackage{graphicx}
\usepackage{epsfig}
\usepackage{verbatim} 
\usepackage{fancybox}
\usepackage{color}
\usepackage{mathdots}
\usepackage{ulem}
\usepackage{enumitem}
\usepackage{youngtab}
\usepackage{subfigure}
\usepackage{bbm}
\usepackage{parskip}
\usepackage[numbers,sort&compress]{natbib}
\usepackage{ytableau}
\usepackage[utf8]{inputenc}
\usepackage{tikz}
\usepackage{simplewick}
\usetikzlibrary{positioning, trees,decorations, decorations.markings, decorations.pathmorphing, arrows, shapes.geometric, snakes}

\newcommand{\Comment}[1]{{}}
\definecolor{darkblue}{rgb}{0.15,0.35,0.55}
\definecolor{reddish}{rgb}{0.65, 0.2, 0.2}
\usepackage[linktocpage=true]{hyperref}
\hypersetup{
colorlinks=true,
citecolor=darkblue,
linkcolor=reddish,
urlcolor=darkblue,
pdfauthor={},
pdftitle={},
pdfsubject={}
}

\newcommand{\be}{\begin{equation}}
\newcommand{\ee}{\end{equation}}
\newcommand{\nn}{\nonumber}
\newcommand{\ccdot}{\! \cdot \!}
\newcommand{\intN}{\int_{\mathcal{M}} }

\numberwithin{equation}{section}

\makeatletter
\def\thickhline{%
  \noalign{\ifnum0=`}\fi\hrule \@height \thickarrayrulewidth \futurelet
   \reserved@a\@xthickhline}
\def\@xthickhline{\ifx\reserved@a\thickhline
               \vskip\doublerulesep
               \vskip-\thickarrayrulewidth
             \fi
      \ifnum0=`{\fi}}
\makeatother

\newlength{\thickarrayrulewidth}
\begin{document}

\renewcommand{\thefootnote}{\fnsymbol{footnote}}
~
\vspace{1.75truecm}
\thispagestyle{empty}
\begin{center}
{\LARGE \bf{
Bootstrap Bounds on Closed Einstein Manifolds 
}}
\end{center} 

\vspace{1cm}
\centerline{\Large James Bonifacio,\footnote{\href{mailto:james.bonifacio@case.edu}{\texttt{james.bonifacio@case.edu}}}
and Kurt Hinterbichler,\footnote{\href{mailto:kurt.hinterbichler@case.edu} {\texttt{kurt.hinterbichler@case.edu}}}
}
\vspace{.5cm}

\centerline{{\it CERCA, Department of Physics,}}
 \centerline{{\it Case Western Reserve University, 10900 Euclid Ave, Cleveland, OH 44106}} 
 \vspace{.25cm}

\vspace{1cm}
\begin{abstract}
\noindent
A compact Riemannian manifold is associated with geometric data given by the eigenvalues of various Laplacian operators on the manifold and the triple overlap integrals of the corresponding eigenmodes.
This geometric data must satisfy certain consistency conditions that follow from associativity and the completeness of eigenmodes.
We show that it is possible to obtain nontrivial bounds on the geometric data of closed Einstein manifolds by using semidefinite programming to study these consistency conditions, in analogy to the conformal bootstrap bounds on conformal field theories. These bootstrap bounds translate to constraints on the tree-level masses and cubic couplings of Kaluza--Klein modes in theories with compact extra dimensions.
We show that in some cases the bounds are saturated by known manifolds.
\end{abstract}

\newpage

\setcounter{tocdepth}{3}
\tableofcontents
\renewcommand*{\thefootnote}{\arabic{footnote}}
\setcounter{footnote}{0}

\section{Introduction}

There are several cases of historically intractable problems in physics which have seen significant recent progress arise from a numerical bootstrap approach.  To employ such an approach, one starts with some general principles, derives from them consistency conditions that the observable quantities of interest must satisfy, and then solves these consistency conditions numerically on a computer. This gives general constraints on the possible values that the observables can take.
Examples of interesting physical systems that have been recently studied using such methods include strongly coupled conformal field theories (CFTs) \cite{Rattazzi:2008pe}, scattering amplitudes \cite{Paulos:2016but,Paulos:2017fhb}, and random matrix models \cite{Lin:2020mme}.  These studies have given information about these systems which is generally out of reach of traditional perturbative methods. 

In this paper we show that it is possible to apply a similar numerical bootstrap approach to obtain nontrivial mathematical results about closed Einstein manifolds.  
A closed Einstein manifold $\mathcal{M}$ is a compact smooth manifold without boundary that is equipped with a metric $g_{mn}$ whose Ricci curvature satisfies $R_{m n} = \lambda g_{m n}$ for some constant $\lambda$.
We limit our consideration to closed Einstein manifolds that are connected and orientable, as well as the quotients of such manifolds by isometry subgroups. 
We find new bounds on the geometric data associated with such manifolds.  Specifically, the bounds involve the spectrum of eigenvalues of various Laplacian operators and the triple overlap integrals of their eigenmodes.

Our bounds come from studying certain consistency conditions involving the geometric data. For a simple example of these consistency conditions, consider the scalar Laplacian, $\Delta$. A scalar eigenfunction $\psi_a$ with eigenvalue $\lambda_a$ satisfies the eigenvalue equation $\Delta \psi_a = \lambda_a \psi_a$. For any three eigenfunctions we can construct the following triple overlap integral:
\be \label{eq:3ptdef}
g_{a_1 a_2 a_3} \equiv \intN \psi_{a_1} \psi_{a_2} \psi_{a_3}.
\ee
The eigenfunctions can be chosen so that they form an orthonormal basis, so we can write the product $\psi_{a_1} \psi_{a_2} $ as a sum over eigenmodes
\be
\psi_{a_1} \psi_{a_2} =\sum_{a} g_{a_1 a_2}{}^{ a}\psi_{a} \,, \label{eq:OPEexpansion}
\ee
where orthonormality gives the coefficients as the triple overlap integrals \eqref{eq:3ptdef}.
By repeatedly applying \eqref{eq:OPEexpansion} to pairs of eigenmodes, integrals involving products of four or more eigenfunctions can be reduced to sums of products of the triple overlap integrals. For example, we can write
\be \label{eq:OPEintegral}
\int_{\mathcal{M}} \psi_{a_1} \psi_{a_2} \psi_{a_3} \psi_{a_4}
=
\sum_a g_{a_1 a_2}{}{}^{ a} g_{a_3 a_4 a}  \, , 
\ee
where we used the completeness relation \eqref{eq:OPEexpansion} on $\psi_{a_1} \psi_{a_2}$ and $\psi_{a_3} \psi_{a_4}$.
Now the key point is that this expansion can be done in other ways by pairing different eigenfunctions. Equating the resulting expressions then gives consistency conditions for the triple overlap integrals. For example, exchanging $a_2 \leftrightarrow a_3$ in Eq.~\eqref{eq:OPEintegral} and equating with the original result gives
\be \label{eq:simpleCC}
\sum_a g_{a_1 a_2}{}{}^{ a} g_{a_3 a_4 a} = \sum_a g_{a_1 a_3 }{}{}^{ a} g_{a_2 a_4 a} \, .
\ee

Consistency conditions of the form \eqref{eq:simpleCC} will be the basic relations that we exploit to get bootstrap bounds on the geometric data.  By considering quartic overlap integrals involving derivatives of scalar eigenfunctions, we will find consistency conditions that are much more complicated and involve various triple overlap integrals and eigenvalues.
The consistency conditions usually cannot be solved directly, but we can extract nontrivial bounds by reformulating them as a semidefinite program.  An example of such a bound is shown 
in Figure~\ref{fig:eigenvalue-bound-1}, which shows numerical upper bounds on the ratio of the first two distinct non-vanishing eigenvalues of the scalar Laplacian on closed Einstein manifolds with non-negative scalar curvature as a function of the smallest Lichnerowicz eigenvalue on transverse traceless tensors.  We obtain many different bounds of this sort on eigenvalues and triple overlap integrals, both numerical and exact.

There is a sharp analogy between this geometry bootstrap and the modern conformal bootstrap \cite{Rattazzi:2008pe} (see Ref.~\cite{Poland:2018epd} for a recent review), which heavily inspires our approach. The geometric data we consider is analogous to the conformal data that defines a CFT: the eigenmodes are analogous to CFT operators, their eigenvalues are analogous to the conformal dimensions, and their triple overlap integrals are analogous to the operator product expansion (OPE) coefficients.  The eigenfunction expansions such as \eqref{eq:OPEexpansion} are like the OPE in a CFT and the consistency conditions such as \eqref{eq:simpleCC} are like the crossing relations that form the basis of the modern CFT bootstrap. See Table~\ref{tab:analogy} for a summary of this analogy.
\begin{table}[ht]
\centering
  \begin{tabular}{ c | c }
   Einstein manifolds & CFTs \\ \thickhline
Laplacian eigenmodes & Primary operators \\
Eigenvalues & Scaling dimensions \\
Triple overlap integrals & OPE coefficients \\
Eigenfunction expansion & OPE \\
Consistency conditions & Crossing equations \\
Lichnerowicz bound & Unitarity bound
\end{tabular}
\caption{An analogy between Einstein manifolds and CFTs.}
\label{tab:analogy}
\end{table}

There are many existing results concerning the spectrum of eigenvalues of Laplacian operators on compact Riemannian manifolds \cite{chavel1984}, but bounds on the triple overlap integrals are less well studied. 
As far as we know, the bounds we find for Einstein manifolds are not known in the mathematical literature.
Often it is difficult to determine explicitly the eigenvalues and eigenfunctions of a given manifold (see Appendix \ref{app:examples} for some exceptions). For example, for compact Calabi--Yau manifolds there are no nontrivial examples for which the spectrum of the scalar Laplacian is known analytically and numerical methods must be used \cite{Headrick:2005ch}. One numerical method for computing the low-lying eigenvalues is to use Donaldson's algorithm to approximate the Ricci-flat K\"ahler metric \cite{Donaldson2005, Douglas:2006rr}, as in Ref.~\cite{Braun:2008jp}.  This is what one might call a Monte Carlo approach, where accurate results are obtained for a specific manifold.  In contrast, the bootstrap approach is able to find bounds that apply to all Einstein manifolds satisfying a few assumptions.

While the results we derive can be considered as purely mathematical results about closed Einstein manifolds, there is also a physical interpretation in terms of Kaluza--Klein theories. Starting with a higher-dimensional theory on a product spacetime, one can integrate over any compact extra dimensions to obtain an equivalent lower-dimensional theory. This lower-dimensional theory will contain towers of particles of various spins which correspond to the excitations of the higher-dimensional fields in the extra dimensions. The masses of these Kaluza--Klein modes are determined by the eigenvalues of Laplacians on the internal manifold and the strengths of their three-point interactions are determined by the triple overlap integrals of the eigenmodes.   In the case where the internal manifold is Einstein, the mathematical bounds on eigenvalues and overlap integrals we obtain thus translate into bounds on the tree-level masses and cubic interactions of these Kaluza--Klein modes. This is the case, for example, in reductions of pure gravity and for Calabi--Yau reductions of supergravity. 

\vspace{.5cm}
\noindent {\bf Conventions:} We use the Einstein summation convention for spatial indices $n, m, \dots$ but not for the indices $a$, $i$, $\mathcal{I}$ labelling eigenfunctions. We (anti)symmetrize with weight one.  The curvature conventions are those of Ref.~\cite{Carroll:2004st}.

\section{Geometry review}
\label{sec:basics}
We start by setting our notation and reviewing some basic Riemannian geometry that we need for the rest of the paper. We also review the geometric consistency conditions discussed in Ref.~\cite{Bonifacio:2019ioc}.

\subsection{Einstein manifolds}

Let $\mathcal{M}$ be an $N$-dimensional Einstein manifold, with metric $g_{mn}$, that is connected, orientable, and closed (compact without boundary). We denote the volume of $\mathcal{M}$ by $V$ and its Ricci curvature by $R_{mn}$. By definition, an Einstein manifold satisfies
\be \label{eq:Einstein}
R_{m n} = \lambda g_{m n}\, ,
\ee
where $\lambda$ is a real constant.  Taking the trace of Eq.~\eqref{eq:Einstein} shows that the Ricci scalar $R$ is also constant, given by $R=N\lambda$. Einstein manifolds with $R=0$ are called Ricci-flat manifolds.

Einstein manifolds are solutions to the vacuum Einstein equations with a possibly nonzero cosmological constant. For this work we are interested in the case of Riemannian Einstein manifolds and their application as the compact spatial dimensions of a Lorentzian Kaluza--Klein product spacetime.
The case $N=1$ would require a separate treatment for many of our formulae and is already completely understood (it is just the circle), so henceforth we restrict to $N\geq 2$.  See Ref.~\cite{Besse:1987pua} for a classic reference on Einstein manifolds.

\subsection{Laplacians}
We need to introduce three different Laplacian operators on the closed Einstein manifold $\mathcal{M}$: the scalar Laplacian, the vector Hodge Laplacian, and the Lichnerowicz Laplacian on tensors. These are self-adjoint elliptic operators (with respect to the appropriate inner products), so we can use standard results of spectral theory on compact manifolds. For example, on any given manifold each Laplacian has a discrete spectrum that is bounded from below and unbounded from above and there exists a basis of real orthonormal eigenmodes. In Kaluza--Klein theories these eigenmodes correspond to different particles and their eigenvalues determine the masses of the particles (see, e.g., Ref.~\cite{Hinterbichler:2013kwa} for details). 

\subsubsection*{Scalar Laplacian}
We denote by $\psi_a$ an orthonormal basis of non-constant real eigenfunctions of the scalar Laplacian on $\mathcal{M}$, where $a$ is a discrete index labelling the different eigenfunctions. These eigenfunctions satisfy
\be
\Delta \psi_a \equiv - \Box \psi_a  = \lambda_a \psi_a, 
\ee
where $\lambda_a >0$ is the corresponding eigenvalue. Orthonormality implies that
\be
\intN \psi_{a_1} \psi_{a_2} = \delta_{a_1 a_2},
\ee
where $\intN$ denotes the integral over $\mathcal{M}$ with the canonical volume form. The normalized constant eigenfunction is $V^{-1/2}$. This is the unique zero mode for the scalar Laplacian and we now treat it separately from the non-constant eigenfunctions (unlike in the introduction).  Completeness tells us that any $L^2$-normalizable scalar function $\phi$ on ${\cal M}$ can be expanded as
\be \phi=\frac{c^0}{ V^{1/2}}+\sum_a c^a \psi_a,\label{scalarcompletee}
\ee
where $c^0=V^{-1/2} \intN \phi$ and $c^a=\intN \phi \psi^a$.

There are special eigenfunctions of the scalar Laplacian called conformal scalars, which are defined as those scalars whose gradients are conformal Killing vectors that are not Killing vectors. We index these by the set $I_{\rm conf.}$. Conformal scalars satisfy the equation
\be
\left( \nabla_m \nabla_n - \frac{1}{N} g_{m n} \Box \right) \psi_a =0 ,\quad a \in I_{\rm conf.},
\ee
and they exist only on the round spheres~\cite{obata1962}.  The conformal scalars are precisely the $L =1$ spherical harmonics on $S^N$ if $N>1$ (see Appendix~\ref{app:examples}).\footnote{There can be many inequivalent Einstein metrics on spheres and exotic spheres~\cite{Boyer:2003pe}. By $S^N$ we will mean the sphere with the standard round metric unless stated otherwise.} 

On Einstein manifolds with $R>0$ the Lichnerowicz bound gives \cite{Lichnerowicz}
\be \label{eq:Lichbound}
\lambda_a \geq \frac{R}{N-1},
\ee
and this is saturated only by conformal scalars. This is analogous to a CFT unitarity bound.

\subsubsection*{Vector Laplacian}
Denote by $Y_{m, i}$ an orthonormal basis of real, transverse eigenvectors of the one-form Hodge Laplacian on $\mathcal{M}$, where $i$ is a discrete index labelling the different eigenvectors. These eigenvectors satisfy
\be
\Delta Y_{m, i} \equiv -\Box Y_{m, i} +R_m{}^n Y_{n, i} = \lambda_i Y_{m, i}, \quad \nabla^m Y_{m, i} = 0,  
\ee
where $\lambda_i \geq 0$ is the corresponding eigenvalue. Orthonormality implies that
\be
\intN Y_{m, i_1} Y^m_{i_2} = \delta_{i_1 i_2}.
\ee
Using completeness and the Hodge decomposition, any one-form $V_m$ on ${\cal M}$ can be expanded as
\be \label{eq:vectorExpand}
V_m=\sum_i c^i Y_{m, i}+ \sum_a c^a \partial_m \psi_a,
\ee
where $c^i=\intN V_m Y^{m}_i$ and $c^a = \lambda_a^{-1} \intN V_m \partial^m \psi_a$.

Killing vectors are special transverse eigenvectors that generate the isometries of the metric. We index them by the set $I_{\rm Killing}$. They satisfy the Killing equation,
\be
\nabla_{(m} Y_{n), i} =0, \quad i \in I_{\rm Killing}.
\ee
There is another lower bound on the eigenvalues $\lambda_i$ for closed Einstein manifolds,
\be
\lambda_i \geq \frac{2R}{N},
\ee
which is saturated only by Killing vectors.
Since $\lambda_{i} \geq 0$ there are no nontrivial Killing vectors on closed Einstein manifolds with $R < 0$.

\subsubsection*{Lichnerowicz Laplacian}
Lastly, we denote by $h_{mn, \mathcal{I}}^{TT}$ an orthonormal basis of real transverse traceless eigentensors of the Lichnerowicz Laplacian on $\mathcal{M}$, where $\mathcal{I}$ is a discrete index labelling the different eigentensors. On an Einstein manifold, these eigentensors satisfy
\be
\Delta_L h^{TT}_{m n, \mathcal{I}} \equiv -\Box h^{TT}_{m n, \mathcal{I}} + \frac{2R}{N} h^{TT}_{m n, \mathcal{I}} -2 R_{m}{}^p{}_n{}^q h^{TT}_{p q \mathcal{I}} = \lambda_{\mathcal{I}} h^{TT}_{m n, \mathcal{I}}, \quad \nabla^m h^{TT}_{m n, \mathcal{I}} = h^{TT}_m{}^{m}= 0, 
\ee
where $\Delta_L$ is the Lichnerowicz Laplacian and $\lambda_{\mathcal{I}}$ is a Lichnerowicz eigenvalue.\footnote{In this paper the Lichnerowicz eigenvalues will always correspond to transverse traceless symmetric 2-tensors.} Orthonormality implies that
\be
\intN h^{TT}_{m n, \mathcal{I}_1} h^{mn, TT}_{\mathcal{I}_2} = \delta_{\mathcal{I}_1 \mathcal{I}_2} .
\ee
Using completeness and the tensor decomposition, any symmetric tensor field $T_{mn}$ on ${\cal M}$ can be expanded as (see, e.g., Ref.~\cite{Hinterbichler:2013kwa} for details)
\begin{align}
T_{mn}&=\sum_{\mathcal{I}}  c^{\mathcal{I}} h_{mn, \mathcal{I}}^{TT} + 2\sum_{i \notin I_{\rm Killing}} c^i \nabla_{(m} Y_{n), i} + \sum_{a \notin I_{\rm conf.} } \tilde{c}^a\left(\nabla_m \nabla_n \psi_a -\frac{1}{N} \nabla^2 \psi_a g_{mn}\right) \nn \\
& + \sum_a \frac{1}{N} c^a \psi_a g_{mn}+ \frac{1}{N V^{1/2}}c^0 g_{mn}, \label{eq:tensorExpand}
\end{align}
where the coefficients are
\begin{subequations}
\begin{align}
 c^{\mathcal{I}} &=\intN T^{mn} h^{TT}_{mn, \mathcal{I}}, \quad c^i  = (\lambda_i -2R/N)^{-1} \intN T^{m n} \nabla_{(m} Y_{n), i} \, , \\ 
\tilde{c}^a & ={N\over \lambda_a \left( (N-1)\lambda_a-R \right)} \intN \left(\nabla_m \nabla_n \psi^a -\frac{1}{N} \nabla^2 \psi^a g_{mn}\right) T^{mn}, \\
 c^a & =\intN \psi^a g^{mn}T_{mn}, \quad  c^0 ={1\over V^{1/2}}\intN g^{mn}T_{mn } \, .
\end{align}
\end{subequations}

There is no general lower bound on Lichnerowicz eigenvalues, so a finite number of them can be negative on any given manifold. For example, the B\"ohm metrics on $S^3 \times S^2$ can have large negative Lichnerowicz eigenvalues \cite{Gibbons:2002th}. However, some lower bounds do exist if we make additional assumptions. For example, on closed Einstein manifolds with $R>0$ that admit a Killing spinor, there is the following lower bound~\cite{Wang91, Gibbons:2002th}:
\be \label{eq:lichbound}
\lambda_{\mathcal{I}} \geq  \left(16-(5-N)^2\right)\frac{R}{4N(N-1)}.
\ee
Examples of odd-dimensional manifolds admitting a Killing spinor are Einstein--Sasaki manifolds. Similarly, the Lichnerowicz eigenvalues are non-negative on closed Ricci-flat manifolds that admit a parallel spinor~\cite{Wang91, Dai2005}. This includes Ricci-flat manifolds with special holonomy, such as Calabi--Yau manifolds \cite{Yau1978}. Closed K\"ahler--Einstein manifolds with $R \geq 0$ also have $\lambda_{\mathcal{I}} \geq 0$ \cite{dai2005stability}.

Another distinguished Lichnerowicz eigenvalue is $\lambda_{\mathcal{I}}=2R/N$. Eigentensors $h_{mn, \mathcal{I}}^{TT}$ with this eigenvalue are called infinitesimal Einstein deformations and correspond to directions in the moduli space of Einstein structures of $\mathcal{M}$, i.e., the space of Einstein metrics on $\mathcal{M}$ modulo diffeomorphisms and volume rescalings. They give rise to massless scalars, called shape moduli, in Kaluza--Klein reductions of gravity. Einstein manifolds with $\lambda_{\mathcal{I}} > 2R/N$ are therefore rigid,\footnote{The converse is not true since infinitesimal Einstein deformations do not always integrate to curves of Einstein structures \cite{Koiso1982}.}  i.e., isolated points in moduli space. Examples of rigid manifolds are the compact symmetric spaces discussed in Appendix \ref{app:examples} \cite{Koiso1980}.

\subsection{Eigenmode expansions}
\label{sec:expansion}

Since the eigenfunctions of the scalar Laplacian are complete, we can expand any product of two eigenfunctions as a sum over eigenfunctions using Eq.~\eqref{scalarcompletee},
\be \label{eq:expansion1}
\psi_{a_1} \psi_{a_2} 
= V^{-1} \delta_{a_1 a_2} + \sum_a g_{a_1 a_2}{}^a \psi_a,
\ee
where the coefficients are given by the triple overlap integrals of eigenfunctions,
\be
g_{a_1 a_2 a_3} \equiv \intN \psi_{a_1} \psi_{a_2} \psi_{a_3}.
\ee 
Triple overlap integrals such as these are natural objects associated to any closed manifold.\footnote{We expect that generically the triple overlap integrals should be non-zero unless there is a symmetry that causes them to vanish. Note that they transform covariantly under eigenspace rotations, whereas eigenvalues are invariant.}

We can also expand products involving derivatives of eigenfunctions,
\begin{align}
\partial_m \psi_{a_1} \partial^m \psi_{a_2} & = V^{-1} \lambda_{a_1} \delta_{a_1 a_2} + \frac{1}{2}\sum_a  \left( \lambda_{a_1} + \lambda_{a_2} -\lambda_a \right) g_{a_1 a_2}{}^a \psi_a, \\
\nabla_m \nabla_n \psi_{a_1} \nabla^m \nabla^n \psi_{a_2} & = V^{-1} \lambda_{a_1} \left(\lambda_{a_1} -\frac{R}{N} \right) \delta_{a_1 a_2} \nn \\
&+ \frac{1}{4} \sum_a \left( \lambda_{a_1} + \lambda_{a_2} -\lambda_a \right) \left( \lambda_{a_1} + \lambda_{a_2} -\lambda_a -\frac{2R}{N}\right) g_{a_1 a_2}{}^a \psi_a.
\end{align}
Similarly, using the expansions for vectors and symmetric tensors in Eqs.~\eqref{eq:vectorExpand} and \eqref{eq:tensorExpand} we obtain
\begin{align}
\psi_{a_1} \partial_m \psi_{a_2} & = -\sum_i g_{a_1 a_2}{}^i Y_{m, i} + \sum_a \frac{\lambda_{a_2} -\lambda_{a_1} + \lambda_a}{2 \lambda_a} g_{a_1 a_2}{}^a \partial_m \psi_a, \\
\partial_{(m} \psi_{a_1} \partial_{n)} \psi_{a_2}  & = \sum_{\mathcal{I}} g_{a_1 a_2}{}^{\mathcal{I}} h_{m n, \mathcal{I}}^{TT} + \sum_{i \notin I_{\rm Killing}} \frac{\lambda_{a_2} -\lambda_{a_1}}{\lambda_i - \frac{2R}{N}} g_{a_1 a_2}{}^i \nabla_{(m} Y_{n),i} \nn \\
& + \sum_{a \notin I_{\rm conf.}} \frac{\left((N-2)\lambda_a^2+2 \lambda_a (\lambda_{a_1}+\lambda_{a_2})-N (\lambda_{a_1}-\lambda_{a_2})^2 \right) g_{a_1 a_2}{}^a}{4 \lambda_a \left((N-1)\lambda_a-R\right)} \left(\nabla_{m} \nabla_n \psi_a - \frac{1}{N} \Box \psi_a g_{mn} \right) \nn \\
&+\frac{ \lambda_{a_1} }{NV}\delta_{a_1 a_2}  g_{mn} + \frac{1}{2N} g_{mn} \sum_a ( \lambda_{a_1}+\lambda_{a_2} -\lambda_a )g_{a_1 a_2}{}^a \psi_a ,
\end{align}
where we have defined two additional triple overlap integrals,
\be
g_{a_1 a_2 i_3 } \equiv \intN \partial^m \psi_{a_1} \psi_{a_2} Y_{m, i_3}, \quad g_{a_1 a_2 \mathcal{I}_3 } \equiv \intN \partial^m \psi_{a_1} \partial^n \psi_{a_2} h_{m n, \mathcal{I}_3}^{TT} .
\ee

\subsection{Consistency conditions}
\label{sec:sumrules}
We now introduce the consistency conditions that are the main tool in the bootstrap. These conditions arise from using completeness to evaluate quartic overlap integrals of eigenfunctions in multiple ways. For example, we have the identity
\be
\intN
\contraction{}{\partial_m \psi_{a_1}}{}{\partial^m \psi_{a_1} }
\contraction{\partial_m \psi_{a_1} \partial^m \psi_{a_1}}{\psi_{a_1}}{}{\psi_{a_1}}
\partial_m \psi_{a_1} \partial^m \psi_{a_1} \psi_{a_1} \psi_{a_1} 
=
\intN
\contraction{}{\partial_m \psi_{a_1}}{\partial^m \psi_{a_1} }{ \psi_{a_1}}
\contraction[8pt]{\partial_m \psi_{a_1}}{ \partial^m \psi_{a_1}}{\psi_{a_1}}{\psi_{a_1}}
\partial_m \psi_{a_1} \partial^m \psi_{a_1} \psi_{a_1} \psi_{a_1},
\ee
where the Wick contraction notation denotes that the indicated pair of fields is replaced by the appropriate eigenmode expansion from Section \ref{sec:expansion}. Evaluating the contractions gives 
\be \label{eq:consistency1}
4 V^{-1} \lambda_{a_1} + \sum_a \left(4 \lambda_{a_1} -3 \lambda_a \right) g_{a_1 a_1 a}^2 = 0.
\ee
This is a consistency condition that must be satisfied by the geometric data on any connected and orientable closed manifold. It appears in a slightly different context in Ref.~\cite{Csaki:2003dt}.

For closed Einstein manifolds we can derive additional consistency conditions by evaluating more complicated integrals. We consider the following two additional identities: 
\begin{align}
\intN
\contraction{}{\partial_m \psi_{a_1}}{}{\partial_n \psi_{a_1} }
\contraction{\partial_m \psi_{a_1} \partial_n \psi_{a_1}}{\partial^m \psi_{a_1}}{}{\partial^n\psi_{a_1}}
\partial_m \psi_{a_1} \partial_n \psi_{a_1} \partial^m \psi_{a_1} \partial^n \psi_{a_1} & = 
\intN
\contraction{}{\partial_m \psi_{a_1}}{\partial_n \psi_{a_1} }{\partial_m \psi_{a_1}}
\contraction[8pt]{\partial_m \psi_{a_1}}{ \partial_n \psi_{a_1}}{\partial^m \psi_{a_1}}{\partial^n\psi_{a_1}}
\partial_m \psi_{a_1} \partial_n \psi_{a_1} \partial^m \psi_{a_1} \partial^n \psi_{a_1}, \\
\intN
\contraction{}{\partial_m \psi_{a_1}}{}{\partial_n \psi_{a_1} }
\contraction{\partial_m \psi_{a_1} \partial_n \psi_{a_1}}{\Delta_L ( \partial^m \psi_{a_1}}{}{\partial^n\psi_{a_1})}
\partial_m \psi_{a_1} \partial_n \psi_{a_1} \Delta_L \left( \partial^m \psi_{a_1} \partial^n \psi_{a_1}\right)  & = 
\intN
\contraction{}{\partial_m \psi_{a_1}}{\partial_n \psi_{a_1} \Delta_L (}{ \partial_m \psi_{a_1}}
\contraction[8pt]{\partial_m \psi_{a_1}}{ \partial_n \psi_{a_1}}{\Delta_L ( \partial^m \psi_{a_1}}{\partial^n\psi_{a_1})}
\partial_m \psi_{a_1} \partial_n \psi_{a_1}\Delta_L \left( \partial^m \psi_{a_1} \partial^n \psi_{a_1} \right).
\end{align}
Including Eq.~\eqref{eq:consistency1}, the resulting consistency conditions are equivalent to\footnote{We have used the fact that $g_{a_1 a_1 a} =0$ if $ a \in I_{\rm conf.}$ and $N>1$, which follows from parity symmetry on $S^N$.} 
\be \label{eq:sumrule1}
V^{-1}  \vec{F}_1+ \frac{1}{\lambda_{a_1}^2}\sum_{\mathcal{I}} \vec{F}_{2} \, g_{a_1 a_1 \mathcal{I}}^2 + \sum_{a \notin I_{\rm conf.}} \left[ \vec{F}_{3}+\frac{ R \vec{F}_{4} }{ (N-1)\lambda_a -R}  \right] g_{a_1 a_1 a}^2 = 0,
\ee
where we have defined
\begin{align}
\vec{F}_1 & =\left(4, \, -16, \, 0\right), \\
\vec{F}_{2} & = \left(0, \, 16N(N-1), \, 16N(N-1) \frac{\lambda_{\mathcal{I}}}{\lambda_{a_1}} \right), \\
\vec{F}_{3} & = \left(4- \frac{3\lambda_{a}}{\lambda_{a_1}}, \, \frac{N\lambda_a}{\lambda_{a_1}}\left(  4N +\frac{(4-3N)\lambda_a}{\lambda_{a_1}} \right), \,  \frac{N\lambda_a}{\lambda_{a_1}} \left(4-\frac{\lambda_a}{\lambda_{a_1}}\right) \left(4N-\frac{(3N-2)\lambda_a}{\lambda_{a_1}} \right) \right), \\
 \vec{F}_{4} & = \left(0, \, \left( 4 +\frac{(N-2)\lambda_a}{\lambda_{a_1}}\right)^2, \,  \frac{\lambda_a}{\lambda_{a_1}} \left( 4+\frac{(N-2)\lambda_a}{\lambda_{a_1}}\right)^2\right).
\end{align}
When $R=0$, these consistency conditions are the sum rules responsible for the good high-energy behavior of the four-point tree amplitudes of massive Kaluza--Klein excitations of the graviton in general relativity (GR) with Ricci-flat compact extra dimensions~\cite{Bonifacio:2019ioc}. In these cases the sums correspond to sums over the exchanged particles in the four-point amplitude. (See Refs.~\cite{Chivukula:2019rij, Chivukula:2019zkt, Chivukula:2020hvi} for related recent work on the scattering of Kaluza--Klein states in a Randall--Sundrum model with one extra dimension.) 

These consistency conditions must be satisfied by the geometric data on any closed Einstein manifold $\mathcal{M}$ that is connected and orientable. In fact, they also hold for the quotients of such manifolds, $\mathcal{M}/\Gamma$, where $\Gamma$ is any subgroup of the isometry group of $\mathcal{M}$. This is because the $\Gamma$-invariant eigenmodes form a closed subsector of the consistency conditions. These quotient spaces include certain non-orientable manifolds, such as $\mathbb{RP}^N$ with even $N$, as well as certain orbifolds. The quotients also include spaces that are not orbifolds. In the following sections we will usually just write ``closed Einstein manifolds" instead of ``quotients of connected and orientable closed Einstein manifolds."

Note that we do not get additional independent consistency conditions from integrals involving more than four eigenfunctions if we consider all consistency conditions with four eigenmodes.

\section{Bootstrap bounds}

In analogy to the conformal bootstrap bounds on conformal data, we can use the consistency conditions from Section~\ref{sec:sumrules} to find bounds on the geometric data of closed Einstein manifolds. To do this we postulate some candidate geometric data, i.e., a collection of eigenvalues and triple overlap integrals, and then search for a constant vector $\vec{\alpha} \in \mathbb{R}^3$ such that the condition
\be \label{eq:sumrule1b}
V^{-1}  \vec{\alpha} \cdot \vec{F}_1+ \frac{1}{\lambda_{a_1}^2} \sum_{\mathcal{I}} \vec{\alpha} \cdot \vec{F}_{2} \, g_{a_1 a_1 \mathcal{I}}^2 + \sum_{a \notin I_{\rm conf.}} \left[\vec{\alpha} \cdot\vec{F}_{3}+\frac{R \, \vec{\alpha} \cdot\vec{F}_{4} }{ (N-1)\lambda_a- R}  \right] g_{a_1 a_1 a}^2 = 0
\ee
can never be satisfied by this data. If such an $\vec{\alpha}$ exists, then the candidate geometric data cannot correspond to any closed Einstein manifold. Although we have far fewer constraints than in typical conformal bootstrap problems, we will see that it is still possible to get some nontrivial bounds. 

\subsection{Eigenvalue bounds}
We begin by looking for bootstrap bounds on the eigenvalues of the scalar Laplacian on closed Einstein manifolds. These translate to bounds on the tree-level masses of Kaluza--Klein modes. 

\subsubsection{Low-lying eigenvalues}
Suppose we are given a closed Einstein manifold $\mathcal{M}$ with $R\geq 0$. Let $\psi_{a_1}$ be an eigenfunction of the scalar Laplacian on $\mathcal{M}$ whose eigenvalue $\lambda_{a_1}$ is the smallest nonzero eigenvalue and let $\lambda_{a_2}$ be the next distinct eigenvalue in order of increasing size. The nonzero scalar eigenvalues therefore satisfy
\be
\lambda_a \in \{\lambda_{a_1} \} \cup [\lambda_{a_2}, \infty), \quad 0 < \lambda_{a_1} < \lambda_{a_2}. 
\ee
We can write the eigenfunction expansion of $\psi_{a_1}^2$ as
\be \label{eq:psi1psi1}
\psi_{a_1}^2 = V^{-1} + \sum_{\substack{ \lambda_a = \lambda_{a_1}}} g_{a_1 a_1}{}^a \psi_a + \sum_{\substack{\lambda_a \geq \lambda_{a_2}}} g_{a_1 a_1}{}^a \psi_a\, .
\ee
Similarly, we let $\lambda_{\mathcal{I}_1}$ be the smallest eigenvalue of the Lichnerowicz Laplacian on $\mathcal{M}$ acting on transverse traceless tensors. 

The three eigenvalues $\lambda_{a_1}$,  $\lambda_{a_2}$, and $\lambda_{\mathcal{I}_1}$ define a candidate low-lying  spectrum on $\mathcal{M}$. 
\begin{subequations} \label{eq:constraints1}
Given such a candidate spectrum, our goal is to determine if it is inconsistent. To do this, we search for a constant vector $\vec{\alpha} \in \mathbb{R}^3$ such that the following conditions are satisfied:
\begin{align}
& \vec{\alpha} \cdot \vec{F}_1  = 1,  \\ 
& \vec{\alpha} \cdot \vec{F}_2 \geq 0, \quad \forall \, \lambda_{\mathcal{I}} \geq\lambda_{\mathcal{I}_1}, \\
& \vec{\alpha} \cdot \vec{F}_k  \geq 0, \quad k=3,4, \quad \forall \, \lambda_a \geq \lambda_{a_2}, \\
& \vec{\alpha} \cdot \vec{F}_k\Big|_{\lambda_a =\lambda_{a_1}} \geq 0,  \quad k= 3,4 . \label{eq:alphacons1d}
\end{align}
\end{subequations}
If such an $\vec{\alpha}$ exists, then the candidate spectrum could not have come from a closed Einstein manifold since it is inconsistent with the consistency conditions \eqref{eq:sumrule1b}. To reach this conclusion we have used the positivity of $g_{a_1 a_1 a}^2$ and $g_{a_1 a_1 \mathcal{I}}^2$, which follows from the reality of the eigenmode basis. Alternatively, if no such $\vec{\alpha}$ can be found, then we conclude nothing. 

The problem of finding an $\vec{\alpha}$ that satisfies Eqs.~\eqref{eq:constraints1} can be formulated as a semidefinite program (SDP). The general procedure for doing this is reviewed in Refs.~\cite{Poland:2011ey, Kos:2014bka} on the conformal bootstrap. Once we have the problem formulated as an SDP, we can solve it using numerical solvers. The resulting bounds are then rigorous up to the precision of the numerics. For the numerical computations in this work we mostly used the specialized program \texttt{SDPB} \cite{Simmons-Duffin:2015qma, Landry:2019qug}. Some of the simpler SDPs can also be solved in \texttt{Mathematica} using the function \texttt{SemidefiniteOptimization}.

\begin{figure}[h!]
\begin{center}
\epsfig{file=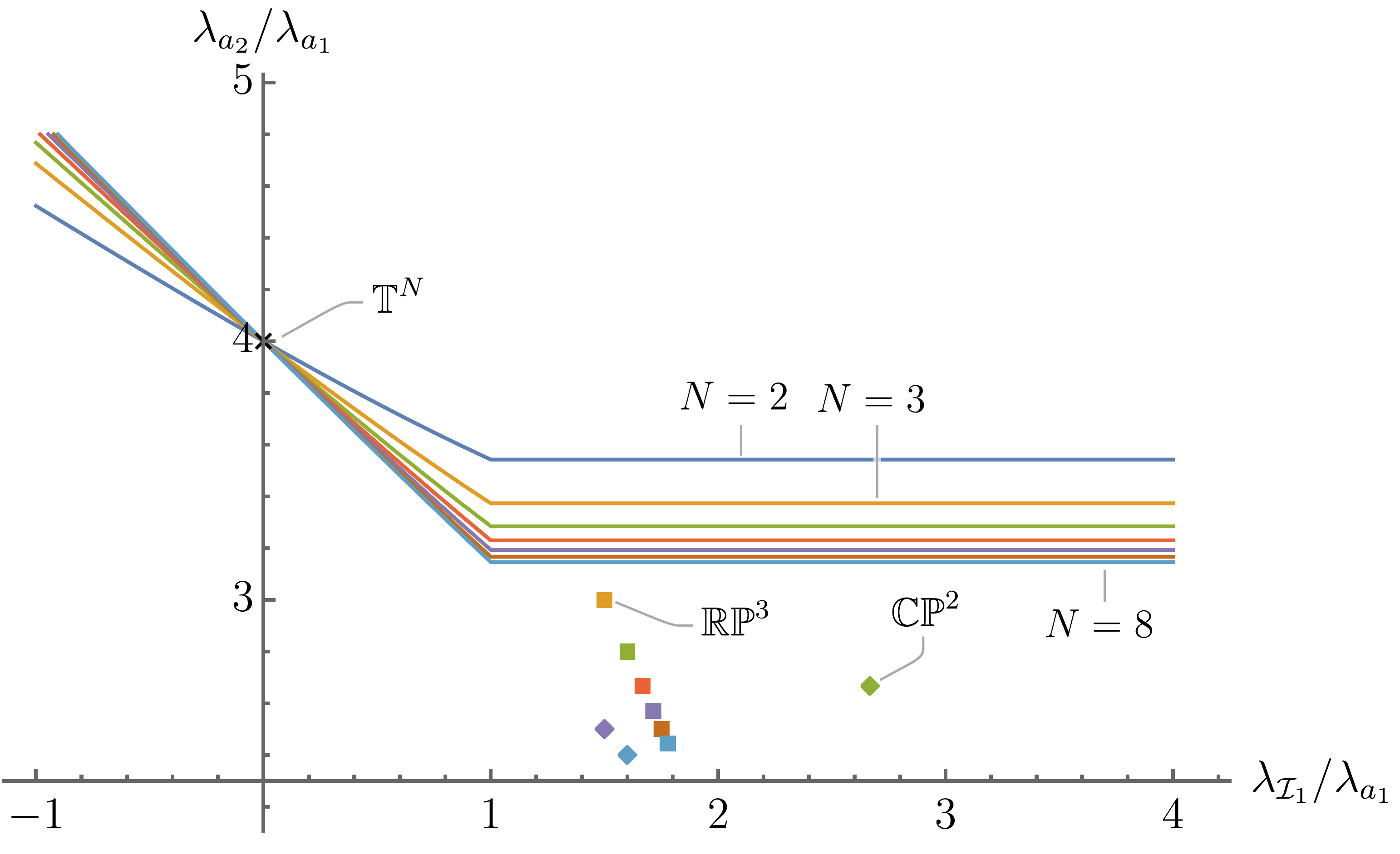, width=12cm}
\caption{Upper bounds on the ratio of the first distinct nonzero eigenvalues of the scalar Laplacian on closed Einstein manifolds with $R \geq 0$ and $N=2, \dots, 8$ for different values of the smallest Lichnerowicz eigenvalue of transverse traceless tensors. The square and diamond markers correspond to $\mathbb{RP}^N$ and $\mathbb{CP}^{N/2}$ with their standard metrics and the cross corresponds to long flat $N$-tori.}
\label{fig:eigenvalue-bound-1}
\end{center}
\end{figure}

A simplifying feature of this problem is that the consistency conditions depend only on the two ratios $\lambda_{a_2}/\lambda_{a_1}$ and $\lambda_{\mathcal{I}_1}/\lambda_{a_1}$, which ensures that the resulting constraints are independent of constant rescalings of the metric.
For each value of $\lambda_{\mathcal{I}_1}/\lambda_{a_1}$ we solve a series of SDPs to obtain an upper bound on $\lambda_{a_2}/\lambda_{a_1}$. These upper bounds are shown in Figure~\ref{fig:eigenvalue-bound-1} for $N=2,  \dots, 8$. 
When $\lambda_{\mathcal{I}_1}$ vanishes we find that $\lambda_{a_2} \leq 4\lambda_{a_1}$; this a special case of the bound found in Ref.~\cite{Bonifacio:2019ioc} and is saturated by flat tori whose first eigenvalues are identical to those of a circle, which we call ``long flat tori." For $\lambda_{\mathcal{I}_1}>0$, the upper bounds decrease linearly until $\lambda_{\mathcal{I}_1} = \lambda_{a_1}$, at which point they exhibit kinks and thereafter become constant. The upper bounds grow linearly for negative $\lambda_{\mathcal{I}_1}$, so we cannot rule out large values for the ratio of the first two nonzero eigenvalues of the scalar Laplacian on manifolds with large negative Lichnerowicz eigenvalues. In the plot we have also marked the points corresponding to long flat tori and the projective spaces $\mathbb{RP}^N$ for $N=3, \dots, 8$ and $\mathbb{CP}^{N/2}$ for $N=4,6,8$ with their standard metrics (there are no transverse traceless tensors on $\mathbb{RP}^2$ and $\mathbb{CP}^1 \approx S^2$). These projective spaces all fall comfortably below the bounds. It would be interesting to find out if there are manifolds that live at the kinks of this plot, just as nontrivial CFTs often appear at such kinks in the conformal bootstrap. 

\begin{figure}[h!]
\begin{center}
\epsfig{file=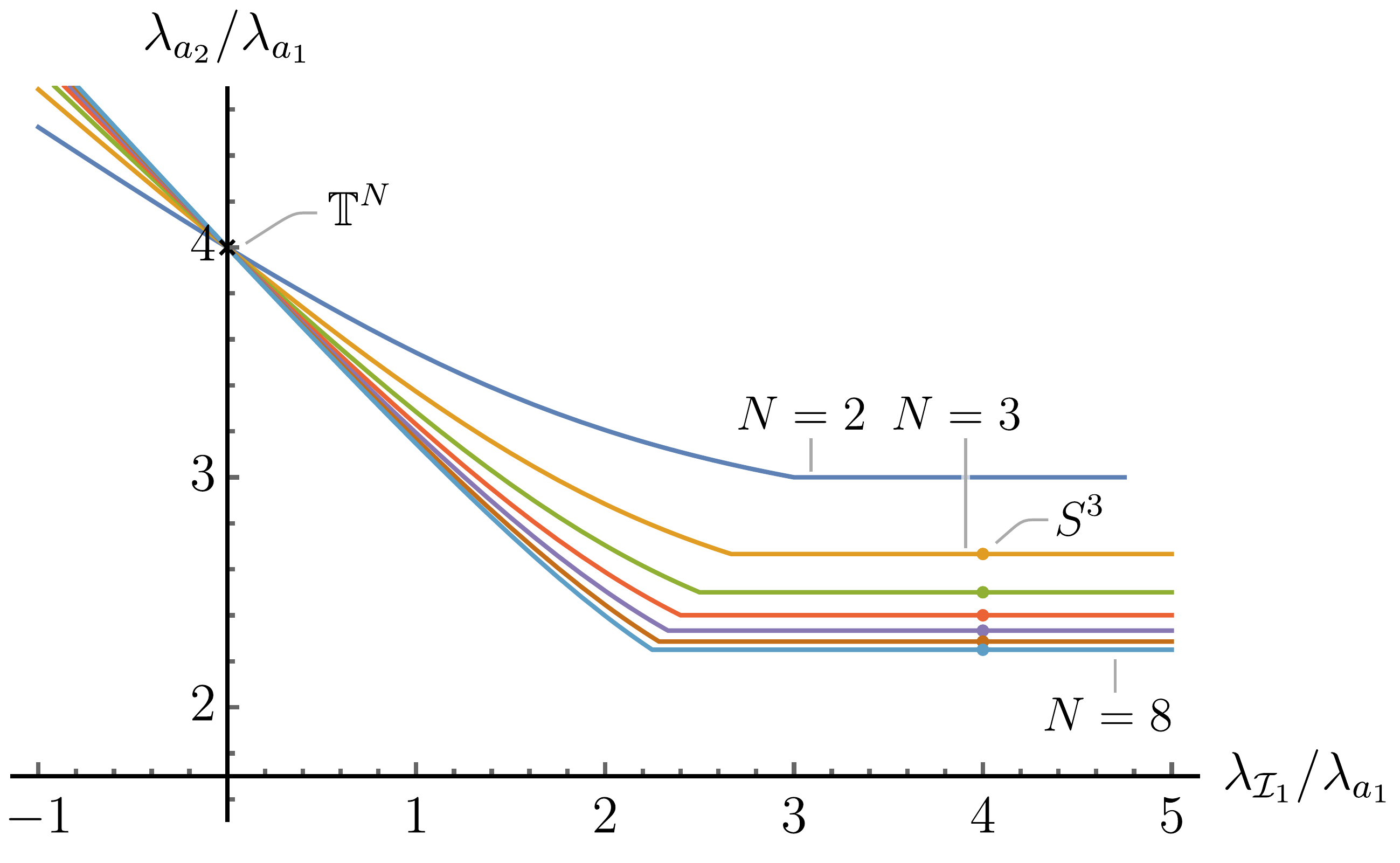, width=12cm}
\caption{Upper bounds on the ratio of the first distinct nonzero eigenvalues of the scalar Laplacian on closed Einstein manifolds with $R \geq 0$ and $N=2, \dots, 8$ for different values of the lightest Lichnerowicz eigenvalue, assuming that $g_{a_1 a_1 a}=0$ when $\lambda_a = \lambda_{a_1}$. 
The filled circles correspond to round $N$-spheres and the cross corresponds to long flat $N$-tori.}
\label{fig:eigenvalue-bound-2}
\end{center}
\end{figure}

An additional assumption we can make is that the triple overlap integrals $g_{a_1 a_1 a}$ vanish when $\lambda_a = \lambda_{a_1}$, so that the first sum in Eq.~\eqref{eq:psi1psi1} vanishes. This follows, for example, if there is a $\mathbb{Z}_2$ symmetry under which $\psi_{a_1}$ is odd, as is the case for parity on $S^N$. This means that we can drop the constraints in Eq.~\eqref{eq:alphacons1d}, so it becomes easier to rule out a point and hence the resulting bounds are stronger. We plot these bounds for $N=2, \dots, 8$ in Figure~\ref{fig:eigenvalue-bound-2}. As with the previous upper bounds, the trend is an approximately linear decrease until some critical $\lambda_{\mathcal{I}_1}/\lambda_{a_1}$ is reached, now given by $2(N+1)/N$, above which the bound is constant. We also marked the points corresponding to $S^N$ for $N=3, \dots, 8$, which lie on the boundaries of the allowed regions and away from the kinks. Round spheres thus have the largest possible ratio of the first two distinct nonzero eigenvalues of the scalar Laplacian amongst closed Einstein manifolds with $R\geq 0$, a $\mathbb{Z}_2$ symmetry, and a sufficiently large value of $\lambda_{\mathcal{I}_1}/\lambda_{a_1}$.

Now let us consider manifolds with either $R > 0$ or $R<0$.  In these cases we can use the rescaling freedom to set $R = \pm 1$ when computing bounds. For $R >0$, the bounds are stronger compared to the $R \geq 0$ bounds we considered previously, but the price we pay is the introduction of an additional scale in the problem.
\begin{subequations}
In either case, a candidate spectrum is determined by $\lambda_{a_1}$, $\lambda_{a_2}$, and $\lambda_{\mathcal{I}_1}$, where for $R>0$ we also ensure consistency with the Lichnerowicz bound \eqref{eq:Lichbound}. Given such a spectrum, we try to rule it out by searching for a constant vector $\vec{\alpha} \in \mathbb{R}^3$ such that
\begin{align}
& \vec{\alpha} \cdot \vec{F}_1  = 1,  \\ 
& \vec{\alpha} \cdot \vec{F}_2  \geq 0, \quad \forall \, \lambda_{\mathcal{I}} \geq \lambda_{\mathcal{I}_1}, \\
& \vec{\alpha} \cdot  \left[ ((N-1)\lambda_a- R)\vec{F}_{3}+R  \vec{F}_{4}   \right]   \geq 0, \quad \forall \, \lambda_a \geq \lambda_{a_2}, \\
& \vec{\alpha} \cdot  \left[ ((N-1)\lambda_a- R)\vec{F}_{3}+R  \vec{F}_{4}   \right]\Big|_{\lambda_a =\lambda_{a_1}}   \geq 0.
\end{align}
\end{subequations}

\begin{figure}[h!]
\begin{center}
\epsfig{file=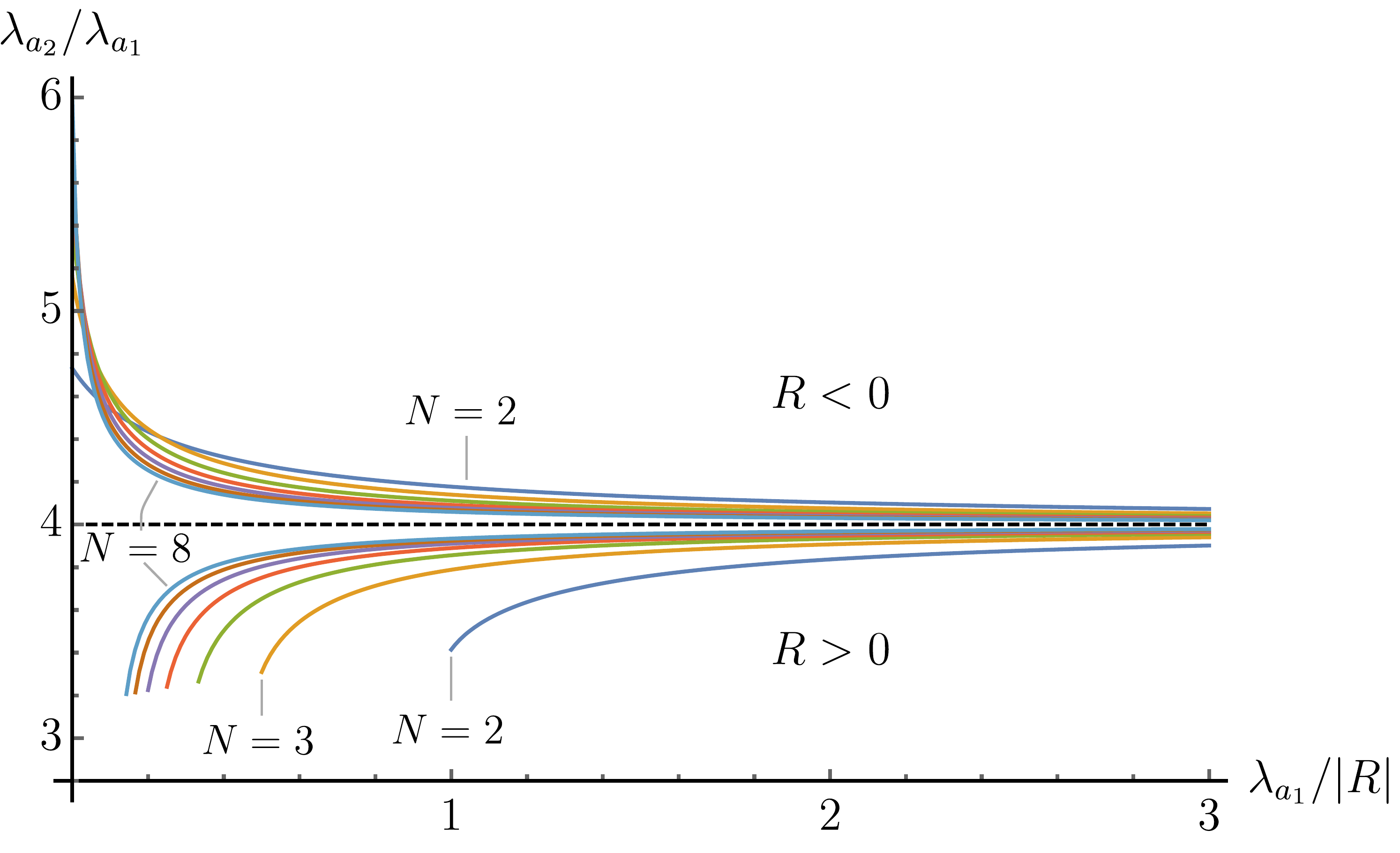, width=12cm}
\caption{Upper bounds on the ratio of the first distinct nonzero eigenvalues of the scalar Laplacian for closed Einstein manifolds with $R \neq 0$ and $N=2, \dots, 8$,  assuming that $\lambda_{\mathcal{I}} \geq 0$. The minimum eigenvalues for $R>0$ are set by the Lichnerowicz bound \eqref{eq:Lichbound}.}
\label{fig:eigenvalue-bound-3}
\end{center}
\end{figure}

Since we now have an extra scale, we must fix $\lambda_{\mathcal{I}_1}$ to make a two-dimensional plot of the allowed eigenvalues $\lambda_{a_1}$ and $\lambda_{a_2}$. In Figure~\ref{fig:eigenvalue-bound-3} we show the upper bounds on $\lambda_{a_2}/\lambda_{a_1}$ for each value of $\lambda_{a_1}/|R|$ on closed Einstein manifolds with $N=2, \dots, 8$ and $R>0$ or $R<0$, assuming that $\lambda_{\mathcal{I}} \geq 0$.  When $\lambda_{a_1}$ is of order the curvature scale, the $R>0$ bounds are stronger than what we found above, but they approach the $R \geq 0$ bound (the dashed line) as $\lambda_{a_1}/R \rightarrow \infty$.  Intuitively, this is because the effects of curvature become negligible for high-frequency modes. The bounds for $R <0$ approach $\lambda_{a_2}/\lambda_{a_1}=4$ from above.

\begin{figure}[h!]
\begin{center}
\epsfig{file=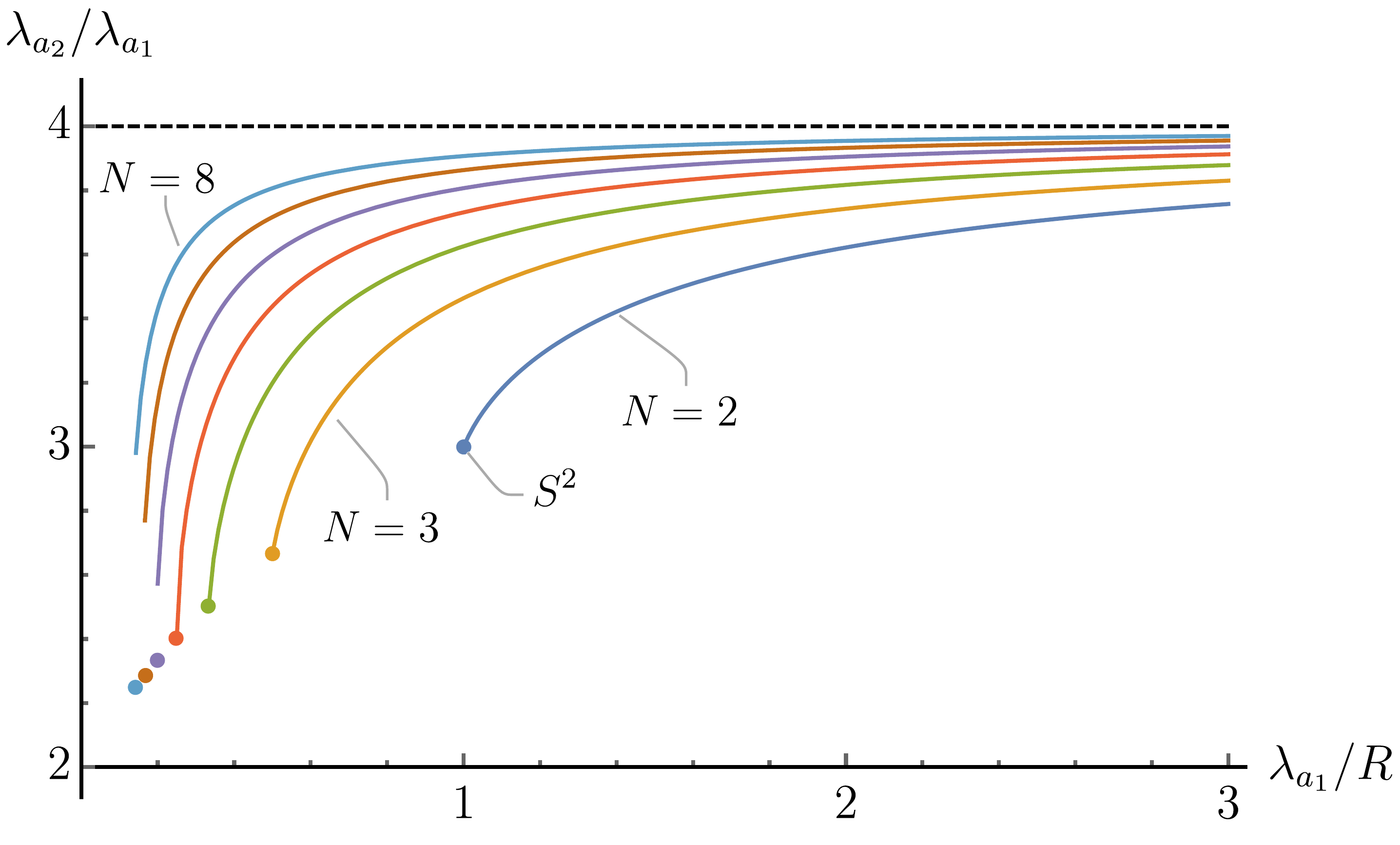, width=12cm}
\caption{Upper bounds on the ratio of the first distinct nonzero eigenvalues of the scalar Laplacian for closed Einstein manifolds with $R > 0$ and $N=2, \dots, 8$, assuming that the Lichnerowicz eigenvalues satisfy Eq.~\eqref{eq:lichbound}. These bounds apply to closed Einstein manifolds that admit Killing spinors, such as Einstein--Sasaki manifolds. The filled circles correspond to $S^N$.}
\label{fig:eigenvalue-bound-4}
\end{center}
\end{figure}

We now show how the bounds change if we make a stronger assumption about the size of the first Lichnerowicz eigenvalue. In Figure~\ref{fig:eigenvalue-bound-4} we show the bounds on $\lambda_{a_2}/\lambda_{a_1}$ for closed Einstein manifolds with $R>0$ and $N=2, \dots, 8$ when $\lambda_{\mathcal{I}}$ satisfies the inequality in Eq.~\eqref{eq:lichbound}, the lower bound implied by the existence of a Killing spinor. We also plot the points corresponding to $S^N$, which saturate the bounds for $N=2, \dots, 5$ but not for $N =6, 7, 8$ (they would if we made the stronger assumption that $\lambda_{\mathcal{I}} \geq 4R/(N-1)$).

\begin{figure}[h!]
\begin{center}
\epsfig{file=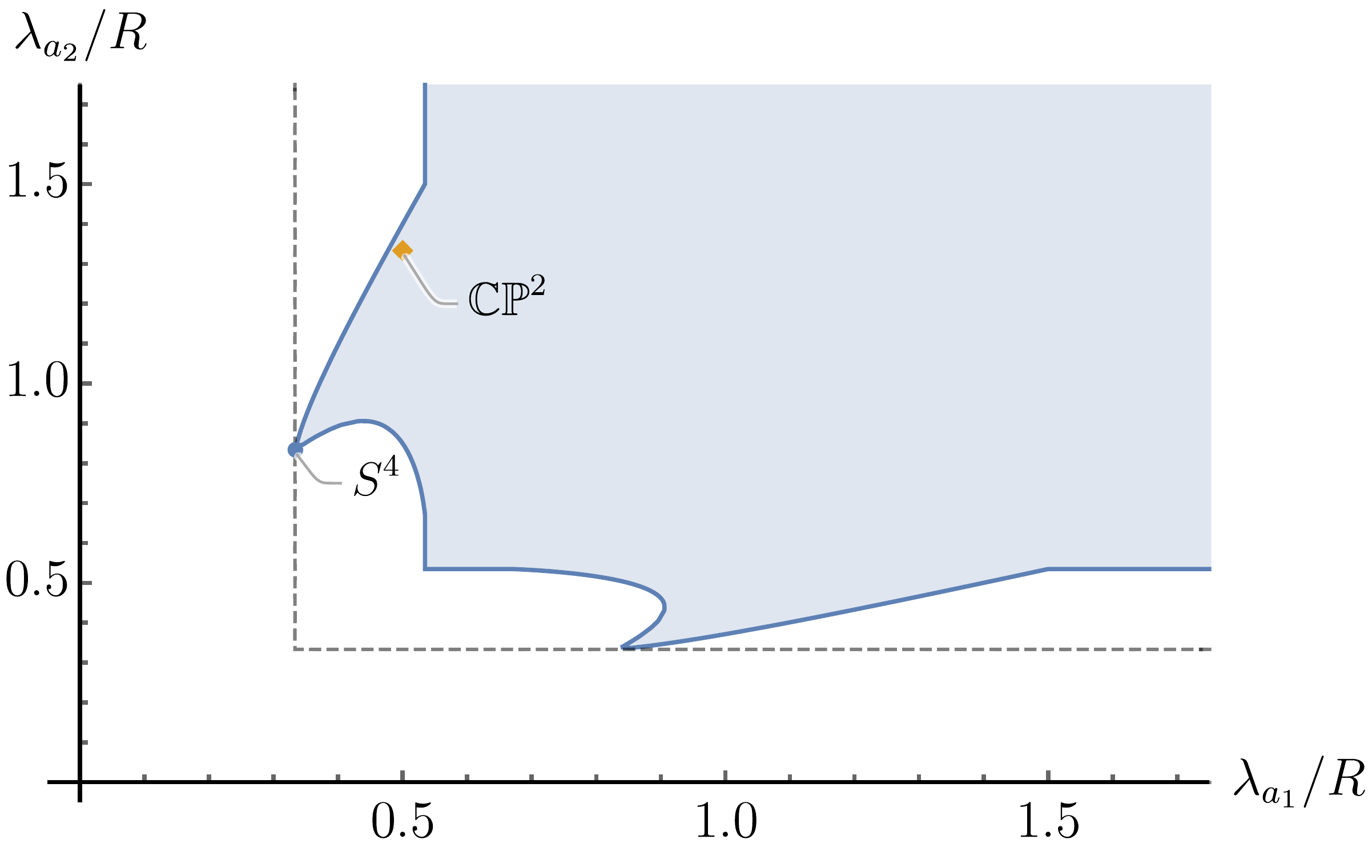, width=12cm}
\caption{The blue region shows the allowed values of the two smallest nonzero eigenvalues of the scalar Laplacian on closed four-dimensional Einstein manifolds with $R>0$, assuming the eigenvalues satisfy the conditions in Eq.~\eqref{eq:gapTo3rd}. The dashed line is the Lichnerowicz bound and the points corresponding to $S^4$ and $\mathbb{CP}^2$ with their standard metrics are marked.}
\label{fig:region-bound-1}
\end{center}
\end{figure}

Next we explore a slightly different type of question about eigenvalues. Suppose we assume a particular value for the third nonzero scalar eigenvalue and ask what values the first and second eigenvalues can take. In particular, let us set $N=4$ and assume the following:
\be \label{eq:gapTo3rd}
\lambda_a \in \{\lambda_{a_1}, \lambda_{a_2} \} \cup \left[\frac{3R}{2}, \infty \right), \quad \lambda_{\mathcal{I}} \geq \frac{4R}{3} \,,
\ee
which is consistent with both $S^4$ and $\mathbb{CP}^2$ with their standard metrics.
With these assumptions, the values of $\lambda_{a_1}/R$ and $\lambda_{a_2}/R$ allowed by the consistency conditions \eqref{eq:sumrule1} are given by the blue region in Figure~\ref{fig:region-bound-1}. We see that the 4-sphere lives at the tip of a pointed part of the allowed region. To compute these bounds we used two copies of the consistency conditions \eqref{eq:sumrule1}, one for $\lambda_{a_1}$ and one for $\lambda_{a_2}$. The plot is invariant under a reflection along the diagonal since $\lambda_{a_1}$ and $\lambda_{a_2}$ are treated symmetrically.

The results so far have been general bounds for all manifolds satisfying a handful of assumptions. We can also try to obtain constraints for specific manifolds by inputting additional information. Suppose that we know the scalar eigenfunctions on a given manifold but not necessarily the Lichnerowicz spectrum. We can then write the consistency conditions as 
\be 
\vec{\alpha} \cdot \vec{F}'_1+ \frac{1}{\lambda_{a_1}^2} \sum_{\mathcal{I}} \vec{\alpha} \cdot \vec{F}_{2} \, g_{a_1 a_1 \mathcal{I}}^2  = 0,
\ee
where $\vec{F}_1'$ combines the first and third terms of Eq.~\eqref{eq:sumrule1b} and is assumed to be some known constant vector for a given choice of $\psi_{a_1}$. We can then find an upper bound on $\lambda_{\mathcal{I}_1}$, the smallest Lichnerowicz eigenvalue on transverse traceless tensors, by finding the smallest $\lambda_{\mathcal{I}_1}$ for which there exists a vector $\vec{\alpha} \in \mathbb{R}^3$ satisfying
\begin{align}
& \vec{\alpha} \cdot \vec{F}'_1  = 1,  \\ 
& \vec{\alpha} \cdot \vec{F}_2 \geq 0, \quad \forall \, \lambda_{\mathcal{I}} \geq\lambda_{\mathcal{I}_1}.
\end{align}
We have carried out this procedure for the compact symmetric spaces of rank one restricted to the subsector of zonal spherical functions (see Appendix~\ref{app:examples}), taking $\psi_{a_1}$ to be the lightest nontrivial zonal spherical function. For these cases, the equations can be solved analytically and therefore the bounds are exact. This gives the following finite bounds for $n>1$:
\begin{itemize} 
\item  For $\mathbb{CP}^n$ we find $\lambda_{\mathcal{I}_1} \leq 8(n+2)$, which is saturated for $n=2$ \cite{Warner82, Boucetta07}. 
\item For $\mathbb{HP}^n$ we find $\lambda_{\mathcal{I}_1} \leq 8(2n+3)$, consistent with the lower bound $\lambda_{\mathcal{I}_1} \geq 16n$ inferred from Ref.~\cite{Koiso1980}. 
\item For $\mathbb{OP}^2$ we find $\lambda_{\mathcal{I}_1} \leq 26$, consistent with the lower bound $\lambda_{\mathcal{I}_1} \geq 16$ inferred from Ref.~\cite{Koiso1980}.
\end{itemize}
 
\subsubsection{General eigenvalues}

The above bounds constrain just the smallest eigenvalues, but we can also look for bounds on general eigenvalues. 
Let $\psi_{a_1}$ now be a generic eigenfunction of the scalar Laplacian, with eigenvalue $\lambda_{a_1}$, and define $\lambda_{a_2}$ as the smallest eigenvalue bigger than $\lambda_{a_1}$, i.e., the scalar eigenvalues satisfy
\be
\lambda_a \in (0, \lambda_{a_1} ] \cup [\lambda_{a_2}, \infty )\, , \quad \lambda_{a_1} < \lambda_{a_2} \,.
\ee
The eigenfunction expansion of $\psi_{a_1}^2$ can then be written as
\be \label{eq:psi1psi2}
\psi_{a_1}^2 = V^{-1} + \sum_{\substack{ \lambda_a \leq \lambda_{a_1}}} g_{a_1 a_1}{}^a \psi_a + \sum_{\substack{\lambda_a \geq \lambda_{a_2} }} g_{a_1 a_1}{}^a \psi_a \, .
\ee
We again define $\lambda_{\mathcal{I}_1}$ as the smallest eigenvalue of the Lichnerowicz Laplacian on transverse traceless tensors.

\begin{figure}[h!]
\begin{center}
\epsfig{file=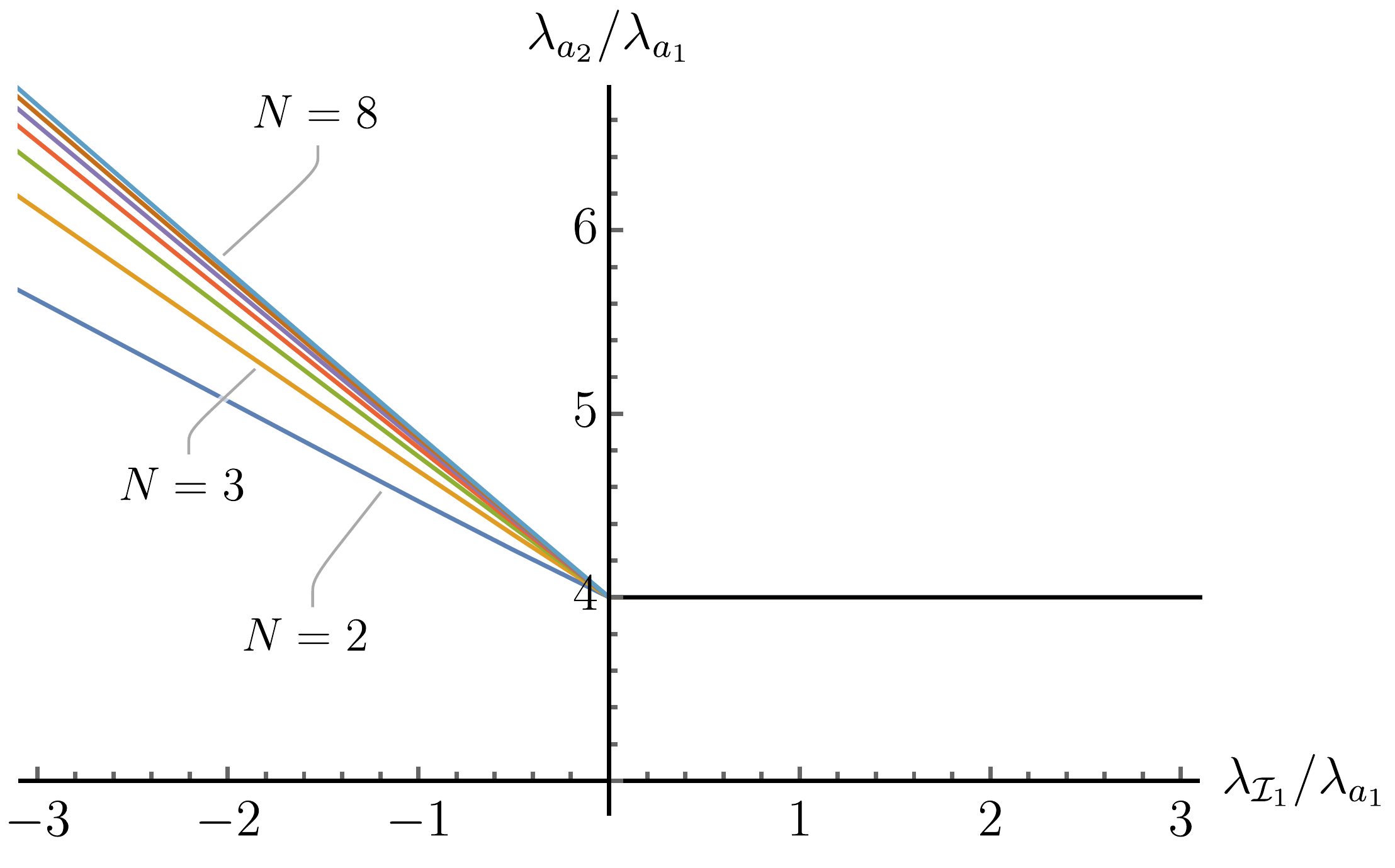, width=12cm}
\caption{Upper bounds on the ratio of any two consecutive nonzero eigenvalues of the scalar Laplacian for closed Einstein manifolds with $R \geq 0$ and $N=2,\dots, 8$ for different values of $\lambda_{\mathcal{I}_1}$. The bounds for $\lambda_{\mathcal{I}_1}\geq0$, given by the solid black line,  are the same for each $N$.}
\label{fig:eigenvalue-bound-5}
\end{center}
\end{figure}

We want to determine how large $\lambda_{a_2}/\lambda_{a_1}$ can be for a given choice of $\lambda_{\mathcal{I}_1}/\lambda_{a_1}$.
\begin{subequations}
For a closed Einstein manifold with $R \geq 0$, we can deduce that a given choice of eigenvalues is not feasible if we can find a constant vector $\vec{\alpha} \in \mathbb{R}^3$ such that
\begin{align}
& \vec{\alpha} \cdot \vec{F}_1  = 1,  \\ 
& \vec{\alpha} \cdot \vec{F}_2 \geq 0, \quad \forall \, \lambda_{\mathcal{I}} \geq\lambda_{\mathcal{I}_1}, \\
& \vec{\alpha} \cdot \vec{F}_k  \geq 0, \quad k=3,4, \quad \forall \,  \lambda_a \geq \lambda_{a_2}, \\
& \vec{\alpha} \cdot \vec{F}_k  \geq 0, \quad k=3,4, \quad \forall \,  \lambda_a \leq \lambda_{a_1} .
\end{align}
\end{subequations}
To be able to formulate this last condition as an SDP we can write it as
\be
\lambda_a^{-3}\vec{\alpha} \cdot \vec{F}_k  \geq 0, \quad k=3,4, \quad \lambda_a = \lambda_{a_1}/(1+x), \quad \forall \,  x \geq 0.
\ee
The resulting upper bounds on $\lambda_{a_2}/\lambda_{a_1}$ are shown in Figure~\ref{fig:eigenvalue-bound-5} for $N=2, \dots, 8$. The bounds grow linearly for $\lambda_{\mathcal{I}_1}<0$ and are constant for $\lambda_{\mathcal{I}_1} \geq 0$, given by $\lambda_{a_2}/\lambda_{a_1} \leq 4$ in all dimensions. This result in the case $\lambda_{\mathcal{I}_1}=0$ was already given in Ref.~\cite{Bonifacio:2019ioc}.

\subsection{Bounds on cubic couplings}
\label{sec:coupling}

We have seen that the consistency conditions can be used to derive bounds on Laplacian eigenvalues. 
Now we show that they can also give upper bounds on the triple overlap integrals of eigenfunctions of Laplacian operators, which correspond to cubic couplings in Kaluza--Klein theories.\footnote{It is also possible to find lower bounds in some cases, but we will not discuss these here.} These couplings must be real and can take either sign, but we will see that sometimes their absolute values are bounded from above by fixed multiples of geometric invariants.

\subsubsection{Spin-2 self-interactions}
\label{sec:g111}

We start by looking for bounds on the integral of the cube of an eigenfunction of the scalar Laplacian, 
\be
g_{a_1 a_1 a_1} \equiv  \intN \psi_{a_1}^3.
\ee  
In Kaluza--Klein reductions of GR down to a $d$-dimensional flat spacetime, the massive spin-2 excitations of the graviton in the lower-dimensional theory interact with one another through tree-level cubic amplitudes of the form \cite{Bonifacio:2019ioc} 
\be
\mathcal{A}(1_h^{a_1},\, 2_h^{a_2}, \, 3_h^{a_3}) =\frac{2}{M_d^{\frac{d-2}{2}}} g_{a_1 a_2 a_3}  \sqrt{V} \left( \epsilon_1 \ccdot \epsilon_2 \, \epsilon_3 \ccdot p_1 + \epsilon_2 \ccdot \epsilon_3 \, \epsilon_1 \ccdot p_2+\epsilon_1 \ccdot \epsilon_3 \, \epsilon_2 \ccdot p_3 \right)^2,
\ee
where $M_d$ is the lower-dimensional Planck mass, $p_i$ are the momenta, and $\epsilon_i \otimes \epsilon_i$ are polarizations.
The quantities $g_{a_1 a_2 a_3} \sqrt{V}$ thus control the strengths of these interactions relative to the strength of gravity. In particular, $g_{a_1 a_1 a_1} \sqrt{V}$ gives the relative strength of the self-interaction.

We assume that the scalar eigenvalues satisfy
\be \label{eq:evalueAssumption}
\lambda_a \in \{ \lambda_{a_1}\} \cup [\lambda_{a_2}, \infty)\, ,
\ee
so that $\lambda_{a_2}$ is second smallest nonzero scalar eigenvalue if it is greater than $\lambda_{a_1}$, otherwise it is the smallest nonzero eigenvalue.
We again define $\lambda_{\mathcal{I}_1}$ as the smallest Lichnerowicz eigenvalue on transverse traceless tensors. We want to use the consistency conditions to find an upper bound on $|g_{a_1 a_1 a_1}| \sqrt{V}$ for a given choice of the low-lying eigenvalues $\lambda_{a_1}$, $\lambda_{a_2}$, and $\lambda_{\mathcal{I}_1}$. This can be achieved with the same strategy used to bound OPE coefficients in the conformal bootstrap~\cite{Caracciolo:2009bx}. If $R \geq 0$, we look for a vector $\vec{\alpha} \in \mathbb{R}^3$ that satisfies the following constraints:
\begin{subequations} 
\label{eq:alphacons3} 
\begin{align}
& \vec{\alpha} \cdot \vec{F}_3\Big|_{\lambda_a =\lambda_{a_1}} =1, \\
& \vec{\alpha} \cdot \vec{F}_4\Big|_{\lambda_a =\lambda_{a_1}} \geq 0, \\
& \vec{\alpha} \cdot \vec{F}_k  \geq 0, \quad k=3,4, \quad \forall \lambda_a \geq \lambda_{a_2}, \\
& \vec{\alpha} \cdot \vec{F}_2  \geq 0, \quad \forall \lambda_{\mathcal{I}} \geq\lambda_{\mathcal{I}_1}.
\end{align}
\end{subequations}
Rearranging Eq.~\eqref{eq:sumrule1b} and using these constraints, we get
\be \label{eq:gbound}
V \sum_{\lambda_a = \lambda_{a_1}} g^2_{a_1 a_1 a}  \leq -  \vec{\alpha} \cdot \vec{F}_1 \implies |g_{a_1 a_1 a_1}| \sqrt{V} \leq \sqrt{-  \vec{\alpha} \cdot \vec{F}_1},
\ee
i.e., an upper bound on the rescaling invariant combination $|g_{a_1 a_1 a_1}| \sqrt{V}$.  For simplicity we use the second inequality in Eq.~\eqref{eq:gbound}, although the first inequality is stronger when $\lambda_{a_1}$ is degenerate. To find the best upper bound, we search for an $\vec{\alpha}$ that maximizes the objective function $\vec{\alpha} \cdot \vec{F}_1$ while satisfying the constraints in Eqs.~\eqref{eq:alphacons3}. This is again a problem that can be formulated as an SDP \cite{Poland:2011ey}.  This SDP, and all others in the remainder of this section, are simple enough that they can be solved analytically using the \texttt{Minimize} function in \texttt{Mathematica}. 

\begin{figure}[h!]
\begin{center}
\epsfig{file=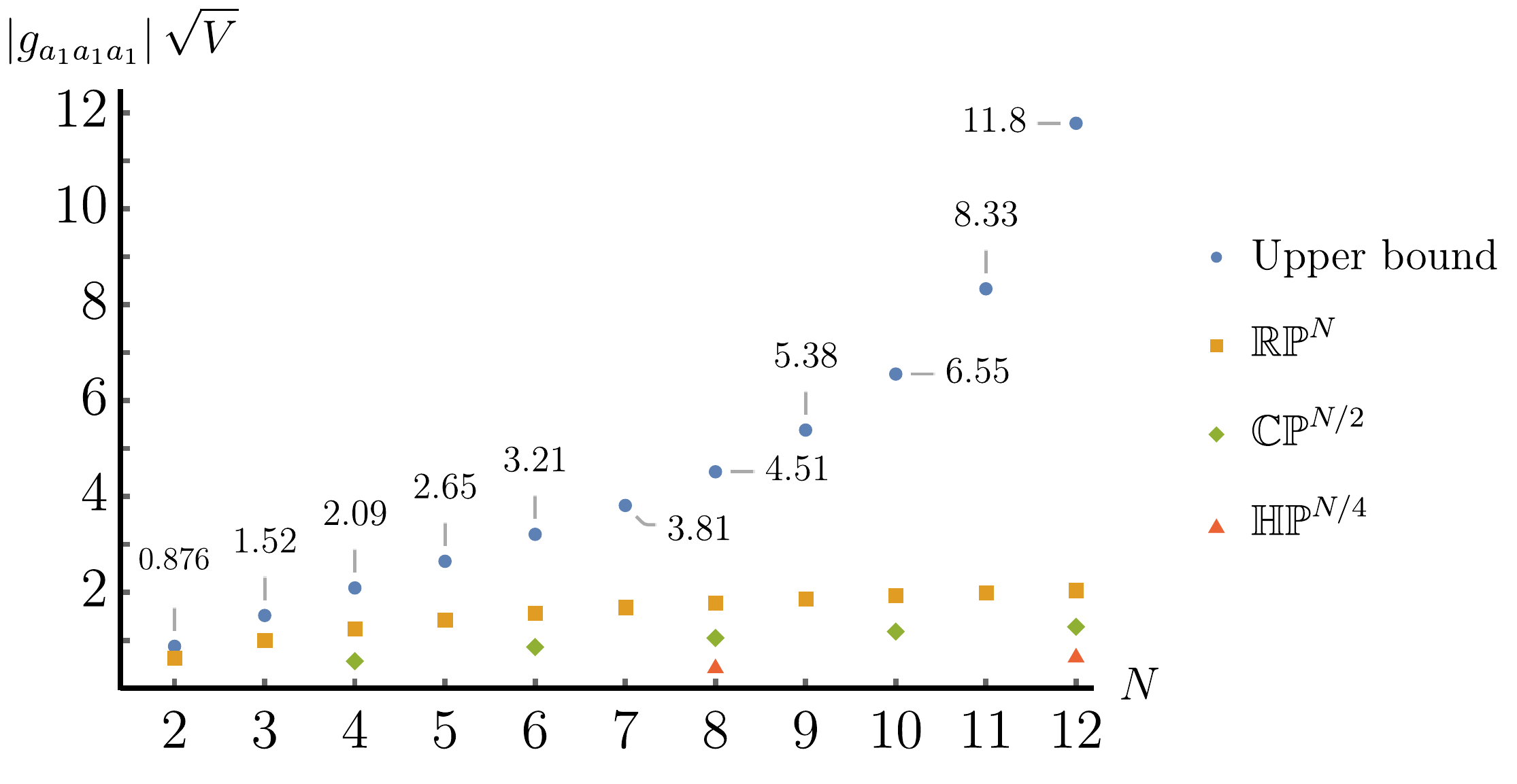, width=14cm}
\caption{Upper bounds on $|g_{a_1 a_1 a_1}|\sqrt{V}$ for closed Einstein manifolds with $R\geq 0$ and $N=2, \dots, 12$, assuming that $\lambda_a \geq \lambda_{a_1}$ and $\lambda_{\mathcal{I}} \geq 0$. The square, diamond, and triangular markers correspond to the lightest nontrivial zonal spherical functions on $\mathbb{RP}^N$, $\mathbb{CP}^{N/2}$, and $\mathbb{HP}^{N/4}$, respectively, with their standard metrics.}
\label{fig:self-coupling-bound-1}
\end{center}
\end{figure}

We start by searching for a bound on $|g_{a_1 a_1 a_1}|\sqrt{V}$ with $\lambda_{a_2} = \lambda_{a_1}$, i.e., assuming only that the nonzero scalar eigenvalues satisfy $\lambda_a \geq \lambda_{a_1}$.
We also assume for now that $\lambda_{\mathcal{I}}\geq0$. With these assumptions, we find finite upper bounds on $|g_{a_1 a_1 a_1}|\sqrt{V}$ for $N \leq 13$. These bounds are independent of $\lambda_{a_1}$, as required by rescaling invariance. We plot the bounds together with their rounded numerical values for $N\leq 12$ in Figure~\ref{fig:self-coupling-bound-1}. For $N=13$ the upper bound is $\approx 28.5$. The exact bounds are algebraic numbers, such as $16/\sqrt{111}$ for $N =3$, but mostly their exact forms are not very enlightening.
We also mark the points corresponding to the lightest nontrivial zonal spherical functions on the projective spaces $\mathbb{RP}^N$, $\mathbb{CP}^{N/2}$, and $\mathbb{HP}^{N/4}$ with their standard metrics. The overlap integrals of these eigenfunctions can be calculated by integrating products of Jacobi polynomials, as described in Appendix~\ref{app:examples}.
The upper bound increases with $N$, so for manifolds satisfying our assumptions we can get a lower bound on the number of extra dimensions if we know the strength of the cubic self-interaction of the lightest Kaluza--Klein state relative to the strength of gravity.

\begin{figure}[h!]
\begin{center}
\epsfig{file=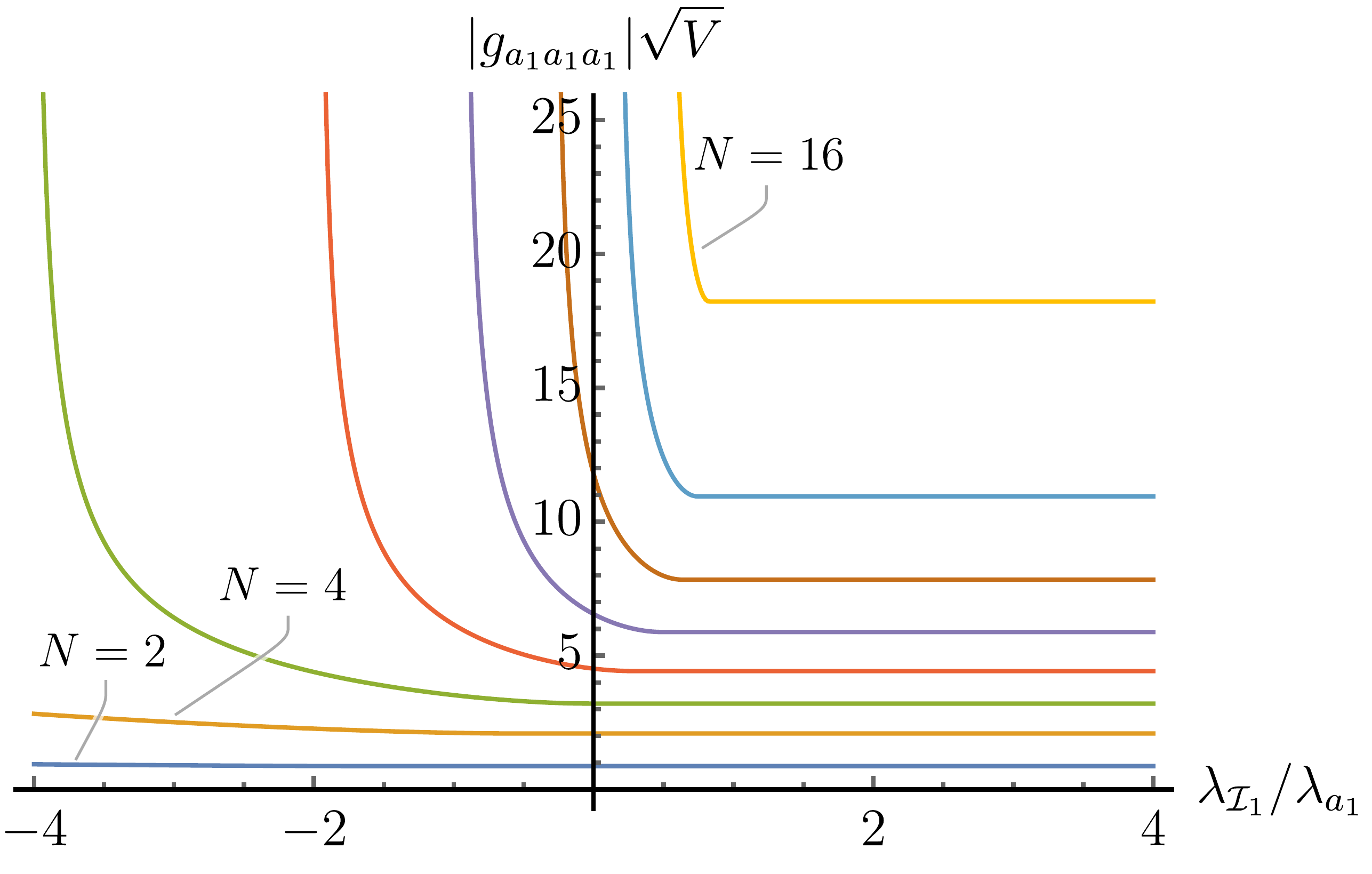, width=12cm}
\caption{Upper bounds on $|g_{a_1 a_1 a_1}| \sqrt{V}$ as a function of $\lambda_{\mathcal{I}_1}/\lambda_{a_1}$ for closed Einstein manifolds with $R \geq0$ and $N=2, 4, \dots, 16$, assuming that $\lambda_a \geq \lambda_{a_1}$ and $\lambda_{\mathcal{I}} \geq \lambda_{\mathcal{I}_1}$.}
\label{fig:self-coupling-bound-2}
\end{center}
\end{figure}

We now generalize the upper bounds on $|g_{a_1 a_1 a_1}| \sqrt{V}$ to allow for different values of $\lambda_{\mathcal{I}_1}/\lambda_{a_1}$. These are shown in Figure~\ref{fig:self-coupling-bound-2} for $N=2,4, \dots, 16$, where the restriction to even $N$ is just to keep the plots readable. The intercepts of the curves with the vertical axis give the bounds of Figure~\ref{fig:self-coupling-bound-1}. There exists a finite upper bound when $N \leq 17$ and $\lambda_{\mathcal{I}_1}/\lambda_{a_1} > w_N$, where $w_N$ is a finite constant. The first few of these constants are $w_2 = -(38+16 \sqrt{6})$, $w_3 \approx -19.380$, $w_4 \approx -9.7272$, $w_5 \approx -5.9714$, and $w_6 = -4$, where the two exact expressions hold to at least 16-digit precision. For large enough values of $\lambda_{\mathcal{I}_1}/\lambda_{a_1}$, the bounds become constant. 

\begin{figure}
\centering
\hspace*{.1cm}{\resizebox{12cm}{!}{%
\begin{tikzpicture}
\node at (0,0) {\includegraphics[width=15cm]{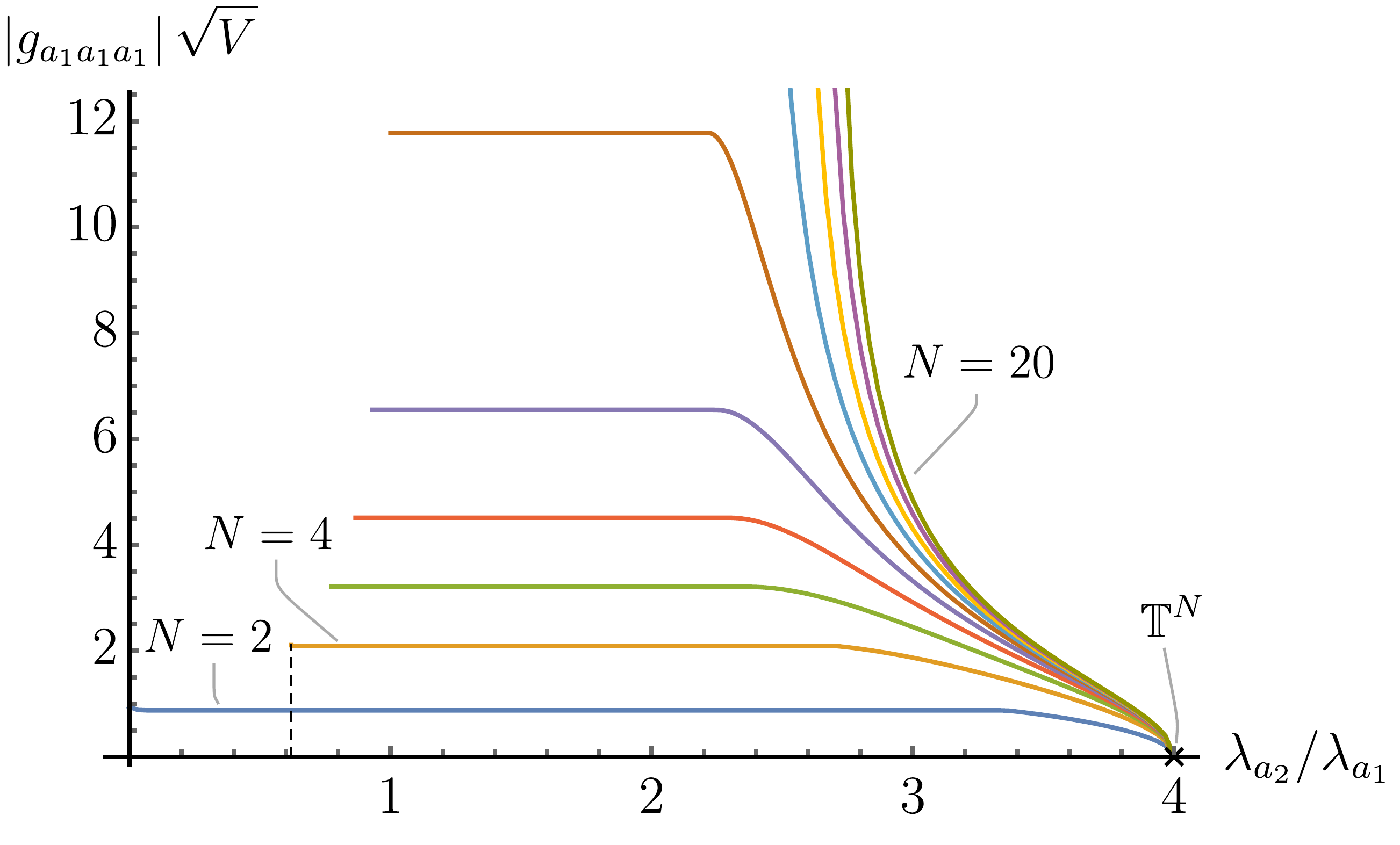}};
\node[scale=1.5] at (-4.42,-4.05) {$x_4$};
\end{tikzpicture}
}}
\caption{Upper bounds on $|g_{a_1 a_1 a_1}| \sqrt{V}$ for closed Einstein manifolds with $R \geq 0$ and $N=2, 4, \dots, 20$, assuming that $\lambda_{\mathcal{I}} \geq 0$ and $\lambda_a \in \{ \lambda_{a_1}\} \cup [\lambda_{a_2}, \infty)$.
}
\label{fig:self-coupling-bound-3}
\end{figure}

Next we explore the effect of varying $\lambda_{a_2}/\lambda_{a_1}$, assuming again that $\lambda_{\mathcal{I}}\geq 0$. 
In Figure~\ref{fig:self-coupling-bound-3} we plot the resulting bounds for $\lambda_{a_2}/\lambda_{a_1} \in (0, 4]$ for manifolds with $R\geq 0$ and $N=2,4, \dots ,10$. When $\lambda_{a_2}/\lambda_{a_1}=1$, we recover the bounds of Figure~\ref{fig:self-coupling-bound-1}. When $\lambda_{a_2}/\lambda_{a_1}=4$, which is the maximum possible value and is achieved by long flat tori, we find that the self-coupling must vanish. This is consistent with the vanishing of all self-couplings on flat tori, which follows from their $U(1)^N$ symmetry. The eigenfunction $\psi_{a_1}$ becomes completely general in the limit $\lambda_{a_2}/ \lambda_{a_1} \rightarrow 0$, so it is not surprising that in this limit there is no finite bound for $N>2$. More precisely, a finite bound exists for $\lambda_{a_2}/\lambda_{a_1} > x_N$, where $x_N$ is a constant that is positive for $N>2$.  For $N \leq 13$ the bounds approach a finite value as $\lambda_{a_2}/\lambda_{a_1} \rightarrow x_N^+$, so to high numerical precision they appear to be discontinuous at these points. It would be interesting if there is an explanation for this in terms of explicit manifolds. We list numerical approximations to $x_N$ for $N=2, \dots, 14$ and the corresponding bounds in Table~\ref{tab:self-coupling-bound-2}. To at least $15$-digit precision we have $x_3=12/(15+2\sqrt{30})$, $x_4=(16 \sqrt{6}-24)/10\sqrt{6}$, and $x_{14}=12/5$. 
\begin{table}[ht]
\centering
  \begin{tabular}{ c | c | c}
    $N$ & $x_N$ & $|g_{a_1 a_1 a_1}|\sqrt{V}$ \\ \thickhline
 2 & 0 & $2\sqrt{2}/3\approx 0.94281$ \\
3 & 0.46235 & 1.5187 \\
4 & 0.62020 & 2.1094 \\
5 & 0.70941 & 2.6671 \\
6 & 0.77426 & 3.2109 \\
7 & 0.82514 & 3.8127 \\
8 & 0.86641 & 4.5146 \\
9 & 0.90072 & 5.3844 \\
10 & 0.92980 & 6.5524 \\
11 & 0.95483 & 8.3291 \\
12 & 0.97666 & 11.781 \\
13 & 0.99588 & 28.459 \\
14 & 2.4000 & $\infty$
\end{tabular}
\caption{Upper bounds on the size of the cubic self-coupling $|g_{a_1 a_1 a_1}| \sqrt{V}$ for closed Einstein manifolds with $R \geq0$ and $\lambda_{\mathcal{I}} \geq 0$ as $\lambda_{a_2}/\lambda_{a_1} \rightarrow x_N^+$, where $\lambda_a \in \{ \lambda_{a_1}\} \cup [\lambda_{a_2}, \infty)$ and $x_N$ is the smallest value of  $\lambda_{a_2}/\lambda_{a_1}$ below which there is no finite bound.}
\label{tab:self-coupling-bound-2}
\end{table}

\subsubsection{Interactions with multiple spin-2 fields}

We now look for bounds on triple overlap integrals of the form 
\be
g_{a_1 a_1 a_2}  \equiv  \intN \psi_{a_1}^2 \psi_{a_2}.
\ee
In Kaluza--Klein reductions of gravity, these integrals give the strengths of the tree-level cubic interactions between two spin-2 particles of different masses. 

We assume again that $R \geq 0$ and that the scalar eigenvalues satisfy Eq.~\eqref{eq:evalueAssumption}.
The procedure is similar to before. 
We choose the vector $\vec{\alpha} \in \mathbb{R}^3$ to satisfy the following constraints:
\begin{subequations} \label{eq:112constraints}
\begin{align}
& \vec{\alpha} \cdot \vec{F}_3\Big|_{\lambda_a =\lambda_{a_2}} =1, \label{eq:norm112} \\
& \vec{\alpha} \cdot \vec{F}_2  \geq 0, \quad \forall \lambda_{\mathcal{I}} \geq\lambda_{\mathcal{I}_1}, \label{eq:LichCondition} \\
& \vec{\alpha} \cdot \vec{F}_k\Big|_{\lambda_a =\lambda_{a_1}} \geq 0, \quad k=3,4, \label{eq:condition1}\\
& \vec{\alpha} \cdot \vec{F}_k  \geq 0, \quad k=3,4, \quad \forall \lambda_a \geq \lambda_{a_2}. \label{eq:condition2}
\end{align}
\end{subequations}
With these constraints satisfied, we search for an $\vec{\alpha} \in \mathbb{R}^3$ that maximizes the objective function $\vec{\alpha} \cdot \vec{F}_1$, giving the bound
\be
V \sum_{\lambda_a = \lambda_{a_2}} g^2_{a_1 a_1 a}  \leq -  \vec{\alpha} \cdot \vec{F}_1 \implies |g_{a_1 a_1 a_2}| \sqrt{V} \leq \sqrt{-  \vec{\alpha} \cdot \vec{F}_1}.
\ee

\begin{figure}[h!]
\begin{center}
\epsfig{file=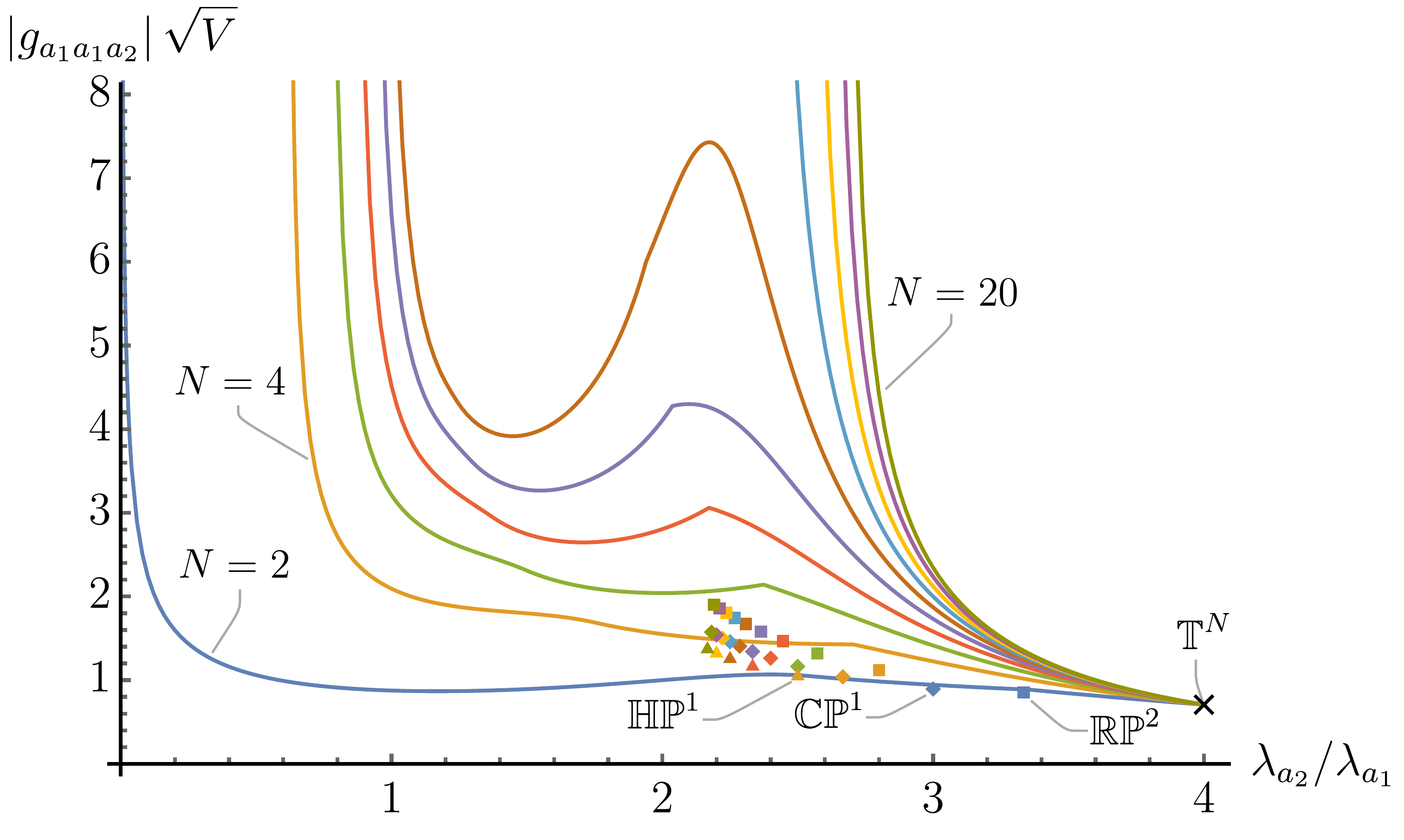, width=12cm}
\caption{Upper bounds on the size of the cubic coupling $|g_{a_1 a_1 a_2}| \sqrt{V}$ for closed Einstein manifolds with $R \geq 0$ and $N=2, 4, \dots, 20$ as we vary $\lambda_{a_2}/\lambda_{a_1}$, where $\lambda_a \in \{ \lambda_{a_1}\} \cup [\lambda_{a_2}, \infty)$ and $\lambda_{\mathcal{I}} \geq 0$. The cross corresponds to the lightest nontrivial modes on long flat $N$-tori. The square, diamond, and triangular markers correspond to the lightest nontrivial zonal spherical functions on $\mathbb{RP}^N$, $\mathbb{CP}^{N/2}$, and $\mathbb{HP}^{N/4}$ with their standard metrics.}
\label{fig:next-lightest-coupling-bound-1}
\end{center}
\end{figure}

Taking $\lambda_{\mathcal{I}} \geq 0$ gives the bounds shown in Figure~\ref{fig:next-lightest-coupling-bound-1} for $\lambda_{a_2}/\lambda_{a_1} \in (0, 4]$ and $N=2, 4, \dots, 20$. The upper bounds do not decrease monotonically as we increase $\lambda_{a_2}/\lambda_{a_1}$, unlike the previous bound. This is possible due to the fact that the normalization condition in Eq.~\eqref{eq:norm112} depends on $\lambda_{a_2}$, so a solution $\vec{\alpha}$ for some $\lambda_{a_2}$ is not necessarily a solution for larger $\lambda_{a_2}$. The bounds all approach $1/\sqrt{2}$ as $\lambda_{a_2}/\lambda_{a_1} \rightarrow 4$, which is the value attained by long flat tori and is marked with a cross in the plot. The bounds diverge for $\lambda_{a_2}/\lambda_{a_1} < x_N$, where $x_N$ are the constants encountered above.
We also mark the points corresponding to the overlap of the two lightest nontrivial zonal spherical functions on $\mathbb{RP}^N$, $\mathbb{CP}^{N/2}$,  and $\mathbb{HP}^{N/4}$ with their standard metrics.

\begin{figure}[h!]
\begin{center}
\epsfig{file=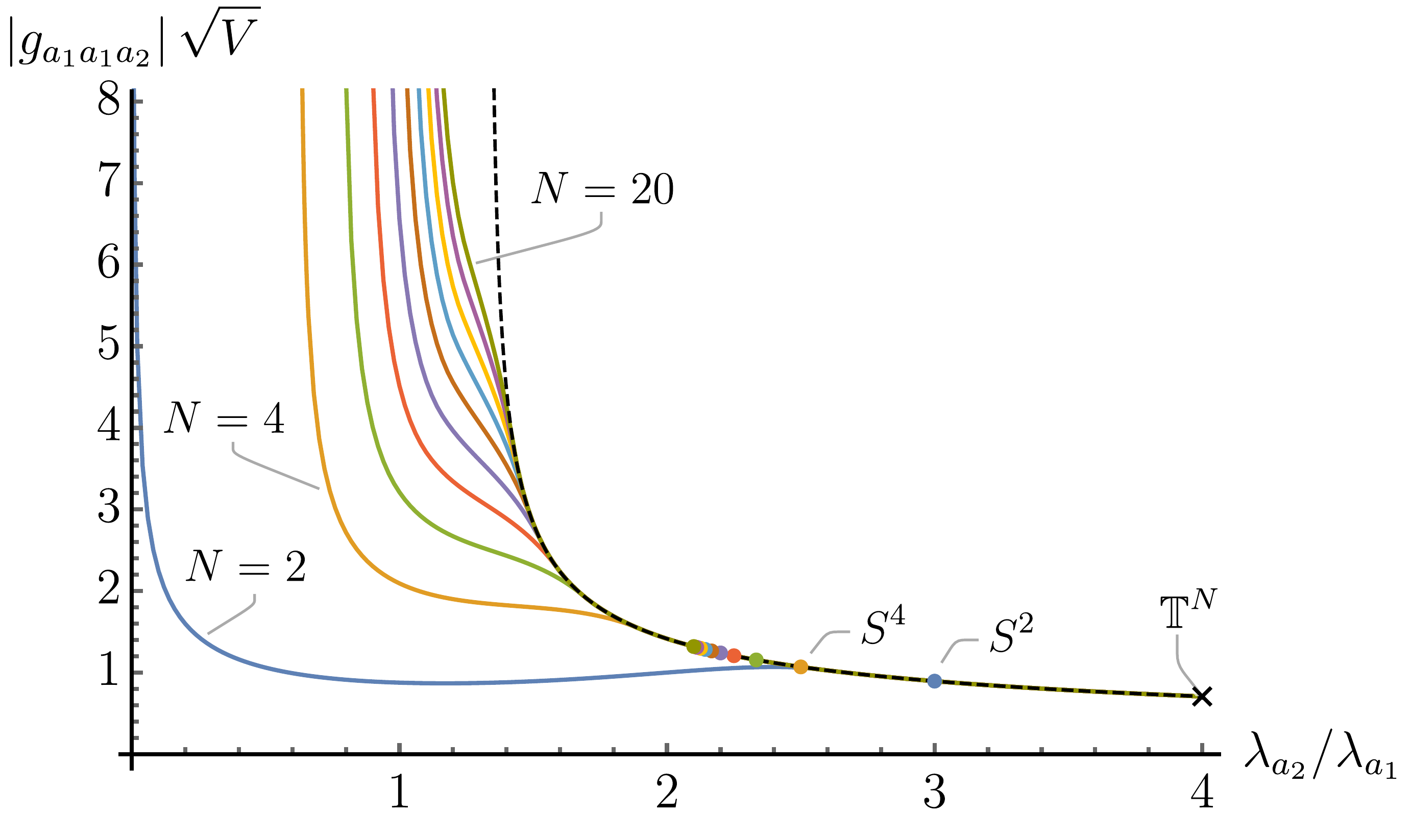, width=12cm}
\caption{Upper bounds on $|g_{a_1 a_1 a_2}| \sqrt{V}$ as we vary $\lambda_{a_2}/\lambda_{a_1}$ for closed Einstein manifolds with $R \geq 0$ and a $\mathbb{Z}_2$ symmetry, assuming that $\lambda_{\mathcal{I}} \geq 0$, $\lambda_a^- \geq \lambda_{a_1}$, and $\lambda_a^+ \geq \lambda_{a_2}$. The cross again corresponds to modes on long flat tori. The filled circles correspond to the overlaps of the lightest nontrivial zonal spherical harmonics on $S^N$. The dashed line is $2 (3 \lambda_{a_2}/\lambda_{a_1}-4)^{-1/2}$, which the bounds approach as $N \rightarrow \infty$. This plot is identical to Figure~\ref{fig:next-lightest-coupling-bound-1} in the region $\lambda_{a_2}/\lambda_{a_1} \leq 1$.}
\label{fig:next-lightest-coupling-bound-2}
\end{center}
\end{figure}

We can get stronger bounds by further imposing that $g_{a_1 a_1 a}$ vanishes if $\lambda_{a} = \lambda_{a_1}$. This is the case when the eigenfunctions with eigenvalue $\lambda_{a_1}$ are odd under a $\mathbb{Z}_2$ symmetry. Suppose also that the eigenfunctions with eigenvalue $\lambda_{a_2}$ are even, so the eigenvalues of the scalar Laplacian satisfy
\be
\lambda_a^- \in [\lambda_{a_1}, \infty ) , \quad \lambda_a^+ \in [ \lambda_{a_2}, \infty ),
\ee
where $\lambda^{\pm}_a$ are the nonzero eigenvalues of the even/odd eigenfunctions.
We implement this by dropping condition \eqref{eq:condition1}. The resulting bounds are shown in Figure~\ref{fig:next-lightest-coupling-bound-2} for $N=2, 4, \dots, 20$. We also mark the points corresponding to the overlap of the lightest nontrivial zonal spherical harmonics on $S^N$, which lie on the boundaries of the allowed regions.

\subsubsection{Spin-2 and scalar interactions}

We now look for bounds on triple overlap integrals involving an eigenfunction of the scalar Laplacian and a transverse traceless eigentensor of the Lichnerowicz Laplacian,
\be
g_{a_1 a_1 \mathcal{I}} \equiv \intN \partial^m \psi_{a_1} \partial^n \psi_{a_1} h_{m n, \mathcal{I}}^{TT} .
\ee 
In Kaluza--Klein reductions of gravity, these integrals control the strengths of the cubic interactions between a massive spin-2 mode and a spin-0 Lichnerowicz mode. 

\begin{subequations}
We look for bounds involving the lightest eigentensor, $h_{mn, \mathcal{I}_1}^{TT}$. We assume that  $\lambda_a \in \{ \lambda_{a_1}\} \cup [\lambda_{a_2}, \infty)$. For the case $R\geq 0$, we require the vector $\vec{\alpha} \in \mathbb{R}^3$ to satisfy the following constraints:
\begin{align}
& \vec{\alpha} \cdot \vec{F}_2\Big|_{\lambda_{\mathcal{I}} =\lambda_{\mathcal{I}_1}} =1, \\
& \vec{\alpha} \cdot \vec{F}_k\Big|_{\lambda_a =\lambda_{a_1}} \geq 0, \quad k=3,4, \\
& \vec{\alpha} \cdot \vec{F}_k  \geq 0, \quad k=3,4, \quad \forall \lambda_a \geq \lambda_{a_2}, \\
& \vec{\alpha} \cdot \vec{F}_2  \geq 0, \quad \forall \lambda_{\mathcal{I}} \geq\lambda_{\mathcal{I}_1}.
\end{align}
\end{subequations}
This then gives the inequality
\be
V \lambda_{a_1}^{-2}\sum_{\lambda_{\mathcal{I}} = \lambda_{\mathcal{I}_1}} g^2_{a_1 a_1 \mathcal{I}_1}  \leq -  \vec{\alpha} \cdot \vec{F}_1 \implies |g_{a_1 a_1 \mathcal{I}_1}| \lambda_{a_1}^{-1} \sqrt{V} \leq \sqrt{-  \vec{\alpha} \cdot \vec{F}_1}.
\ee
Maximizing the objective function $\vec{\alpha} \cdot \vec{F}_1$ subject to the above constraints then gives the best upper bound on the combination $|g_{a_1 a_1 \mathcal{I}_1}| \lambda_{a_1}^{-1} \sqrt{V}$.

\begin{figure}[h!]
\begin{center}
\epsfig{file=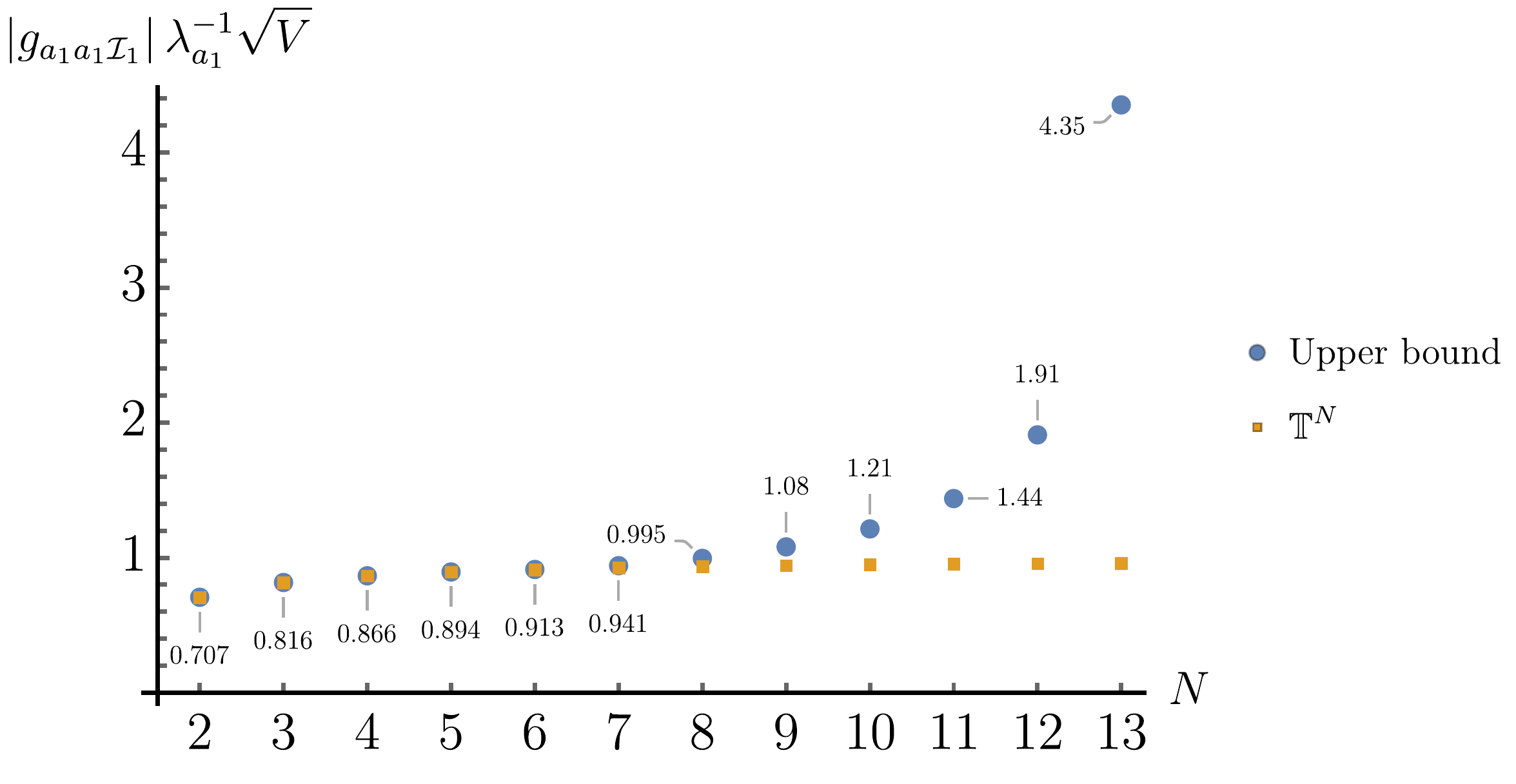, width=14cm}
\caption{Upper bounds on $|g_{a_1 a_1 \mathcal{I}_1}| \lambda_{a_1}^{-1} \sqrt{V}$ for closed Einstein manifolds with $R \geq 0$ and $N=2, \dots, 13$, where  $\lambda_{\mathcal{I}_1}=0$ and $\lambda_a \geq \lambda_{a_1}$. Some values for flat $N$-tori are also plotted.}
\label{fig:lichnerowicz-bound-1}
\end{center}
\end{figure}

The simplest case to consider is $\lambda_{\mathcal{I}_1}=0$ and $\lambda_{a_2} = \lambda_{a_1}$, so we have $\lambda_{\mathcal{I}} \geq 0$ and $\lambda_a \geq \lambda_{a_1}$. This is relevant, for example, to Ricci-flat manifolds with shape moduli and a parallel spinor. 
The resulting upper bounds are finite for $N\leq 13$ and are shown in Figure \ref{fig:lichnerowicz-bound-1}. We also plot the couplings for certain modes on flat $N$-tori,\footnote{For example, on the square flat $N$-torus we can take 
\be
\psi_{a_1} =\frac{\sqrt{2} \cos \theta_N}{(\sqrt{2\pi})^{N}}, \quad h_{mn, \mathcal{I}_1}^{TT}=\frac{{\rm diag}(1,\dots,1,1-N)}{\sqrt{(2\pi)^N N(N-1)}},
\ee
where $\theta_k \in [0, 2 \pi)$ parameterizes the $k$\textsuperscript{th} circle in $(S^1)^N$. By elongating the $N$\textsuperscript{th} circle we get $\lambda_{a_2}=4\lambda_{a_1}$.} 
\be \label{eq:toribound}
|g_{a_1 a_1 \mathcal{I}_1}| \lambda_{a_1}^{-1} \sqrt{V} = \sqrt{1-\frac{1}{N}}.
\ee
Interestingly, these tori modes saturate the upper bounds for $N=2, \dots, 6$, but not for $N\geq 7$.  The bounds do not change if we assume instead that $R=0$.

\begin{figure}
\centering
\hspace*{.1cm}{\resizebox{12cm}{!}{%
\begin{tikzpicture}
\node at (0,0) {\includegraphics[width=15cm]{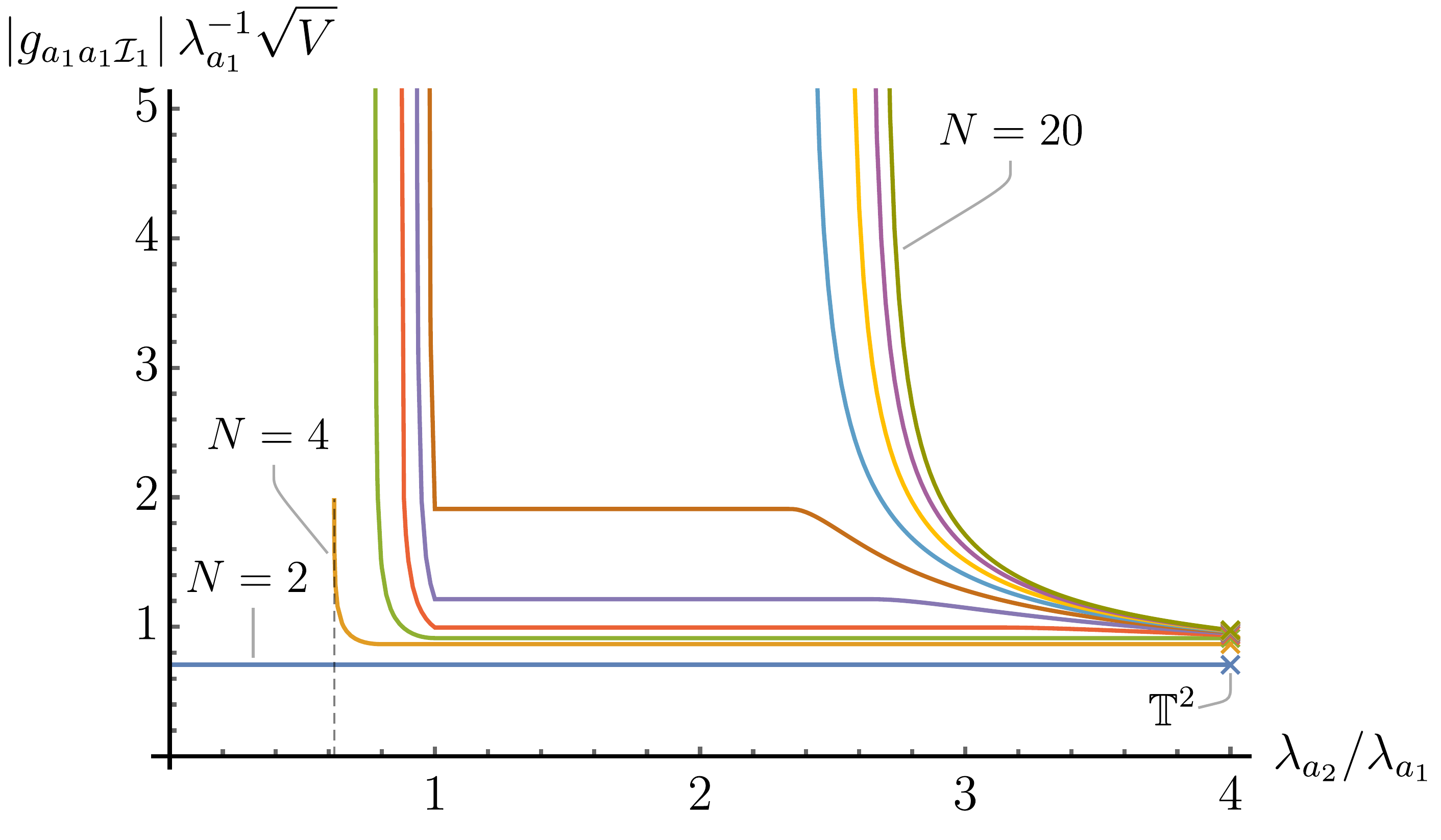}};
\node[scale=1.5] at (-4.05,-3.93) {$x_4$};
\end{tikzpicture}
}}
\caption{Upper bounds on $|g_{a_1 a_1 \mathcal{I}_1}| \lambda_{a_1}^{-1} \sqrt{V}$ as we vary $\lambda_{a_2}/\lambda_{a_1}$ for closed Einstein manifolds with $R\geq 0$, $N=2, 4, \dots, 20$, and a lightest Lichnerowicz eigentensor with $\lambda_{\mathcal{I}_1}=0$, where $\lambda_a \in \{ \lambda_{a_1} \} \cup [\lambda_{a_2}, \infty)$. The crosses at $\lambda_{a_2}/\lambda_{a_1}=4$ correspond to modes on long flat $N$-tori.}
\label{fig:lichnerowicz-bound-2}
\end{figure}

We can also find bounds when we vary $\lambda_{a_2}/\lambda_{a_1}$, keeping $\lambda_{\mathcal{I}_1}$ fixed at zero. We show these in Figure~\ref{fig:lichnerowicz-bound-2} for $N=2, 4, \dots, 20$. We also mark the points corresponding to the maximum couplings on flat $N$-tori given by Eq.~\eqref{eq:toribound} at $\lambda_{a_2}/\lambda_{a_1}=4$, which saturate the bounds for every $N$. A finite bound exists for $\lambda_{a_2}/\lambda_{a_1}> x_N$, where $x_N$ are the constants from Section~\ref{sec:g111}. For $N\leq 4$ the bound as $\lambda_{a_2}/\lambda_{a_1} \rightarrow x_N$ is finite, given by $1/\sqrt{2}$, $\approx 1.0491$, or $\approx 1.9773$ for $N=2, 3,$ or 4, respectively.

\begin{figure}
\centering
\hspace*{.1cm}{\resizebox{12cm}{!}{%
\begin{tikzpicture}
\node at (0,0) {\includegraphics[width=18cm]{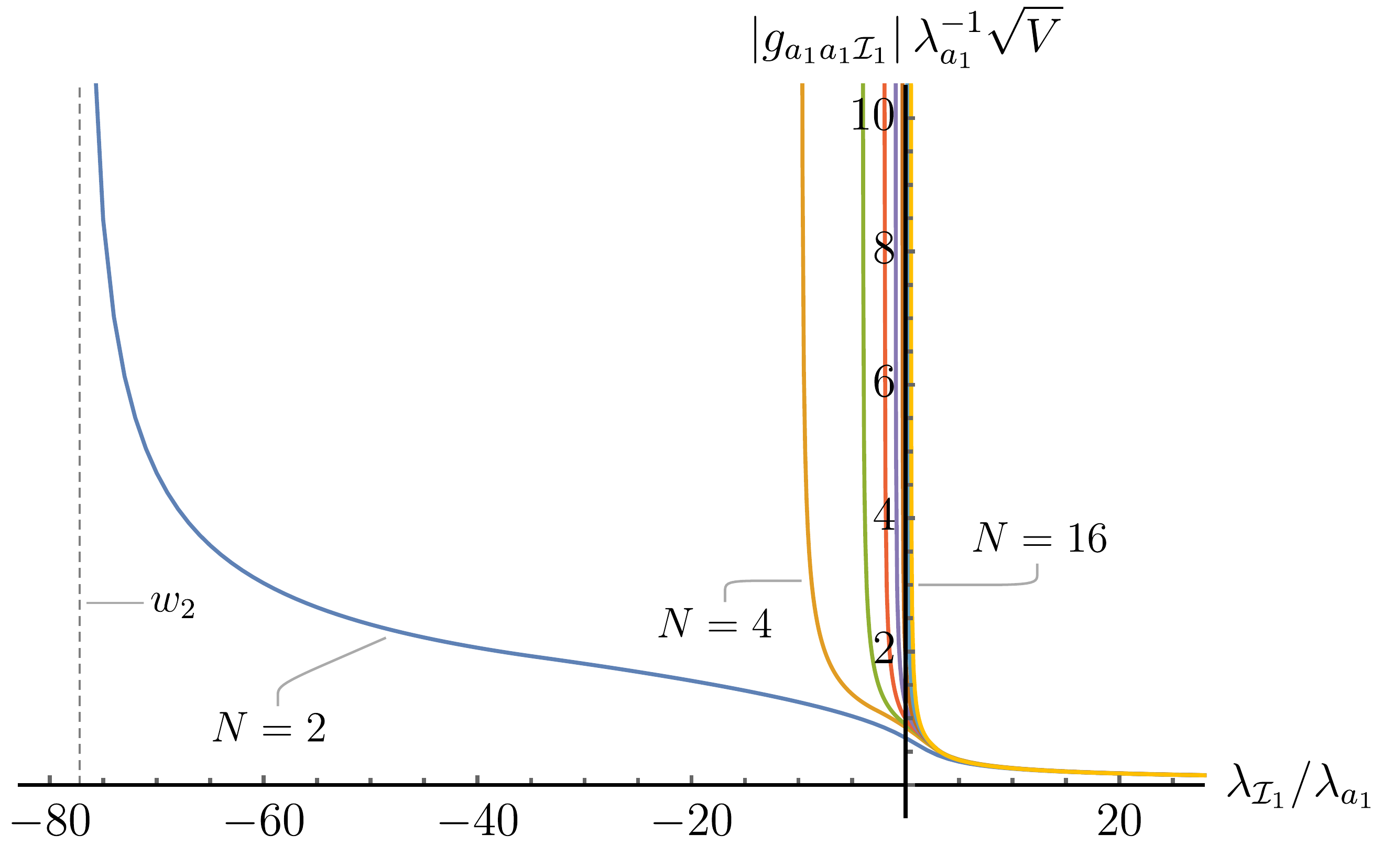}};
\node at (-2.5,2) {\includegraphics[width=6cm]{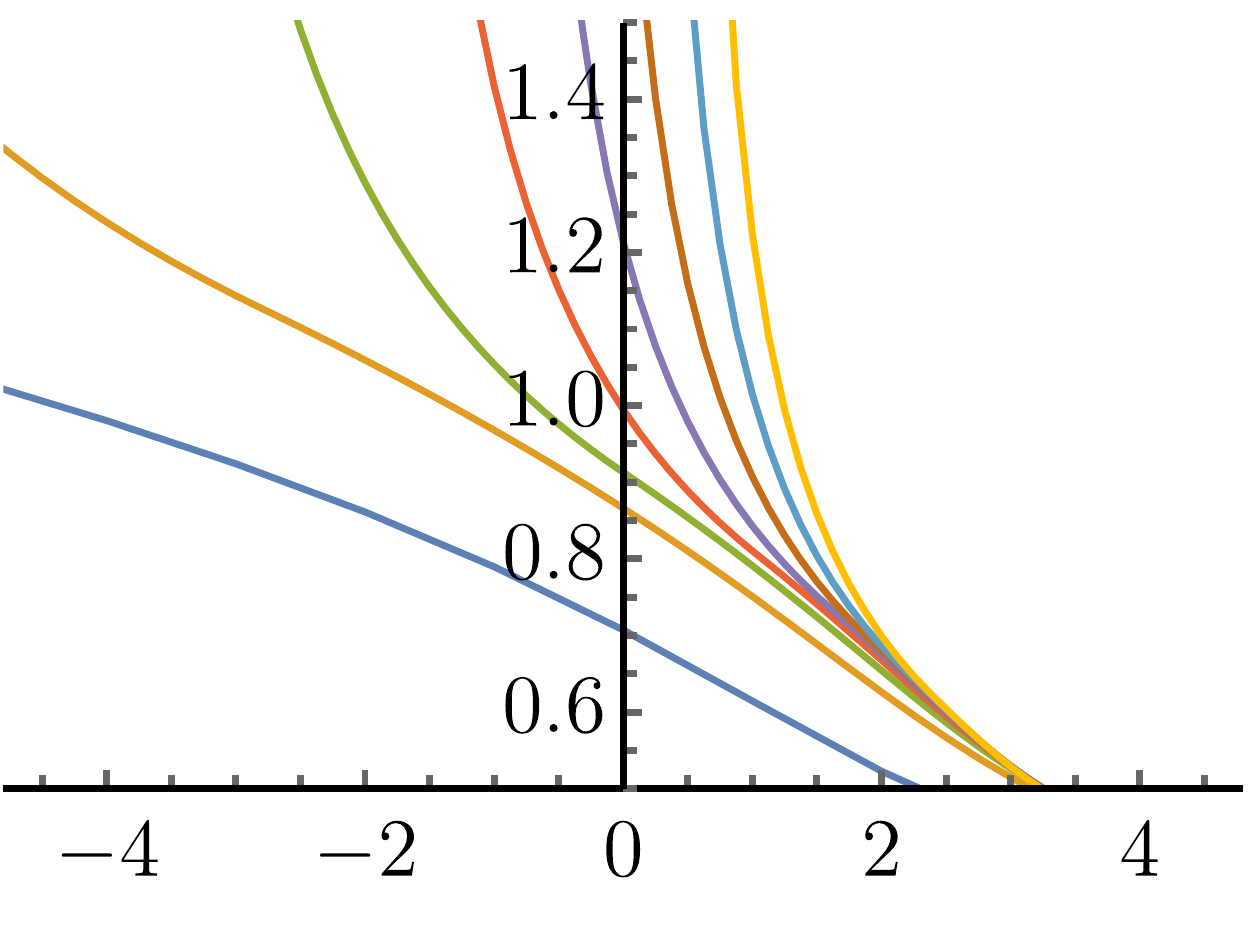}};
\draw[thick, dashed] (-5.5,-.2) -- (-5.5,4.15) -- (.5,4.15) -- (.5,-.2) -- cycle;
\draw[thick, dashed] (2.2,-4.015) -- (2.2,-3.257) -- (3.29,-3.257) -- (3.29,-4.015) -- cycle;
\end{tikzpicture}
}}
\caption{Upper bounds on $|g_{a_1 a_1 \mathcal{I}_1}| \lambda_{a_1}^{-1} \sqrt{V}$ as we vary $\lambda_{\mathcal{I}_1}/\lambda_{a_1}$ for closed  Einstein manifolds with $R\geq 0$ and $N=2, 4, \dots, 16$, assuming that $\lambda_a \geq\lambda_{a_1}$ and $\lambda_{\mathcal{I}} \geq\lambda_{\mathcal{I}_1}$.}
\label{fig:lichnerowicz-bound-3}
\end{figure}

We can similarly find bounds when we vary $\lambda_{\mathcal{I}_1}/\lambda_{a_1}$, assuming that $\lambda_a \geq \lambda_{a_1}$. We show these bounds for closed Einstein manifolds with $R\geq0$ and $N=2, 4, \dots, 16$ in Figure~\ref{fig:lichnerowicz-bound-3}. A finite bound exists when $N\leq 17$ and $\lambda_{\mathcal{I}_1}/\lambda_{a_1} > w_N$, where $w_N$ are the constants from Section~\ref{sec:g111}.

\section{Consistency conditions with two fixed eigenfunctions}

The consistency conditions we have been studying so far involve one fixed scalar eigenfunction. These correspond to sum rules for the scattering of identical massive spin-2 Kaluza--Klein particles in dimensional reductions of pure GR when $R=0$. We now study bounds obtained from more general consistency conditions involving two different fixed scalar eigenfunctions, which are related to four-point amplitudes with two different Kaluza--Klein external states. These are the analogue of mixed correlator bootstrap equations for CFTs \cite{Kos:2014bka}. 

The consistency conditions involving two fixed scalar eigenfunctions, $\psi_{a_1}$ and $\psi_{a_2}$, are given explicitly in Appendix~\ref{app:sumrules}, where we also explain how to derive them. When combined with the earlier consistency conditions, they can be put in the following form:
\begin{align} \label{eq:sumrule2}
&V^{-1}  \vec{F}_1+ \frac{1}{\lambda_{a_1}^2}\sum_{\mathcal{I}}
\begin{pmatrix}
g_{a_1 a_1 \mathcal{I}}  &  g_{a_2 a_2 \mathcal{I}}
\end{pmatrix}
\vec{F}_2
\begin{pmatrix}
g_{a_1 a_1 \mathcal{I}} \\
 g_{a_2 a_2 \mathcal{I}}
\end{pmatrix} 
+ \frac{1}{\lambda_{a_1}^2}\sum_{\mathcal{I}}  \vec{F}_{3}  g_{a_1 a_2 \mathcal{I}}^2
+ \sum_{i \notin I_{\rm Killing} } \frac{1}{\lambda_i}\left[ \vec{F}_{4}  + \frac{R\,  \vec{F}_{5} }{N \lambda_i -2 R } \right] g_{a_1 a_2 i}^2 \nn \\
&+ \frac{1}{\lambda_{a_1}}\sum_{i \in I_{\rm Killing} } \vec{F}_{6} \, g_{a_1 a_2 i}^2  +\sum_{a \notin I_{\rm conf.}} 
\begin{pmatrix}
g_{a_1 a_1 a} & g_{a_2 a_2 a}
\end{pmatrix}
\left[ \vec{F}_7
+
\frac{R\, \vec{F}_8}{ (N-1)\lambda_a- R}
\right]
\begin{pmatrix}
g_{a_1 a_1 a} \\
 g_{a_2 a_2 a}
\end{pmatrix} \nn \\
& + \sum_{a \notin I_{\rm conf.}}\frac{\lambda_{a_1}^2}{\lambda_a^2} \left[ \vec{F}_{9}+\frac{ R \, \vec{F}_{10} }{ (N-1)\lambda_a- R}  \right] g_{a_1 a_2 a}^2 +\sum_{a \in I_{\rm conf.}} \vec{F}_{11}g_{a_1 a_2 a}^2 = 0,
\end{align}
where $\vec{F}_k$ are 14-component vectors if $k \in \{1, 3, 4, 5, 6, 9, 10, 11 \}$ and 14-component vectors of $2 \times 2$ matrices if $k \in \{2, 7, 8\}$, with elements that are polynomial functions of the eigenvalue ratios
\be
\frac{\lambda_a}{\lambda_{a_1}}, \quad \frac{\lambda_{a_2}}{\lambda_{a_1}}, \quad \frac{\lambda_i}{\lambda_{a_1}}, \quad \frac{\lambda_{\mathcal{I}}}{\lambda_{a_1}}.
\ee
We include these $\vec{F}_k$'s explicitly in the ancillary notebook.
We have assumed that $a_1 \neq a_2$ by dropping terms proportional to $\delta_{a_1 a_2}$.

We can determine whether a given spectrum is consistent with these consistency conditions by again formulating the problem as an SDP, following the approach of the mixed correlator conformal bootstrap \cite{Kos:2014bka}. 

\subsection{Bounds on the third eigenvalue}

First we investigate how large we can make the third nonzero scalar eigenvalue relative to the first two eigenvalues. Suppose that the scalar eigenvalues satisfy
\be
\lambda_a \in \{ \lambda_{a_1}, \lambda_{a_2} \} \cup [\lambda_{a_3}, \infty)\, ,
\ee
and let $\psi_{a_k}$ be an eigenfunction with eigenvalue $\lambda_{a_k}$ for $k=1, 2$. We also take $\lambda_i \geq \lambda_{i_1}$ and $\lambda_{\mathcal{I}} \geq \lambda_{\mathcal{I}_1}$. 

Now consider closed Einstein manifolds with $R\geq 0$ and a candidate low-lying spectrum defined by the five eigenvalues $\lambda_{a_1}$, $\lambda_{a_2}$, $\lambda_{a_3}$, $\lambda_{i_1}$, and $\lambda_{\mathcal{I}_1}$.
We can rule out such a spectrum by finding an $\vec{\alpha} \in \mathbb{R}^{14}$ satisfying the following constraints:
\begin{subequations} \label{eq:mixedPMP1}
\begin{align}
& \vec{\alpha} \cdot \vec{F}_1  = 1,  \\ 
& \vec{\alpha} \cdot \vec{F}_k \succeq 0, \quad k=2,3, \quad \forall \, \lambda_{\mathcal{I}} \geq\lambda_{\mathcal{I}_1}, \\
& \vec{\alpha} \cdot \vec{F}_k  \geq 0, \quad k=4,5, \quad \forall \, \lambda_i \geq \lambda_{i_1}, \\
& \vec{\alpha} \cdot \vec{F}_k   \succeq 0, \quad k=7,8, 9, 10, \quad \forall \, \lambda_a \geq \lambda_{a_3}, \\
& \vec{\alpha} \cdot \vec{F}_k\Big|_{\lambda_a =\lambda_{a_1},\, \lambda_{a_2}} \succeq 0,  \quad k= 7,8, 9,10, \label{eq:mixed-7,8,9,10}
\end{align}
\end{subequations}
where $M \succeq 0$ means that the matrix $M$ is positive semidefinite.
If $\lambda_{a_1} = \lambda_{a_2}$, we additionally require that
\be
 \vec{\alpha} \cdot \vec{F}_6  \geq 0, \quad \forall \, \lambda_i \geq 0,
\ee
which ensures that the contributions from any Killing vectors are non-negative. Similarly, if we want to include round spheres then we must also ensure that the contributions from conformal scalars are non-negative,
\be
\vec{\alpha} \cdot \vec{F}_{11} \geq 0, \quad \forall \, \lambda_{a} > 0.
\ee
In practice, it is simpler to just manually append the sphere data to any exclusion plots if it is excluded by the other constraints.
Following Refs.~\cite{Kos:2014bka, Kos:2016ysd}, we can also implement the symmetry of $g_{a_1 a_2 a_3}$ by replacing the eight constraints in Eq.~\eqref{eq:mixed-7,8,9,10} with the four constraints
\begin{align} \label{eq:mixed-7,9}
\vec{F}_7\Big|_{\lambda_a =\lambda_{a_1}} + \frac{\lambda_{a_1}^2}{\lambda_{a_2}^2} \vec{F}_9 \Big|_{\lambda_a =\lambda_{a_2}}
\begin{pmatrix}
0 &0 \\
0 & 1
\end{pmatrix}
 \succeq 0, 
& \quad 
\vec{F}_7\Big|_{\lambda_a =\lambda_{a_2}} + \vec{F}_9 \Big|_{\lambda_a =\lambda_{a_1}}
\begin{pmatrix}
1 &0 \\
0 & 0
\end{pmatrix}
 \succeq 0, \\
\label{eq:mixed-8,10}
\vec{F}_8\Big|_{\lambda_a =\lambda_{a_1}} + \frac{\lambda_{a_1}^2}{\lambda_{a_2}^2} \vec{F}_{10} \Big|_{\lambda_a =\lambda_{a_2}}
\begin{pmatrix}
0 &0 \\
0 & 1
\end{pmatrix}
 \succeq 0, 
& \quad 
\vec{F}_8\Big|_{\lambda_a =\lambda_{a_2}} + \vec{F}_{10} \Big|_{\lambda_a =\lambda_{a_1}}
\begin{pmatrix}
1 &0 \\
0 & 0
\end{pmatrix}
 \succeq 0,
\end{align}
which can result in stronger bounds. 

\begin{figure}[h!]
\begin{center}
\epsfig{file=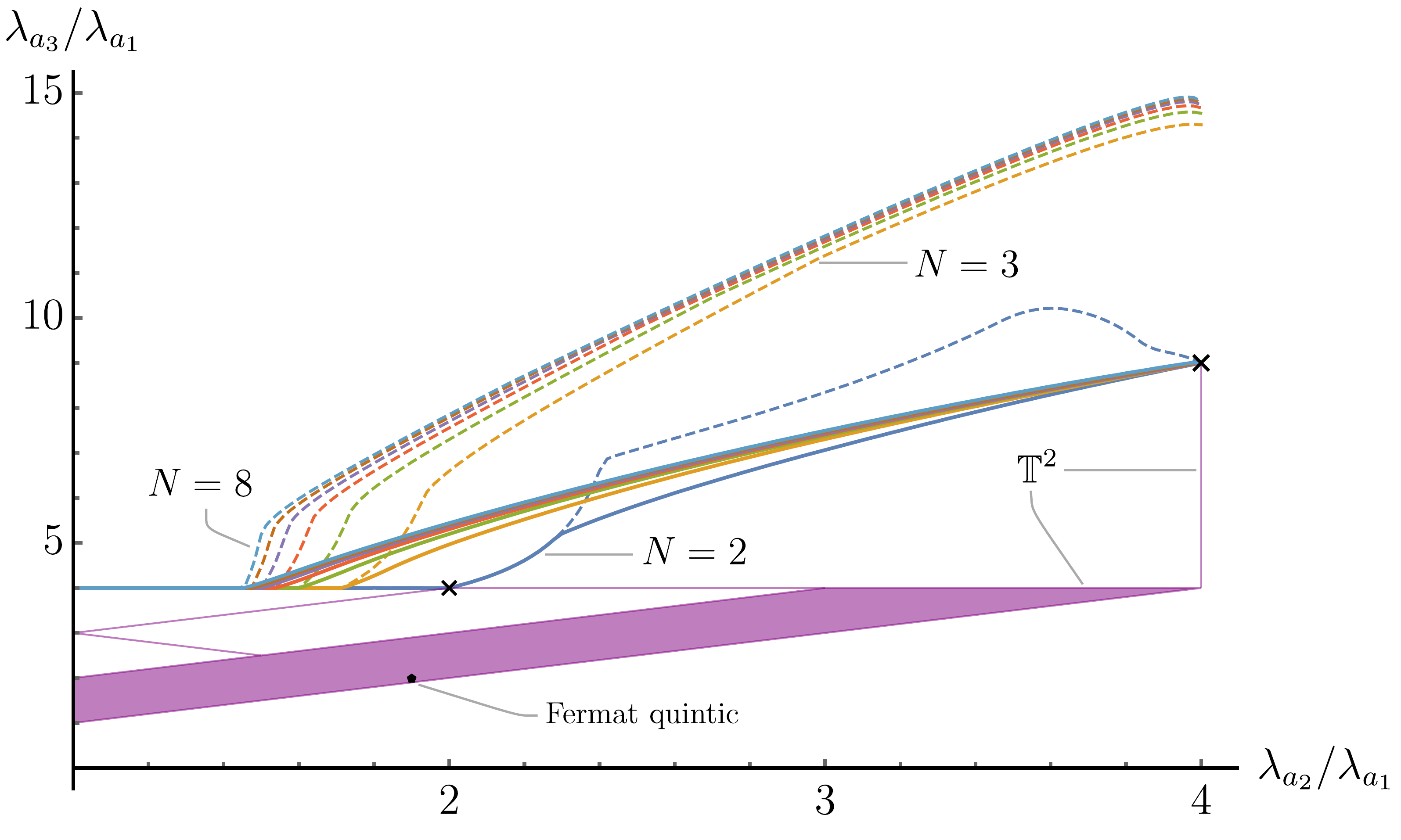, width=15cm}
\caption{
Upper bounds on $\lambda_{a_3}/\lambda_{a_1}$ on closed Einstein manifolds with $R \geq 0$ and $N=2, \dots, 8$, where $\lambda_a  \in \{ \lambda_{a_1}, \lambda_{a_2} \} \cup [\lambda_{a_3}, \infty)$ and we assume that $\lambda_{\mathcal{I}} \geq 0$. The bounds represented by the solid lines take into account the symmetry of $g_{a_1 a_2 a_3}$, while the bounds represented by the dashed lines do not. The purple region, including the various purple lines, covers the image of the moduli space of flat 2-tori. The crosses mark the square $2$-torus and very long flat $N$-tori, while the pentagon marks the eigenvalues of the Fermat quintic threefold found numerically in Ref.~\cite{Braun:2008jp}.}
\label{fig:mixed-bound-1}
\end{center}
\end{figure}

We take $\lambda_{\mathcal{I}} \geq 0$ and $\lambda_{i}\geq 0$.
For each  $\lambda_{a_2}/\lambda_{a_1}\in [1,4]$ we find an upper bound on $\lambda_{a_3}/\lambda_{a_1}$ as shown in Figure~\ref{fig:mixed-bound-1} for $N=2,3, \dots, 8$, where the solid lines are the bounds obtained when the symmetry of $g_{a_1 a_2 a_3}$ is implemented through Eqs.~\eqref{eq:mixed-7,9} and \eqref{eq:mixed-8,10}  and the dashed lines instead use Eq.~\eqref{eq:mixed-7,8,9,10}.  
The purple region, including the various purple lines, corresponds to the moduli space of flat 2-tori.\footnote{More precisely, this region covers the image of the moduli space. The intersection with the lines $\lambda_{a_1} =\lambda_{a_2}$ and $\lambda_{a_2} = \lambda_{a_3}$ should be excluded since we take the eigenvalues to be distinct.} The cross at $(4,9)$ corresponds to flat $N$-tori whose first three distinct nonzero eigenvalues coincide with those of a circle, which we call ``very long flat tori"---for $N=2$, these are the tori lying in the region of the canonical fundamental domain with $|\tau| \geq 3$---and the cross at $(2,4)$ corresponds to the square flat 2-torus. We also mark the Fermat quintic threefold using the eigenvalues found numerically in Ref.~\cite{Braun:2008jp}.
When using \texttt{SDPB} to produce these bounds, we found cases where to obtain the optimal bound it was necessary to disallow the program from terminating due to detection of a primal feasible solution (cf. the discussion in Ref.~\cite{Chester:2019ifh}). 

\begin{figure}[h!]
\begin{center}
\epsfig{file=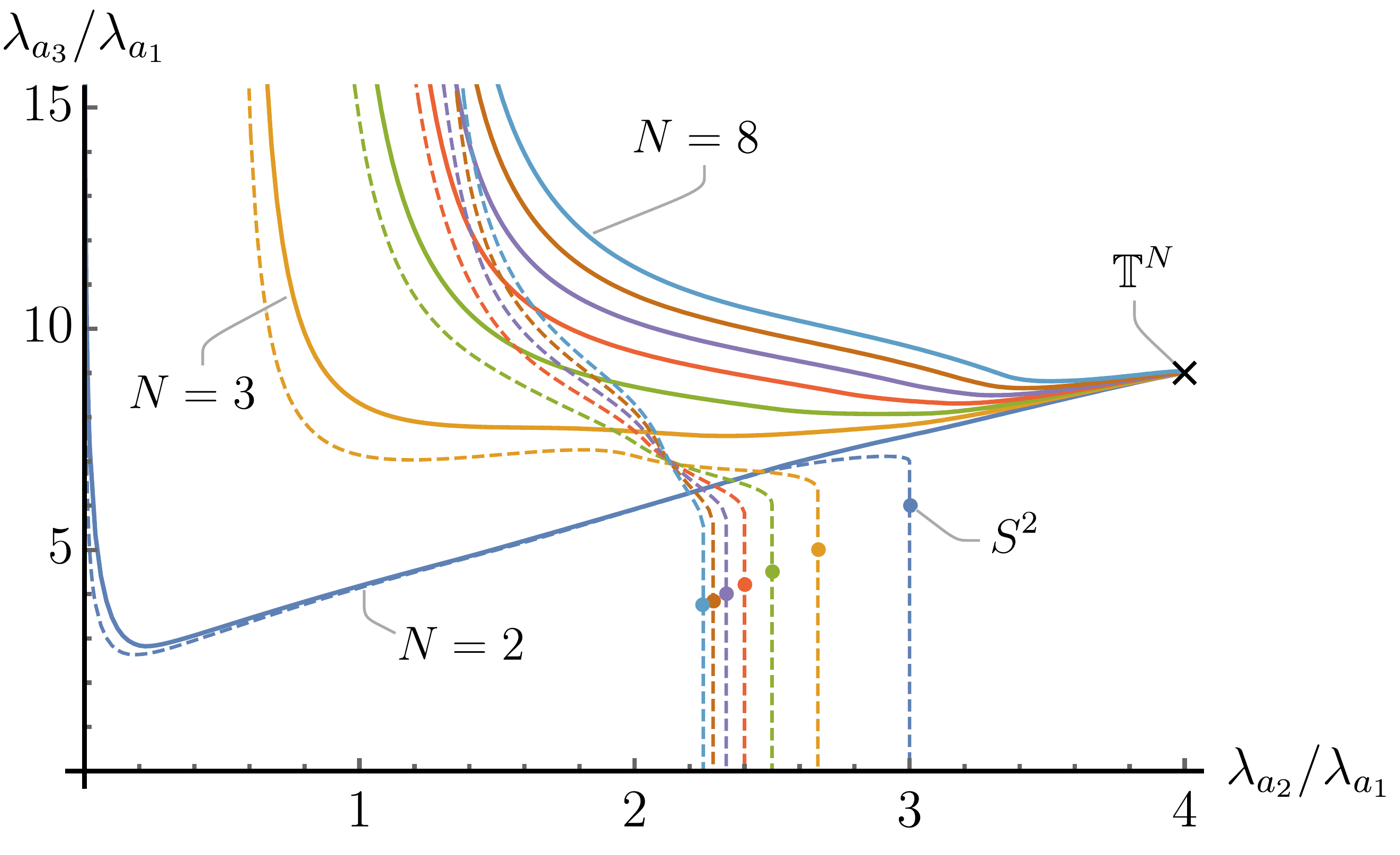, width=12cm}
\caption{Upper bounds on $\lambda_{a_3}/\lambda_{a_1}$ on closed Einstein manifolds with $R \geq 0$ and a $\mathbb{Z}_2$ symmetry for $N=2, \dots, 8$, where $\lambda_a^- \in \{ \lambda_{a_1} \} \cup [\lambda_{a_3}, \infty)$,  $\lambda_a^+ \geq \lambda_{a_2}$, and either $\lambda_{\mathcal{I}}\geq 0$ (solid lines) or $\lambda_{\mathcal{I}}/\lambda_{a_1} \geq 4$ (dashed lines). The filled circles correspond to the smallest distinct nonzero eigenvalues on $S^N$ and the cross corresponds to very long flat tori.}
\label{fig:mixed-bound-2}
\end{center}
\end{figure}

Suppose now that the manifold has a $\mathbb{Z}_2$ symmetry such that $\psi_{a_1}$ is odd and $\psi_{a_2}$ is even and let the nonzero scalar eigenvalues satisfy
\be
\lambda_a^- \in \{ \lambda_{a_1} \} \cup [\lambda_{a_3}, \infty)\, , \quad \lambda_a^+ \in  [\lambda_{a_2}, \infty)\, ,
\ee
where $\lambda_a^{\pm}$ are the eigenvalues of the even/odd eigenfunctions. The consistency conditions \eqref{eq:sumrule2} allow us to probe the ratio of the first two odd eigenvalues, $\lambda_{a_3}/\lambda_{a_1}$. 
We again consider closed Einstein manifolds with $R\geq 0$.
To find bounds we solve the SDP obtained from the subset of the conditions in Eqs.~\eqref{eq:mixedPMP1} that are consistent with the $\mathbb{Z}_2$ symmetry (imposing the symmetry of $g_{a_1 a_2 a_3}$ does not help in this case). We obtain an upper bound on $\lambda_{a_3}/\lambda_{a_1}$ for each $\lambda_{a_2}/\lambda_{a_1} \in (0, 4]$ as shown in Figure~\ref{fig:mixed-bound-2}, where we take $\lambda_{i}\geq 0$ and either $\lambda_{\mathcal{I}} \geq 0$ (solid lines) or $\lambda_{\mathcal{I}}/\lambda_{a_1} \geq 4$ (dashed lines). The very long flat $N$-tori at position $(4,9)$ saturate the $\lambda_{\mathcal{I}} \geq 0$ bounds, as marked by the cross, and the round spheres lie at the edge of the $\lambda_{\mathcal{I}}/\lambda_{a_1} \geq 4$ bounds, as shown by the filled circles.

\subsection{Bounds on vector couplings}
Now we use the consistency conditions \eqref{eq:sumrule2} to find bounds on the triple overlap integrals $g_{a_1 a_2 i}$. These integrals were not accessible with the earlier consistency conditions since they are antisymmetric in $a_1$ and $a_2$. Here we will focus on the case where the manifold has $R \geq 0$ and the transverse eigenvector corresponds to a Killing vector, so that $\lambda_i = 2 R/N$. This is possible when the manifold has a continuous symmetry. The quantity $g_{a_1 a_2 i}$ for $i \in I_{\rm Killing}$ can be non-vanishing only if $\lambda_{a_1} = \lambda_{a_2}$. In the case of GR with a Ricci-flat internal manifold with a continuous symmetry, this quantity measures the Kaluza--Klein charge of the massive spin-2 states, $q= \sqrt{2V}g_{a_1 a_2 i}M_d^{(2-d)/2}$.

\begin{figure}[h!]
\begin{center}
\epsfig{file=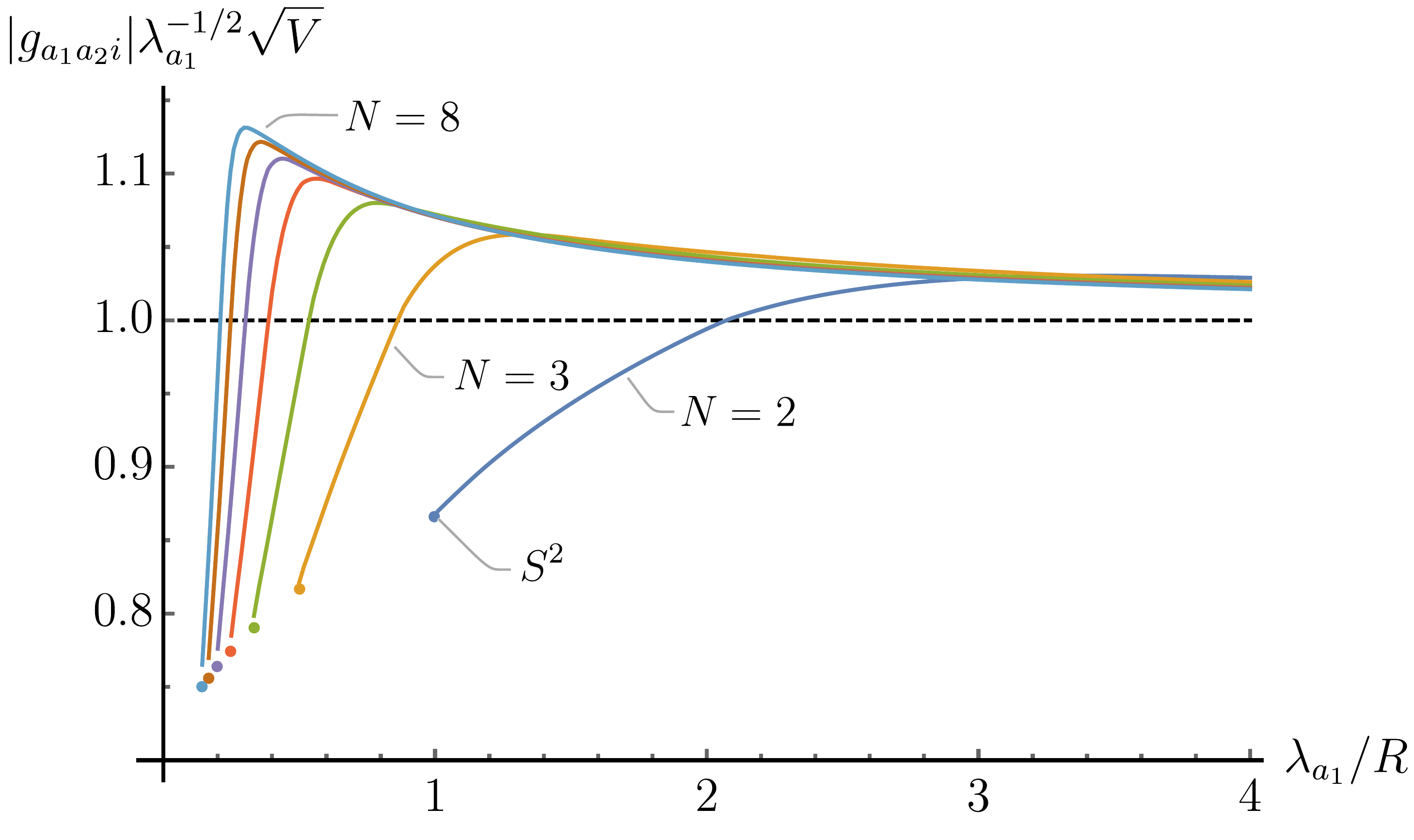, width=12cm}
\caption{Upper bounds on $|g_{a_1 a_2 i}|\lambda_{a_1}^{-1/2} \sqrt{V}$ for closed Einstein manifolds with $R >0$ and $N=2, \dots ,8$, where $\lambda_i = 2R/N$, $\lambda_{a_1} =\lambda_{a_2}$ is the smallest nonzero eigenvalue of the scalar Laplacian, and we have assumed that $\lambda_{\mathcal{I}} \geq 0$. The markers correspond to the round spheres and the dashed line is the bound for Ricci-flat manifolds with $N \leq 13$. }
\label{fig:charge-bound-1}
\end{center}
\end{figure}

Following the same approach as in Section~\ref{sec:coupling}, we can use the consistency conditions to obtain upper bounds on the following scale invariant combination:
\be
\sum_{i \in I_{\rm Killing}}g_{a_1 a_2 i}^2 \frac{V}{\lambda_{a_1}}.
\ee
In Figure~\ref{fig:charge-bound-1} we plot the upper bounds on the square root of this quantity for closed Einstein manifolds with $R>0$ and $N=2, \dots, 8$, where we take $\lambda_{a_1}=\lambda_{a_2}$ as the smallest nonzero eigenvalue of the scalar Laplacian and we assume that $\lambda_{\mathcal{I}} \geq 0$. 
We also mark points corresponding to certain modes on the round $N$-spheres,\footnote{These have $|g_{a_1 a_2 i}|\lambda_{a_1}^{-1/2} \sqrt{V}=\sqrt{(N+1)/2N}$ and can be obtained from $Y^m_i \propto (1,0, \dots, 0)$, corresponding to rotations in the $\theta_1$ direction, and $\psi_{a_1} = \bar{Y}_{1,1, \dots, 1}$, $\psi_{a_2} = \bar{Y}_{-1,1, \dots, 1}$, using the notation of Appendix \ref{app:examples}.} which lie just below the upper bounds. For $N>13$, the bounds diverge above some finite value of $\lambda_{a_1}/R$. The dashed line gives the upper bound for Ricci-flat manifolds with $N \leq 13$; this bound is saturated by flat tori, which is related to the saturation of the weak gravity conjecture by Kaluza--Klein modes in toroidal compactifications of gravity \cite{Heidenreich:2015nta}.

\section{Discussion}

We have developed a bootstrap approach to find bounds on the geometric data of closed Einstein manifolds. Specifically, our bounds constrain the eigenvalues and triple overlap integrals of eigenfunctions of various Laplacians. 
These bounds are found by starting with associativity conditions that must be satisfied by various higher-point integrals of eigenfunctions, then formulating them as semidefinite programming problems which can be solved numerically or analytically on a computer.  This approach mirrors the conformal bootstrap. As far as we know, the bounds we find are new results about Einstein manifolds that have not appeared in the mathematical literature.
Physically, these bounds can be thought of as constraints on the possible masses and three-point couplings of massive Kaluza--Klein modes in dimensional reductions of gravity.

There are several interesting directions that could be explored in the future. We have presented examples that illustrate some of the different types of bounds that can be deduced from the consistency conditions, but there should be many other examples that are worth exploring.
It would also be worthwhile to try to find more powerful versions of the constraints studied here by looking for additional consistency conditions coming from higher-derivative integrands. Having a systematic or recursive method for finding these would be useful. This may not be possible without further assumptions, such as restricting to quotients of maximally symmetric manifolds (which still includes nontrivial spaces such as compact hyperbolic manifolds). On a technical level, the obstruction is due to the inevitable appearance of Riemann tensors in higher-derivative integrands. More conceptually, this is connected to the restrictions imposed by the consistent propagation of higher-spin particles in nontrivial gravitational backgrounds \cite{Rahman:2020qal}. Perhaps string theory backgrounds satisfy additional consistency conditions that give more powerful bootstrap constraints.  

Our results take the form of general bounds that apply to many Einstein manifolds. It would be interesting to see if consistency conditions are powerful enough to isolate rigid manifolds in ``bootstrap islands" with only a small number of assumptions about the eigenvalues above the lightest modes, like what happens in the CFT bootstrap for strongly coupled CFTs such as the 3D Ising model at criticality \cite{Kos:2014bka, Kos:2016ysd}, and whether bootstrapping can be an efficient method to calculate the geometric data of manifolds of interest. This seems likely to require additional consistency conditions beyond those studied here. Other interesting directions would be to look for consistency conditions involving higher $p$-forms or fermions, as would occur in dimensional reductions of supergravity; to search for consistency conditions for spaces other than closed manifolds, such as bounded domains in Euclidean space; and to incorporate symmetry constraints from more complicated isometry subgroups.

\paragraph{Acknowledgements:} We would like to thank Cliff Cheung for early discussions about the idea of this paper. We also thank Anthony Ashmore, Latham Boyle, and Sergei Dubovsky for helpful discussions and Mohamed Boucetta, Uwe Semmelmann, and Gregor Weingart  for helpful correspondence.  The authors acknowledge support from DOE grant DE-SC0019143 and Simons Foundation Award Number 658908.

\appendix

\section{Compact rank one symmetric spaces}
\label{app:examples}
Here we briefly review some useful results about the geometric data of a few simple Einstein manifolds. We consider round spheres and the projective spaces over the reals $\mathbb{R}$, the complex numbers $\mathbb{C}$, the quaternions $\mathbb{H}$, and the octonians $\mathbb{O}$, each with their standard metrics. These are the compact symmetric spaces of rank one. 

\subsection{Eigenvalues and eigenfunctions}

\subsubsection*{Round spheres}
Consider the $N$-dimensional round sphere with unit radius, $S^N$. This is the set of points that are unit distance from the origin in $\mathbb{R}^{N+1}$ with the metric induced from the standard Euclidean metric. It can also be described as the symmetric space $SO(N+1)/SO(N)$. 
For $(x_1, \dots, x_{N+1}) \in \mathbb{R}^{N+1}$, we can parameterize the sphere with spherical coordinates 
\be
x_1 = \prod_{k=1}^N \sin \theta_k , \quad x_{\alpha} = \cos\theta_{\alpha-1}\prod_{k=\alpha}^N \sin \theta_k, \quad \alpha=2, \dots, N+1, 
\ee
where $0 \leq \theta_1 < 2 \pi$ and $0 \leq \theta_k \leq \pi$ for $2\leq k \leq N$.
The metric in these coordinates is
\begin{align}
& ds^2_{S^1}= d\theta_1^2, \quad ds^2_{S^N} =\sin^2\theta_{N} \, ds^2_{S^{N-1}} +   d \theta_N^2,  \quad N\geq 2.
\end{align}
The Ricci scalar and volume are
\be
R = N(N-1), \quad V = \frac{2 \pi^{\frac{N+1}{2}} }{\Gamma\left[(N+1)/2\right]}.
\ee

The eigenfunctions of the scalar Laplacian on $S^N$, i.e, the (higher-dimensional) spherical harmonics, are obtained from the harmonic homogeneous polynomials in $\mathbb{R}^{N+1}$ restricted to the sphere \cite{Berger1971}. They can be written in spherical coordinates in terms of associated Legendre polynomials \cite{Higuchi:1986wu}
\be
Y_{M_1, \dots, M_N} = \frac{e^{ i M_1 \theta_1}}{\sqrt{2 \pi}}  \prod_{k=2}^{N} \sqrt{\frac{(2M_k+k-1)(M_k+M_{k-1}+k-2)!}{2(M_k-M_{k-1})!}} \sin^{\frac{2-k}{2}}\theta_k P_{M_k+\frac{k-2}{2}}^{-(M_{k-1}+\frac{k-2}{2})} \left( \cos \theta_k \right),
\ee
where $M_k$ for $k=1, \dots, N$ are integers satisfying
\be
|M_1| \leq M_2 \leq \dots \leq M_{N-1} \leq M_N \equiv L.
\ee
The spherical harmonics with a given value of $L$ transform into each other under rotations, forming the rank-$L$ symmetric traceless tensor representation of $SO(N+1)$.
Under parity they transform by an overall factor of $(-1)^{L}$.
In general they are complex, but we can define real spherical harmonics as (restricting to $N >1$)
\be
\overline{Y}_{M_1, \dots, M_{N-1}, L} \equiv 
\begin{cases}
\left( Y_{-M_1, M_2, \dots, M_{N-1}, L}+ (-1)^{M_1} Y_{M_1, M_2, \dots, M_{N-1}, L} \right)/\sqrt{2} \quad & {\rm if} \quad M_1>0,  \\
Y_{M_1, \dots,M_{N-1}, L} \quad & {\rm if} \quad M_1=0,\\
i \left( Y_{M_1, M_2, \dots, M_{N-1}, L}- (-1)^{M_1} Y_{-M_1, M_2, \dots, M_{N-1}, L} \right)/\sqrt{2} \quad &  {\rm if} \quad M_1<0.
\end{cases}
\ee
The corresponding eigenvalues are
\be \label{eq:SphereEvalue}
\lambda = L(L+N-1).
\ee

A special subset of the spherical harmonics are the zonal spherical harmonics. These are eigenfunctions that are invariant under the $SO(N)$ rotations leaving a given point fixed. They are an example of zonal spherical functions, which are defined for general homogeneous spaces and always form a closed subsector of the consistency conditions. Taking the fixed point as $(0, \dots, 0, 1) \in \mathbb{R}^{n+1}$, the zonal harmonics can be written in terms of $\vec{x}^{\,2} \equiv x_1^2 + \dots + x_N^2$ and $x_{N+1}$.
For example, the first few nontrivial zonal spherical harmonics on $S^3$ with unit $L^2$-norm are
\begin{align}
L=1: & \quad  \frac{\sqrt{2}x_4 }{\pi}= \frac{\sqrt{2}\cos \theta_3}{\pi} , \\
L=2: & \quad \frac{3 x_4^2- \vec{x}^{\,2}}{\sqrt{2} \pi}= \frac{ \left( 1+2 \cos 2 \theta_3 \right)}{\sqrt{2} \pi} , \\
L =3: & \quad  \frac{2 \sqrt{2}x_4 \left( x_4^2-\vec{x}^{\,2} \right)}{\pi}=\frac{\sqrt{2}\left( \cos \theta_3 + \cos 3 \theta_3\right)}{\pi} .
\end{align}
In general, the normalized zonal spherical harmonics are the spherical harmonics with $M_k =0$ for $k=1, \dots, N-1$ and can be written in terms of Gegenbauer polynomials,
\be
Y_{0, \dots, 0, L} = \sqrt{\frac{(2 L+N-1) \Gamma[N-1]\Gamma\left[(N-1)/2\right]\Gamma[L+1]}{4 \pi^{(N+1)/2} \Gamma[L+N-1]}}C_{L}^{\frac{N-1}{2}}\left(\cos\theta_N\right).
\ee

Vector and tensor eigenmodes on $S^N$ can be obtained from contractions of certain mixed-symmetry constant tensors with the  coordinates of $\mathbb{R}^{N+1}$ \cite{Chodos:1983zi}. The eigenvalues of the Hodge Laplacian on transverse vectors are 
\be
\lambda = L(L+N-1)+N-2, \quad L \in \mathbb{Z}_{\geq 1},
\ee
and the eigenvalues of the Lichnerowicz Laplacian on symmetric transverse traceless 2-tensors are
\be
\lambda= L(L+N-1)+2(N-1), \quad L \in \mathbb{Z}_{\geq 2}, \quad N>2.
\ee

\subsubsection*{Real projective space}
The $N$-dimensional real projective space $\mathbb{RP}^N$ is $S^N/\mathbb{Z}_2$, where $\mathbb{Z}_2$ acts as the antipodal map on $S^N$. The eigenmodes on $\mathbb{RP}^N$ are thus proportional to the parity-even eigenmodes on $S^N$, i.e., the scalar spherical harmonics and transverse traceless eigentensors with even $L$ and the transverse eigenvectors with odd $L$.

\subsubsection*{Complex projective space}

The complex projective space $\mathbb{C P}^n$ with the Fubini--Study metric is the symmetric space $U(n+1)/(U(n) \times U(1))$.
Equivalently, it is the quotient space $S^{2n+1}/S^1$ obtained from the Hopf fibration. It has real dimension $N=2n$.
The Ricci scalar and volume are
\be
R = 4n(n+1), \quad V = \frac{\pi^n}{n!}.
\ee
The complex projective line $\mathbb{C P}^1$ is homothetic to $S^2$, i.e., isometric to a constant rescaling of $S^2$.

The eigenfunctions of the scalar Laplacian on $\mathbb{C P}^n$ are obtained by projecting the spherical harmonics on $S^{2n+1} \subset \mathbb{C}^{n+1} $ that are invariant under $U(1) =S^1$, where the action of $U(1)$ on $(z_1, \dots, z_{n+1}) \in \mathbb{C}^{n+1}$ is multiplication by a phase \cite{Berger1971}. The $U(1)$ invariants are thus $z_i \bar{z}_j$. From Eq.~\eqref{eq:SphereEvalue}, we get that the eigenvalues of the scalar Laplacian on $\mathbb{CP}^n$ are
\be
\lambda_k= 4k(k+n), \quad k \in \mathbb{Z}_{\geq 0}.
\ee
The zonal spherical functions fixing the point $(0, \dots, 0, 1) \in \mathbb{C}^{n+1}$ are the eigenfunctions that are further invariant under the action of $U(n)$ on $(z_1, \dots, z_n)$, so they can be written in terms of $|\vec{z}\, |^2 \equiv z_1 \bar{z}_1 + \dots + z_n \bar{z}_n$ and $|z_{n+1} |^2$. For example, the first few nontrivial zonal spherical functions on the complex projective plane $\mathbb{CP}^2$ with unit $L^2$-norm are 
\begin{align}
k=1: & \quad  \frac{2}{\pi} \left(|\vec{z}\, |^2-2 |z_{n+1} |^2 \right), \\
k=2: & \quad  \frac{\sqrt{6}}{\pi} \left( |\vec{z}\, |^4-6|\vec{z}\, |^2|z_{n+1} |^2+3 |z_{n+1} |^4\right), \\
k=3: & \quad  \frac{2\sqrt{2}}{\pi}\left(|\vec{z}\, |^6-12 |\vec{z}\, |^4|z_{n+1} |^2+18|\vec{z}\, |^2|z_{n+1} |^4-4|z_{n+1} |^6 \right) .
\end{align}

The smallest Lichnerowicz eigenvalue on transverse traceless symmetric 2-tensors on $\mathbb{CP}^n$ with the Fubini-Study metric is $\lambda_{\mathcal{I}_1}=32$ for $n=2$ and $\lambda_{\mathcal{I}_1}=8n$ for $n\geq 3$ \cite{Warner82, Boucetta07}.\footnote{We thank Uwe Semmelmann and Gregor Weingart for sharing with us an extended version of Ref.~\cite{semmelmann2020}, which helped us identify an error in the quoted value of $\lambda_{\mathcal{I}_1}$ for $\mathbb{CP}^n$ with $n\geq 3$ in an earlier version of this paper.}

\subsubsection*{Quaternionic projective space}
The quaternionic projective space $\mathbb{HP}^n$ with its standard metric is the quaternionic-K\"ahler symmetric space ${\rm Sp}(n+1)/({\rm Sp}(n) \times {\rm Sp}(1))$. Equivalently, it is the quotient space $S^{4n+3}/S^3$ obtained from the quaternionic Hopf fibration. It has real dimension $N=4n$. The Ricci scalar and volume are
\be
R = 16 n(n+2) \, , \quad V = \frac{\pi^{2n}}{(2n+1)!}\, .
\ee
The quaternionic projective line $\mathbb{HP}^1$ is homothetic to $S^4$. 

The eigenfunctions of the scalar Laplacian on $\mathbb{HP}^n$ are obtained by projecting the spherical harmonics on $S^{4n+3} \subset \mathbb{H}^{n+1}$ that are invariant under ${\rm Sp}(1) =S^3$, where the action of ${\rm Sp}(1)$ on $(q_1, \dots, q_{n+1}) \in \mathbb{H}^{n+1}$ is right multiplication by unit quaternions. Writing a quaternion as a pair of complex numbers, $q_i = z_i +  w_i j$, with quaternion conjugate $\bar{q}_i = \bar{z}_i-w_i j$, the ${\rm Sp}(1)$ invariants of interest are $z_i \bar{z}_j+w_i \bar{w}_j$ \cite{grinberg83}. From Eq.~\eqref{eq:SphereEvalue}, we get that the eigenvalues of the scalar Laplacian on $\mathbb{HP}^n$ are
\be
\lambda_k = 4 k(k+2n+1), \quad k \in \mathbb{Z}_{\geq 0}.
\ee
The zonal spherical functions on $\mathbb{HP}^n$ fixing the point $(0, \dots, 0, 1) \in \mathbb{H}^{n+1}$ are the eigenfunctions that are further invariant under the action of ${\rm Sp}(n)$ on $ (q_1, \dots, q_n)$, so they can be written in terms of $|\vec{q}\, |^2 \equiv q_1 \bar{q}_1 + \dots + q_n \bar{q}_n$ and $|q_{n+1}|^2 \equiv q_{n+1} \bar{q}_{n+1}$. For example, the first few nontrivial zonal spherical functions on the quaternionic projective plane $\mathbb{HP}^2$ with unit $L^2$-norm are
\begin{align}
k=1: & \quad \frac{2 \sqrt{105}}{\pi^2}\left(|\vec{q}\, |^2-2 |q_{n+1}|^2\right), \\
k=2: & \quad  \frac{6 \sqrt{3}}{\pi^2}\left(3|\vec{q}\, |^4-15|\vec{q}\, |^2|q_{n+1}|^2+10|q_{n+1}|^4 \right), \\
k=3: & \quad  \frac{2 \sqrt{462}}{\pi^2}\left(|\vec{q}\, |^6-9|\vec{q}\, |^4 |q_{n+1}|^2+15|\vec{q}\, |^2|q_{n+1}|^4-5|q_{n+1}|^6\right) .
\end{align}

Relatively little has been published about the Lichnerowicz eigenvalues of transverse traceless tensors on $\mathbb{HP}^n$ for $n>1$. 
It is possible to get lower bounds on the first such eigenvalue using representation theory. From the results of Refs.~\cite{Koiso1980, koiso1978} we get $\lambda_{\mathcal{I}} \geq 16 n$.

\subsubsection*{Octonionic projective plane}
The octonionic projective plane (or Cayley plane) $\mathbb{OP}^2$ with its standard metric is the symmetric space $F_4/{\rm Spin}(9)$. It has real dimension $N=16$. In our normalization its Ricci scalar, volume, and scalar Laplacian eigenvalues are \cite{Berger1971, Wolf1976}
\be
R = 144 \, , \quad V = \frac{3! (4\pi)^{8}}{11!} \, , \quad  \lambda_k = k(k+11), \quad k \in \mathbb{Z}_{\geq 0} \,,
\ee
and from Refs.~\cite{koiso1978, Koiso1980} we get $\lambda_{\mathcal{I}} \geq 16$.
The octonionic projective line $\mathbb{OP}^1$ is homothetic to $S^8$ and $\mathbb{OP}^n$ does not exist for $n>2$ due to the non-associativity of the octonions. 

\subsection{Triple overlap integrals}
\label{app:overlaps}

The zonal spherical functions of the above manifolds can all be written in terms of Jacobi polynomials, $P_k^{(\alpha, \beta)}(x)$, which are a generalization of the Gegenbauer polynomials: the parameter $k=0,1,2,\dots$  labels the different eigenfunctions, the argument of the polynomial is $x = \cos d $, where $d$ is the geodesic distance between a given point on the manifold and the fixed point, and the other parameters are given by $\alpha = (N-2)/2$ and $\beta = \alpha, -1/2, 0, 1$, or $3$ for $S^N$, $\mathbb{RP}^{N}$, $\mathbb{CP}^{N/2}$, $\mathbb{HP}^{N/4}$, or $\mathbb{OP}^2$, respectively \cite{Gangolli1967}. Labelling the zonal spherical functions by $k_i$, this gives the following useful formula for their normalized triple overlap integrals, after substituting the appropriate values of $\alpha$ and $\beta$:
\be
\sqrt{V} g_{k_1 k_2 k_3} = \frac{n_{0}^{\alpha, \beta}}{n_{k_1}^{\alpha, \beta} n_{k_2}^{\alpha, \beta} n_{k_3}^{\alpha, \beta}} \int^1_{-1} d x \, (1-x)^{\alpha} (1+x)^{\beta} P_{k_1}^{(\alpha, \beta)}(x) P_{k_2}^{(\alpha, \beta)}(x) P_{k_3}^{(\alpha, \beta)}(x)\, ,
\ee
where $P_0^{(\alpha, \beta)}(x) =1$ and the normalization constants are given by
\begin{align}
\left( n_{k}^{\alpha, \beta} \right)^2& = \int^1_{-1} d x \, (1-x)^{\alpha} (1+x)^{\beta}\left[ P_{k}^{(\alpha, \beta)}(x)\right]^2 = \frac{2^{\alpha + \beta+1}}{2k+\alpha+\beta+1} \frac{\Gamma[ k+\alpha+1] \Gamma[ k+\beta+1]}{\Gamma[ k+\alpha+\beta+1] k!}\,.
\end{align}

\section{Deriving consistency conditions}
\label{app:sumrules}

In this appendix we show how to derive the eight consistency conditions involving two distinct fixed eigenfunctions, generalizing the discussion in Section~\ref{sec:sumrules}.

From the identity\footnote{Recall that the Wick contraction denotes that the indicated pair of fields is to be replaced by the appropriate eigenmode expansion from Section \ref{sec:expansion}. }
\be
\intN \contraction{}{\psi_{a_1} }{}{\psi_{a_1}}
\contraction{\psi_{a_1} \psi_{a_1}}{\psi_{a_2}}{}{\psi_{a_2}}
\psi_{a_1}\psi_{a_1}\psi_{a_2}\psi_{a_2}
=
\intN \contraction{}{\psi_{a_1} }{\psi_{a_1}}{\psi_{a_2}}
\contraction[2ex]{\psi_{a_1}}{ \psi_{a_1}}{\psi_{a_2}}{\psi_{a_2}}
\psi_{a_1}\psi_{a_1}\psi_{a_2}\psi_{a_2},
\ee
we get the consistency condition
\be
\sum_a \left(g_{a_1 a_1}{}^a  g_{a_2 a_2 a}-g_{a_1 a_2 a}^2 \right)+V^{-1}\left(1-\delta_{a_1 a_2} \right)=0.
\ee
From the two identities involving two-derivative integrands,
\be
\intN \contraction{}{\psi_{a_1} }{}{\psi_{a_2}}
\contraction{\psi_{a_1} \psi_{a_2}}{\partial_m\psi_{a_1}}{}{\partial^m\psi_{a_2}}
\psi_{a_1}\psi_{a_2}\partial_m\psi_{a_1}\partial^m\psi_{a_2}
=
\intN \contraction{}{\psi_{a_1} }{\psi_{a_2}}{\partial_m \psi_{a_1}}
\contraction[2ex]{\psi_{a_1}}{ \psi_{a_2}}{\partial_m \psi_{a_1}}{\partial^m \psi_{a_2}}
\psi_{a_1}\psi_{a_2}\partial_m\psi_{a_1}\partial^m\psi_{a_2}
=
\intN \contraction[2ex]{}{\psi_{a_1}}{\psi_{a_2} \partial_m \psi_{a_1}}{\partial^m \psi_{a_2}}
\contraction{\psi_{a_1}}{ \psi_{a_2}}{}{\partial_m \psi_{a_1}}
\psi_{a_1}\psi_{a_2}\partial_m\psi_{a_1}\partial^m\psi_{a_2},
\ee
we get the consistency conditions
\begin{align}
\sum_a \left( 2\left(\lambda_{a_1}+\lambda_{a_2}-\lambda_a \right) g_{a_1 a_2 a}^2-\lambda_a g_{a_1 a_1}{}{}^a g_{a_2 a_2 a}\right) +4V^{-1} \lambda_{a_1} \delta_{a_1 a_2} &=0, \\
4 \sum_i g_{a_1 a_2 i}^2 + \sum_a \lambda_a^{-1} \left((\lambda_{a_1}-\lambda_{a_2})^2+2\lambda_a (\lambda_{a_1}+\lambda_{a_2})-3\lambda_a^2 \right) g_{a_1 a_2 a}^2 +4 V^{-1}\lambda_{a_1} \delta_{a_1 a_2} &=0.
\end{align}
From the three identities involving four-derivative integrands,
\begin{align}
\intN \contraction{}{\partial_{(m} \psi_{a_1} }{}{\partial_{n)} \psi_{a_2}}
\contraction{\partial_{(m}\psi_{a_1} \partial_{n)}\psi_{a_2}}{\partial^{(m}\psi_{a_1}}{}{\partial^{n)}\psi_{a_2}}
\partial_{(m}\psi_{a_1}  \partial_{n)}\psi_{a_2}\partial^{(m}\psi_{a_1}  \partial^{n)}\psi_{a_2}
& =
\intN \contraction{}{\partial_{(m} \psi_{a_1} }{\partial_{n)} \psi_{a_2}}{\partial^{(m}\psi_{a_1}}
\contraction[2ex]{\partial_{(m}\psi_{a_1}}{ \partial_{n)}\psi_{a_2}}{\partial^{(m}\psi_{a_1}}{\partial^{n)}\psi_{a_2}}
\partial_{(m}\psi_{a_1}  \partial_{n)}\psi_{a_2}\partial^{(m}\psi_{a_1}  \partial^{n)}\psi_{a_2}, \\
\intN \contraction{}{\partial_{m} \psi_{a_1} }{}{\partial_{n} \psi_{a_1}}
\contraction{\partial_{m}\psi_{a_1} \partial_{n}\psi_{a_1}}{\partial^{m}\psi_{a_2}}{}{\partial^{n}\psi_{a_2}}
\partial_{m}\psi_{a_1}  \partial_{n}\psi_{a_1}\partial^{m}\psi_{a_2}  \partial^{n}\psi_{a_2}
& =
\intN \contraction{}{\partial_{m} \psi_{a_1} }{\partial_{n} \psi_{a_1}}{\partial^{m}\psi_{a_2}}
\contraction[2ex]{\partial_{m}\psi_{a_1}}{ \partial_{n}\psi_{a_1}}{\partial^{m}\psi_{a_2}}{\partial^{n}\psi_{a_2}}
\partial_{m}\psi_{a_1}  \partial_{n}\psi_{a_1}\partial^{m}\psi_{a_2}  \partial^{n}\psi_{a_2}, \\
\intN \contraction{}{\partial_{m} \psi_{a_1} }{}{ \psi_{a_2}}
\contraction{\partial_{m} \psi_{a_1}  \psi_{a_2}}{ \Delta(\partial^m  \psi_{a_1} }{}{\psi_{a_2})}
\partial_{m} \psi_{a_1}  \psi_{a_2} \Delta(\partial^{m} \psi_{a_1}  \psi_{a_2})
& =
\intN \contraction{}{\partial_{m} \psi_{a_1}}{ \psi_{a_2}}{ \Delta(\partial^m  \psi_{a_1}}
\contraction[2ex]{\partial_{m} \psi_{a_1}}{  \psi_{a_2}}{ \Delta(\partial^m  \psi_{a_1} }{\psi_{a_2})}
\partial_{m} \psi_{a_1}  \psi_{a_2} \Delta(\partial^{m} \psi_{a_1}  \psi_{a_2}),
\end{align}
we get the consistency conditions
\begin{align}
& 8N \sum_{\mathcal{I}} g_{a_1 a_2 \mathcal{I}}^2-\sum_a\left[(N-2) \left( \lambda_{a_1} +\lambda_{a_2}-\lambda_a\right)^2 g_{a_1 a_2 a}^2 +N (2\lambda_{a_1} -\lambda_a) (2\lambda_{a_2} -\lambda_a)g_{a_1 a_1}{}^a g_{a_2 a_2 a}\right]\nn \\
& +4N^2\sum_{i \notin I_{\rm Killing}} \frac{(\lambda_{a_1} -\lambda_{a_2})^2g_{a_1 a_2 i}^2}{N \lambda_i -2 R}+\sum_{a\notin I_{\rm conf.}} \frac{\left( N(\lambda_{a_1}-\lambda_{a_2})^2-2\lambda_a(\lambda_{a_1}+\lambda_{a_2})-(N-2)\lambda_a^2\right)^2}{2\lambda_a \left( (N-1)\lambda_a -R\right)} g_{a_1 a_2 a}^2 \nn \\ 
&-4V^{-1}\lambda_{a_1}\lambda_{a_2} \left( (N-2)\delta_{a_1 a_2} +N\right)=0, \\
&4 N\sum_{\mathcal{I}} g_{a_1 a_1}{}^{\mathcal{I}} g_{a_2 a_2 \mathcal{I}}+ \sum_a \left[ \left(2\lambda_{a_1}-\lambda_a \right)\left(2\lambda_{a_2}-\lambda_a \right)g_{a_1 a_1}{}{}^a g_{a_2 a_2 a}-N\left( \lambda_{a_1}+\lambda_{a_2}-\lambda_a \right)^2 g_{a_1 a_2 a}^2\right] \nn \\
&+\sum_{a\notin I_{\rm conf.}} \frac{\lambda_a \left(4\lambda_{a_1}+(N-2)\lambda_a \right)\left(4\lambda_{a_2}+(N-2)\lambda_a \right)g_{a_1 a_1}{}{}^a g_{a_2 a_2 a}}{4\left((N-1)\lambda_a -R\right)} \nn \\
&+4V^{-1}\lambda_{a_1}\lambda_{a_2}(1-N\delta_{a_1 a_2})=0,\\
&4 \sum_i \lambda_i g_{a_1 a_2 i}^2+\sum_a \left[ \left(\lambda_a +\lambda_{a_1}-\lambda_{a_2} \right)^2 g_{a_1 a_2 a}^2 -(2\lambda_{a_1}-\lambda_a)(2\lambda_{a_1}+2\lambda_{a_2}-\lambda_a)g_{a_1 a_1}{}{}^a g_{a_2 a_2 a}\right] \nn \\
&-4 V^{-1}\lambda_{a_1}(\lambda_{a_1}+\lambda_{a_2})=0.
\end{align}
The remaining two consistency conditions are not straightforward to derive, even with guidance from the scattering amplitude results of Ref.~\cite{Bonifacio:2019ioc}.  They involve integrands that have six derivatives, which must be chosen carefully to cancel terms containing the Riemann tensor. One of them we get from
\begin{align}
&\intN \left[
\contraction{}{\partial_{m} \psi_{a_1} }{}{\partial_n \psi_{a_1}}
\contraction{\partial_{m} \psi_{a_1} \partial_n \psi_{a_1}}{ \Delta_L(\partial^m  \psi_{a_2} }{}{\partial^n\psi_{a_2})}
\partial_{m} \psi_{a_1}  \partial_n \psi_{a_1} \Delta_L(\partial^{m} \psi_{a_2} \partial^n \psi_{a_2})
+2 \contraction{}{\partial_{(m} \psi_{a_1} }{}{\partial_{n)} \psi_{a_2}}
\contraction{\partial_{(m} \psi_{a_1} \partial_{n)} \psi_{a_2}}{ \Delta_L(\partial^{(m}  \psi_{a_1} }{}{\partial^{n)}\psi_{a_2})}
\partial_{(m} \psi_{a_1}  \partial_{n)} \psi_{a_2} \Delta_L(\partial^{(m} \psi_{a_1} \partial^{n)} \psi_{a_2})
\right]
 \nn \\
=&
\intN \left[
\contraction{}{\partial_{m} \psi_{a_1}}{ \partial_n \psi_{a_1}}{ \Delta(\partial^m  \psi_{a_2}}
\contraction[2ex]{\partial_{m} \psi_{a_1}}{  \partial_n \psi_{a_1}}{ \Delta_L(\partial^m  \psi_{a_2} }{\partial^n\psi_{a_2})}
\partial_{m} \psi_{a_1} \partial_n \psi_{a_1} \Delta_L(\partial^{m} \psi_{a_2}\partial^n  \psi_{a_2}) 
+2 \contraction{}{\partial_{(m} \psi_{a_1}}{ \partial_{n)} \psi_{a_2}}{ \Delta(\partial^{(m}  \psi_{a_1}}
\contraction[2ex]{\partial_{(m} \psi_{a_1}}{  \partial_{n)} \psi_{a_2}}{ \Delta_L(\partial^{(m}  \psi_{a_1} }{\partial^{n)}\psi_{a_2})}
\partial_{(m} \psi_{a_1} \partial_{n)} \psi_{a_2} \Delta_L(\partial^{(m} \psi_{a_1}\partial^{n)}  \psi_{a_2})
\right],
\end{align}
where, after expanding the $\Delta_L$'s and adding a total derivative, the right-hand side can be written as
\begin{align}
\intN \Bigg[ &
2(\lambda_{a_1} + \lambda_{a_2})  \contraction{}{\partial_{m} \psi_{a_1} }{}{\partial^{m} \psi_{a_2}}
\contraction{\partial_{m}\psi_{a_1} \partial^{m}\psi_{a_2}}{\partial_{n}\psi_{a_1}}{}{\partial^{n}\psi_{a_2}}
\partial_{m}\psi_{a_1}  \partial^{m}\psi_{a_2}\partial_{n}\psi_{a_1}  \partial^{n}\psi_{a_2}
+ (\lambda_{a_1}+\lambda_{a_2}) \contraction{}{\partial_{m} \psi_{a_1} }{}{\partial^{m} \psi_{a_1}}
\contraction{\partial_{m}\psi_{a_1} \partial^{m}\psi_{a_1}}{\partial_{n}\psi_{a_2}}{}{\partial^{n}\psi_{a_2}}
\partial_{m}\psi_{a_1}  \partial^{m}\psi_{a_1}\partial_{n}\psi_{a_2}  \partial^{n}\psi_{a_2}  \nn \\
& -2 \contraction{}{\partial_{m} \psi_{a_1} }{}{\nabla_p \partial^{m} \psi_{a_1}}
\contraction{\partial_{m}\psi_{a_1} \nabla_p\partial^{m}\psi_{a_1}}{\partial_{n}\psi_{a_2}}{}{\nabla^p\partial^{n}\psi_{a_2}}
\partial_{m}\psi_{a_1}  \nabla_p\partial^{m}\psi_{a_1}\partial_{n}\psi_{a_2} \nabla^p \partial^{n}\psi_{a_2}
 - \contraction{}{\nabla_p(\partial_{m} \psi_{a_1} }{}{\partial^{m} \psi_{a_2})}
\contraction{\nabla_p(\partial_{m}\psi_{a_1} \partial^{m}\psi_{a_2})}{\nabla^p(\partial_{n}\psi_{a_1}}{}{\partial^{n}\psi_{a_2})}
\nabla_p (\partial_{m}\psi_{a_1}  \partial^{m}\psi_{a_2}) \nabla^p(\partial_{n}\psi_{a_1}  \partial^{n}\psi_{a_2})
\Bigg].
\end{align}
The remaining sum rule comes from
\begin{align}
& \intN \left[
\contraction{}{\partial_{m} \psi_{a_1} }{}{\partial_n \psi_{a_1}}
\contraction{\partial_{m} \psi_{a_1} \partial_n \psi_{a_1}}{ \Delta_L(\partial^m  \psi_{a_2} }{}{\partial^n\psi_{a_2})}
\partial_{m} \psi_{a_1}  \partial_n \psi_{a_1} \Delta_L(\partial^{m} \psi_{a_2} \partial^n \psi_{a_2})
-
\contraction{}{\Delta(\partial_{m} \psi_{a_1} }{}{ \psi_{a_2})}
\contraction{\Delta(\partial_{m} \psi_{a_1}  \psi_{a_2})}{ \Delta(\partial^m  \psi_{a_1} }{}{\psi_{a_2})}
\Delta(\partial_{m} \psi_{a_1}  \psi_{a_2}) \Delta(\partial^{m} \psi_{a_1}  \psi_{a_2})
 \right] \nn \\
=& \intN \left[
\contraction{}{\partial_{m} \psi_{a_1}}{ \partial_n \psi_{a_1}}{ \Delta_L(\partial^m  \psi_{a_2}}
\contraction[2ex]{\partial_{m} \psi_{a_1}}{  \partial_n \psi_{a_1}}{ \Delta_L(\partial^m  \psi_{a_2} }{\partial^n\psi_{a_2})}
\partial_{m} \psi_{a_1} \partial_n \psi_{a_1} \Delta_L(\partial^{m} \psi_{a_2}\partial^n  \psi_{a_2}) 
-
\contraction{}{\Delta(\partial_{m} \psi_{a_1}}{ \psi_{a_2})}{ \Delta(\partial^m  \psi_{a_1}}
\contraction[2ex]{\Delta(\partial_{m} \psi_{a_1}}{  \psi_{a_2})}{ \Delta(\partial^m  \psi_{a_1} }{\psi_{a_2})}
\Delta(\partial_{m} \psi_{a_1}  \psi_{a_2} )\Delta(\partial^{m} \psi_{a_1}  \psi_{a_2})
\right],
\end{align}
where, after expanding the Laplacians and adding a total derivative, the right-hand side can be written as
\begin{align}
\intN \Bigg[ &
\contraction{}{\nabla_{m}\partial_n \psi_{a_1}  }{}{  \nabla^{m}\partial^n\psi_{a_1}}
\contraction{\nabla_{m}\partial_n \psi_{a_1}  \nabla^{m}\partial^n\psi_{a_1}}{\partial_{p}\psi_{a_2}}{}{\partial^{p}\psi_{a_2}}
\nabla_{m}\partial_n \psi_{a_1}  \nabla^{m}\partial^n\psi_{a_1}\partial_{p}\psi_{a_2}  \partial^{p}\psi_{a_2} 
+
4 \contraction{}{\nabla_{m}\partial_n \psi_{a_1}  }{}{  \nabla^{m}\partial^n\psi_{a_2}}
\contraction{\nabla_{m}\partial_n \psi_{a_1}  \nabla^{m}\partial^n\psi_{a_2}}{\partial_{p}\psi_{a_1}}{}{\partial^{p}\psi_{a_2}}
\nabla_{m}\partial_n \psi_{a_1}  \nabla^{m}\partial^n\psi_{a_2}\partial_{p}\psi_{a_1}  \partial^{p}\psi_{a_2}
+
 \contraction{}{\nabla_{m}\partial_n \psi_{a_2}  }{}{  \nabla^{m}\partial^n\psi_{a_2}}
\contraction{\nabla_{m}\partial_n \psi_{a_2}  \nabla^{m}\partial^n\psi_{a_2}}{\partial_{p}\psi_{a_1}}{}{\partial^{p}\psi_{a_1}}
\nabla_{m}\partial_n \psi_{a_2}  \nabla^{m}\partial^n\psi_{a_2}\partial_{p}\psi_{a_1}  \partial^{p}\psi_{a_1}
  \nn \\
&+6
\contraction{}{\partial_{m} \psi_{a_1} }{}{\nabla_p \partial^{m} \psi_{a_1}}
\contraction{\partial_{m}\psi_{a_1} \nabla_p\partial^{m}\psi_{a_1}}{\partial_{n}\psi_{a_2}}{}{\nabla^p\partial^{n}\psi_{a_2}}
\partial_{m}\psi_{a_1}  \nabla_p\partial^{m}\psi_{a_1}\partial_{n}\psi_{a_2} \nabla^p \partial^{n}\psi_{a_2}
+4\left(\frac{R}{N} - \lambda_{a_1} \right)
 \contraction{}{\partial_{m} \psi_{a_1} }{}{\partial^{m} \psi_{a_2}}
\contraction{\partial_{m}\psi_{a_1} \partial^{m}\psi_{a_2}}{\partial_{n}\psi_{a_1}}{}{\partial^{n}\psi_{a_2}}
\partial_{m}\psi_{a_1}  \partial^{m}\psi_{a_2}\partial_{n}\psi_{a_1}  \partial^{n}\psi_{a_2} \nn \\
&- (\lambda_{a_1}^2 +2\lambda_{a_2}\lambda_{a_1} +2 \lambda_{a_2}^2)
\contraction{}{\partial_{m} \psi_{a_1} }{}{\partial^{m} \psi_{a_1}}
\contraction{\partial_{m}\psi_{a_1} \partial^{m}\psi_{a_1}}{\psi_{a_2}}{}{\psi_{a_2}}
\partial_{m}\psi_{a_1}  \partial^{m}\psi_{a_1}\psi_{a_2} \psi_{a_2} 
+2\lambda_{a_1} \lambda_{a_2}
\contraction{}{\partial_{m} \psi_{a_1} }{}{\partial^{m} \psi_{a_2}}
\contraction{\partial_{m}\psi_{a_1} \partial^{m}\psi_{a_2}}{\psi_{a_1}}{}{\psi_{a_2}}
\partial_{m}\psi_{a_1}  \partial^{m}\psi_{a_2}\psi_{a_1} \psi_{a_2} 
- \lambda_{a_1}^2
\contraction{}{\partial_{m} \psi_{a_2} }{}{\partial^{m} \psi_{a_2}}
\contraction{\partial_{m}\psi_{a_2} \partial^{m}\psi_{a_2}}{\psi_{a_1}}{}{\psi_{a_1}}
\partial_{m}\psi_{a_2}  \partial^{m}\psi_{a_2}\psi_{a_1} \psi_{a_1} \nn \\
& + 4(\lambda_{a_1} + \lambda_{a_2})
\contraction{}{\partial_{m} \psi_{a_1} }{}{\nabla_n \partial^{m} \psi_{a_1}}
\contraction{\partial_{m}\psi_{a_1} \nabla_n\partial^{m}\psi_{a_1}}{\psi_{a_2}}{}{\partial^n\psi_{a_2}}
\partial_{m}\psi_{a_1}  \nabla_n \partial^{m}\psi_{a_1}\psi_{a_2} \partial^n \psi_{a_2}
+ \frac{2R}{N}
\contraction{}{\partial_{m} \psi_{a_1} }{}{\partial^{m} \psi_{a_1}}
\contraction{\partial_{m}\psi_{a_1} \partial^{m}\psi_{a_1}}{\partial_{n}\psi_{a_2}}{}{\partial^{n}\psi_{a_2}}
\partial_{m}\psi_{a_1}  \partial^{m}\psi_{a_1}\partial_{n}\psi_{a_2}  \partial^{n}\psi_{a_2}
\Bigg].
\end{align}
These two identities give the consistency conditions
\small
\begin{align}
 & 4N\sum_i \lambda_i^2g_{a_1 a_2 i}^2+8N \sum_{\mathcal{I}} \lambda_{\mathcal{I}} g_{a_1 a_2 \mathcal{I}}^2 -4NV^{-1} \lambda_{a_1} \left( (1+2\delta_{a_1 a_2}) \lambda_{a_1}^2 +3 \lambda_{a_1} \lambda_{a_2} +2 \lambda_{a_2}^2\right)\nn \\
&+\sum_a \Big[ N  \left(-4\lambda_{a_1}(\lambda_{a_1} +\lambda_{a_2})(\lambda_{a_1} +2\lambda_{a_2})+4(\lambda_{a_1} +\lambda_{a_2})(2\lambda_{a_1} +\lambda_{a_2})\lambda_a-4(\lambda_{a_1} +\lambda_{a_2})\lambda_a^2+ \lambda_a^3\right) g_{a_1 a_1}{}^a g_{a_2 a_2 a} \nn \\
&+2 \left( -2N\lambda_{a_1}^2 (\lambda_{a_1} +\lambda_{a_2})+((1+4N)\lambda_{a_1}^2 +2\lambda_{a_1}\lambda_{a_2} +\lambda_{a_2}^2 )\lambda_a -2(\lambda_{a_1} +\lambda_{a_2})\lambda_a^2 +\lambda_a^3 \right) g_{a_1 a_2 a}^2 \Big] \nn \\
&+4N^2 \sum_{i \notin I_{\rm Killing}} \frac{\lambda_i (\lambda_{a_1} -\lambda_{a_2})^2 g_{a_1 a_2 i}^2}{N\lambda_i-2R}
+\sum_{a\notin I_{\rm conf.}} \frac{\left( N(\lambda_{a_1}-\lambda_{a_2})^2-2\lambda_a(\lambda_{a_1}+\lambda_{a_2})-(N-2)\lambda_a^2\right)^2g_{a_1 a_2 a}^2}{2((N-1)\lambda_a-R)} =0, \\
&16N\sum_{\mathcal{I}} \lambda_{\mathcal{I}} g_{a_1 a_1}{}^{\mathcal{I}} g_{a_2 a_2 \mathcal{I}}-16 N \sum_i \lambda_i^2 g_{a_1 a_2 i}^2 +16NV^{-1} \lambda_{a_1} \left(\lambda_{a_1}^2-2 (\delta_{a_1 a_2}-1)  \lambda_{a_1} \lambda_{a_2} + \lambda_{a_2}^2\right) \nn \\
&+\sum_a\Big[ 4 N  \left(2(\lambda_{a_1}+\lambda_{a_2})(\lambda_{a_1}^2-2\lambda_{a_1}\lambda_{a_2}-\lambda^2_{a_2})- (3\lambda_{a_1}^2 -10 \lambda_{a_1} \lambda_{a_2}-5\lambda_{a_2}^2)\lambda_a-4(\lambda_{a_1}+\lambda_{a_2})\lambda_a^2+\lambda_a^3 \right) g_{a_1 a_2 a}^2 \nn \\
& +\left(16 N\lambda_{a_1}(\lambda_{a_1}+\lambda_{a_2})^2-8(3N\lambda_{a_1}^2+(3N-2)\lambda_{a_1}\lambda_{a_2}+N\lambda_{a_2}^2 )\lambda_a+8(N-1)(\lambda_{a_1}+\lambda_{a_2})\lambda_a^2 -2(N-2)\lambda_a^3 \right) g_{a_1 a_1}{}^a g_{a_2 a_2 a} \Big] \nn \\
&+ \sum_{a\notin I_{\rm conf.}} \frac{\lambda_a^2 \left(4\lambda_{a_1}+(N-2)\lambda_a \right)\left(4\lambda_{a_2}+(N-2)\lambda_a \right)g_{a_1 a_1}{}^a g_{a_2 a_2 a}  }{(N-1)\lambda_a- R} =0.
\end{align}
\normalsize
These eight consistency conditions reduce to those given in Eq.~\eqref{eq:sumrule1} when $a_1=a_2$. 

\renewcommand{\em}{}
\bibliographystyle{utphys}
\addcontentsline{toc}{section}{References}
\bibliography{geometry-bootstrap-arxiv-v3}

\providecommand{\href}[2]{#2}\begingroup\raggedright\begin{thebibliography}{10}

\bibitem{Rattazzi:2008pe}
R.~Rattazzi, V.~S. Rychkov, E.~Tonni, and A.~Vichi, ``{Bounding scalar operator
  dimensions in 4D CFT},''
  \href{http://dx.doi.org/10.1088/1126-6708/2008/12/031}{{\em JHEP} {\bf 12}
  (2008)  031},
\href{http://arxiv.org/abs/0807.0004}{{\tt arXiv:0807.0004 [hep-th]}}.

\bibitem{Paulos:2016but}
M.~F. Paulos, J.~Penedones, J.~Toledo, B.~C. van Rees, and P.~Vieira, ``{The
  S-matrix bootstrap II: two dimensional amplitudes},''
  \href{http://dx.doi.org/10.1007/JHEP11(2017)143}{{\em JHEP} {\bf 11} (2017)
  143},
\href{http://arxiv.org/abs/1607.06110}{{\tt arXiv:1607.06110 [hep-th]}}.

\bibitem{Paulos:2017fhb}
M.~F. Paulos, J.~Penedones, J.~Toledo, B.~C. van Rees, and P.~Vieira, ``{The
  S-matrix bootstrap. Part III: higher dimensional amplitudes},''
  \href{http://dx.doi.org/10.1007/JHEP12(2019)040}{{\em JHEP} {\bf 12} (2019)
  040},
\href{http://arxiv.org/abs/1708.06765}{{\tt arXiv:1708.06765 [hep-th]}}.

\bibitem{Lin:2020mme}
H.~W. Lin, ``{Bootstraps to strings: solving random matrix models with
  positivite},'' \href{http://dx.doi.org/10.1007/JHEP06(2020)090}{{\em JHEP}
  {\bf 06} (2020)  090}, \href{http://arxiv.org/abs/2002.08387}{{\tt
  arXiv:2002.08387 [hep-th]}}.

\bibitem{Poland:2018epd}
D.~Poland, S.~Rychkov, and A.~Vichi, ``{The Conformal Bootstrap: Theory,
  Numerical Techniques, and Applications},''
  \href{http://dx.doi.org/10.1103/RevModPhys.91.015002}{{\em Rev. Mod. Phys.}
  {\bf 91} (2019)  015002},
\href{http://arxiv.org/abs/1805.04405}{{\tt arXiv:1805.04405 [hep-th]}}.

\bibitem{chavel1984}
I.~Chavel, B.~Randol, and J.~Dodziuk, {\em Eigenvalues in Riemannian Geometry}.
\newblock Pure and Applied Mathematics. Elsevier Science, 1984.

\bibitem{Headrick:2005ch}
M.~Headrick and T.~Wiseman, ``{Numerical Ricci-flat metrics on K3},''
  \href{http://dx.doi.org/10.1088/0264-9381/22/23/002}{{\em Class. Quant.
  Grav.} {\bf 22} (2005)  4931--4960},
  \href{http://arxiv.org/abs/hep-th/0506129}{{\tt arXiv:hep-th/0506129}}.

\bibitem{Donaldson2005}
S.~K. Donaldson, ``{Some numerical results in complex differential geometry},''
  \href{http://arxiv.org/abs/math/0512625}{{\tt arXiv:math/0512625 [math.DG]}}.

\bibitem{Douglas:2006rr}
M.~R. Douglas, R.~L. Karp, S.~Lukic, and R.~Reinbacher, ``{Numerical Calabi-Yau
  metrics},'' \href{http://dx.doi.org/10.1063/1.2888403}{{\em J. Math. Phys.}
  {\bf 49} (2008)  032302},
\href{http://arxiv.org/abs/hep-th/0612075}{{\tt arXiv:hep-th/0612075
  [hep-th]}}.

\bibitem{Braun:2008jp}
V.~Braun, T.~Brelidze, M.~R. Douglas, and B.~A. Ovrut, ``{Eigenvalues and
  Eigenfunctions of the Scalar Laplace Operator on Calabi-Yau Manifolds},''
  \href{http://dx.doi.org/10.1088/1126-6708/2008/07/120}{{\em JHEP} {\bf 07}
  (2008)  120},
\href{http://arxiv.org/abs/0805.3689}{{\tt arXiv:0805.3689 [hep-th]}}.

\bibitem{Carroll:2004st}
S.~M. Carroll, {\em {Spacetime and Geometry}}.
\newblock Cambridge University Press, 7, 2019.

\bibitem{Bonifacio:2019ioc}
J.~Bonifacio and K.~Hinterbichler, ``{Unitarization from Geometry},''
  \href{http://dx.doi.org/10.1007/JHEP12(2019)165}{{\em JHEP} {\bf 12} (2019)
  165},
\href{http://arxiv.org/abs/1910.04767}{{\tt arXiv:1910.04767 [hep-th]}}.

\bibitem{Besse:1987pua}
A.~L. Besse, {\em {Einstein Manifolds}}.
\newblock Springer-Verlag, Berlin, Heidelberg, New York,
1987.
\newblock

\bibitem{Hinterbichler:2013kwa}
K.~Hinterbichler, J.~Levin, and C.~Zukowski, ``{Kaluza-Klein Towers on General
  Manifolds},'' \href{http://dx.doi.org/10.1103/PhysRevD.89.086007}{{\em Phys.
  Rev. D} {\bf 89} (2014) no.~8, 086007},
  \href{http://arxiv.org/abs/1310.6353}{{\tt arXiv:1310.6353 [hep-th]}}.

\bibitem{obata1962}
M.~Obata, ``{Certain conditions for a Riemannian manifold to be isometric with
  a sphere},'' \href{http://dx.doi.org/10.2969/jmsj/01430333}{{\em J. Math.
  Soc. Japan} {\bf 14} (1962) no.~3, 333--340}.

\bibitem{Boyer:2003pe}
C.~P. Boyer, K.~Galicki, and J.~Kollar, ``{Einstein metrics on spheres},''
\href{http://arxiv.org/abs/math/0309408}{{\tt arXiv:math/0309408 [math-dg]}}.

\bibitem{Lichnerowicz}
A.~Lichnerowicz, ``Propagateurs et commutateurs en relativit\'e g\'en\'erale,''
  \href{http://dx.doi.org/10.1007/BF02684612}{{\em Publications Math\'ematiques
  de l'IH\'ES} {\bf 10} (1961) no.~1, 5--56}.

\bibitem{Gibbons:2002th}
G.~W. Gibbons, S.~A. Hartnoll, and C.~N. Pope, ``{Bohm and Einstein-Sasaki
  metrics, black holes and cosmological event horizons},''
  \href{http://dx.doi.org/10.1103/PhysRevD.67.084024}{{\em Phys. Rev.} {\bf
  D67} (2003)  084024},
\href{http://arxiv.org/abs/hep-th/0208031}{{\tt arXiv:hep-th/0208031
  [hep-th]}}.

\bibitem{Wang91}
M.~Y. Wang, ``{Preserving Parallel Spinors under Metric Deformations},'' {\em
  Indiana University Mathematics Journal} {\bf 40} (1991) no.~3, 815--844.
  \url{http://www.jstor.org/stable/24896310}.

\bibitem{Dai2005}
X.~Dai, X.~Wang, and G.~Wei, ``{On the stability of Riemannian manifold with
  parallel spinors},'' \href{http://dx.doi.org/10.1007/s00222-004-0424-x}{{\em
  Inventiones mathematicae} {\bf 161} (2005) no.~1, 151--176}.

\bibitem{Yau1978}
S.-T. Yau, ``{On The Ricci Curvature of a Compact Kähler Manifold and the
  Complex Monge-Ampére Equation, I},''
  \href{http://dx.doi.org/10.1002/cpa.3160310304}{{\em Communications on Pure
  and Applied Mathematics} {\bf 31} (1978) no.~3, 339--411}.

\bibitem{dai2005stability}
X.~Dai, X.~Wang, and G.~Wei, ``{On the Stability of Kähler-Einstein
  Metrics},'' \href{http://arxiv.org/abs/math/0504527}{{\tt arXiv:math/0504527
  [math.DG]}}.

\bibitem{Koiso1982}
N.~Koiso, ``{Rigidity and infinitesimal deformability of Einstein metrics},''
  \href{http://dx.doi.org/10.18910/8420}{{\em Osaka J. Math.} {\bf 19} (1982)
  no.~3, 643--668}.

\bibitem{Koiso1980}
N.~Koiso, ``{Rigidity and stability of Einstein metrics---the case of compact
  symmetric spaces},'' \href{http://dx.doi.org/10.18910/11783}{{\em Osaka J.
  Math.} {\bf 17} (1980) no.~1, 51--73}.

\bibitem{Csaki:2003dt}
C.~Csaki, C.~Grojean, H.~Murayama, L.~Pilo, and J.~Terning, ``{Gauge theories
  on an interval: Unitarity without a Higgs},''
  \href{http://dx.doi.org/10.1103/PhysRevD.69.055006}{{\em Phys. Rev. D} {\bf
  69} (2004)  055006}, \href{http://arxiv.org/abs/hep-ph/0305237}{{\tt
  arXiv:hep-ph/0305237}}.

\bibitem{Chivukula:2019rij}
R.~Sekhar~Chivukula, D.~Foren, K.~A. Mohan, D.~Sengupta, and E.~H. Simmons,
  ``{Scattering amplitudes of massive spin-2 Kaluza-Klein states grow only as
  ${\cal O}(s)$},'' \href{http://dx.doi.org/10.1103/PhysRevD.101.055013}{{\em
  Phys. Rev.} {\bf D101} (2020) no.~5, 055013},
\href{http://arxiv.org/abs/1906.11098}{{\tt arXiv:1906.11098 [hep-ph]}}.

\bibitem{Chivukula:2019zkt}
R.~Sekhar~Chivukula, D.~Foren, K.~A. Mohan, D.~Sengupta, and E.~H. Simmons,
  ``{Sum Rules for Massive Spin-2 Kaluza-Klein Elastic Scattering
  Amplitudes},'' \href{http://dx.doi.org/10.1103/PhysRevD.100.115033}{{\em
  Phys. Rev.} {\bf D100} (2019) no.~11, 115033},
\href{http://arxiv.org/abs/1910.06159}{{\tt arXiv:1910.06159 [hep-ph]}}.

\bibitem{Chivukula:2020hvi}
R.~S. Chivukula, D.~Foren, K.~A. Mohan, D.~Sengupta, and E.~H. Simmons,
  ``{Massive Spin-2 Scattering Amplitudes in Extra-Dimensional Theories},''
  \href{http://dx.doi.org/10.1103/PhysRevD.101.075013}{{\em Phys. Rev. D} {\bf
  101} (2020) no.~7, 075013}, \href{http://arxiv.org/abs/2002.12458}{{\tt
  arXiv:2002.12458 [hep-ph]}}.

\bibitem{Poland:2011ey}
D.~Poland, D.~Simmons-Duffin, and A.~Vichi, ``{Carving Out the Space of 4D
  CFTs},'' \href{http://dx.doi.org/10.1007/JHEP05(2012)110}{{\em JHEP} {\bf 05}
  (2012)  110},
\href{http://arxiv.org/abs/1109.5176}{{\tt arXiv:1109.5176 [hep-th]}}.

\bibitem{Kos:2014bka}
F.~Kos, D.~Poland, and D.~Simmons-Duffin, ``{Bootstrapping Mixed Correlators in
  the 3D Ising Model},'' \href{http://dx.doi.org/10.1007/JHEP11(2014)109}{{\em
  JHEP} {\bf 11} (2014)  109},
\href{http://arxiv.org/abs/1406.4858}{{\tt arXiv:1406.4858 [hep-th]}}.

\bibitem{Simmons-Duffin:2015qma}
D.~Simmons-Duffin, ``{A Semidefinite Program Solver for the Conformal
  Bootstrap},'' \href{http://dx.doi.org/10.1007/JHEP06(2015)174}{{\em JHEP}
  {\bf 06} (2015)  174},
\href{http://arxiv.org/abs/1502.02033}{{\tt arXiv:1502.02033 [hep-th]}}.

\bibitem{Landry:2019qug}
W.~Landry and D.~Simmons-Duffin, ``{Scaling the semidefinite program solver
  SDPB},''
\href{http://arxiv.org/abs/1909.09745}{{\tt arXiv:1909.09745 [hep-th]}}.

\bibitem{Warner82}
N.~P. Warner, ``{The Spectra of Operators on $CP^n$},''
  \href{http://dx.doi.org/10.1098/rspa.1982.0128}{{\em Proceedings of the Royal
  Society of London. Series A, Mathematical and Physical Sciences} {\bf 383}
  (1982) no.~1784, 217--230}.

\bibitem{Boucetta07}
M.~Boucetta, ``{Spectra and symmetric eigentensors of the Lichnerowicz
  Laplacian on $P^n(\mathbb{C})$},'' \href{http://arxiv.org/abs/0712.2830}{{\tt
  arXiv:0712.2830 [math-ph]}}.

\bibitem{Caracciolo:2009bx}
F.~Caracciolo and V.~S. Rychkov, ``{Rigorous Limits on the Interaction Strength
  in Quantum Field Theory},''
  \href{http://dx.doi.org/10.1103/PhysRevD.81.085037}{{\em Phys. Rev.} {\bf
  D81} (2010)  085037},
\href{http://arxiv.org/abs/0912.2726}{{\tt arXiv:0912.2726 [hep-th]}}.

\bibitem{Kos:2016ysd}
F.~Kos, D.~Poland, D.~Simmons-Duffin, and A.~Vichi, ``{Precision Islands in the
  Ising and $O(N)$ Models},''
  \href{http://dx.doi.org/10.1007/JHEP08(2016)036}{{\em JHEP} {\bf 08} (2016)
  036}, \href{http://arxiv.org/abs/1603.04436}{{\tt arXiv:1603.04436
  [hep-th]}}.

\bibitem{Chester:2019ifh}
S.~M. Chester, W.~Landry, J.~Liu, D.~Poland, D.~Simmons-Duffin, N.~Su, and
  A.~Vichi, ``{Carving out OPE space and precise $O(2)$ model critical
  exponents},'' \href{http://dx.doi.org/10.1007/JHEP06(2020)142}{{\em JHEP}
  {\bf 06} (2020)  142}, \href{http://arxiv.org/abs/1912.03324}{{\tt
  arXiv:1912.03324 [hep-th]}}.

\bibitem{Heidenreich:2015nta}
B.~Heidenreich, M.~Reece, and T.~Rudelius, ``{Sharpening the Weak Gravity
  Conjecture with Dimensional Reduction},''
  \href{http://dx.doi.org/10.1007/JHEP02(2016)140}{{\em JHEP} {\bf 02} (2016)
  140}, \href{http://arxiv.org/abs/1509.06374}{{\tt arXiv:1509.06374
  [hep-th]}}.

\bibitem{Rahman:2020qal}
R.~Rahman, ``{The Involutive System of Higher-Spin Equations},''
  \href{http://dx.doi.org/10.1016/j.nuclphysb.2021.115325}{{\em Nucl. Phys. B}
  {\bf 964} (2021)  115325}, \href{http://arxiv.org/abs/2004.13041}{{\tt
  arXiv:2004.13041 [hep-th]}}.

\bibitem{Berger1971}
M.~Berger, P.~Gauduchon, and E.~Mazet,
  \href{http://dx.doi.org/10.1007/BFb0064646}{{\em Le Spectre D'une
  Vari{\'e}t{\'e} Riemannienne}}.
\newblock Springer Berlin Heidelberg, 1971.

\bibitem{Higuchi:1986wu}
A.~Higuchi, ``{Symmetric Tensor Spherical Harmonics on the $N$ Sphere and Their
  Application to the De Sitter Group SO($N$,1)},''
\href{http://dx.doi.org/10.1063/1.527513}{{\em J. Math. Phys.} {\bf 28} (1987)
  1553}.

\bibitem{Chodos:1983zi}
A.~Chodos and E.~Myers, ``{Gravitational Contribution to the Casimir Energy in
  Kaluza-Klein Theories},''
  \href{http://dx.doi.org/10.1016/0003-4916(84)90039-3}{{\em Annals Phys.} {\bf
  156} (1984)  412}.

\bibitem{semmelmann2020}
U.~Semmelmann and G.~Weingart, ``Stability of compact symmetric spaces,''
  \href{http://arxiv.org/abs/2012.07328}{{\tt arXiv:2012.07328 [math.DG]}}.

\bibitem{grinberg83}
E.~L. Grinberg, ``{Spherical Harmonics and Integral Geometry on Projective
  Spaces},'' \href{http://dx.doi.org/10.2307/1999378}{{\em Transactions of the
  American Mathematical Society} {\bf 279} (1983) no.~1, 187--203}.

\bibitem{koiso1978}
N.~Koiso, ``{Nondeformability of Einstein metrics},''
  \href{http://dx.doi.org/10.18910/6578}{{\em Osaka Journal of Mathematics}
  {\bf 15} (1978) no.~2, 419--433}.

\bibitem{Wolf1976}
J.~A. Wolf and R.~S. Cahn, ``{Zeta Functions and Their Asymptotic Expansions
  for Compact Symmetric Spaces of Rank One},'' {\em Commentarii mathematici
  Helvetici} {\bf 51} (1976)  1--22.

\bibitem{Gangolli1967}
R.~Gangolli, ``{Positive definite kernels on homogeneous spaces and certain
  stochastic processes related to L\'evy's brownian motion of several
  parameters},'' {\em Annales de l'I.H.P. Probabilit\'es et statistiques} {\bf
  3} (1967) no.~2, 121--226.

\end{thebibliography}\endgroup

\end{document}